\documentclass[prb,amsmath,amssymb,reprint]{revtex4-1}

\usepackage{graphicx}
\usepackage{subfigure}
\usepackage{dcolumn}

\newcommand{\rmi}{\mathrm{i}}
\newcommand{\rmd}{\mathrm{d}}
\newcommand{\rme}{\mathrm{e}}
\newcommand{\rmh}{\mathrm{h}}
\newcommand{\bi}{\mathbf}
\newcommand{\fr}[2]{\frac{\displaystyle #1}{\displaystyle #2}}
\newcommand{\en}{\epsilon}
\newcommand{\rbr}[1]{\left( #1 \right)}
\newcommand{\cbr}[1]{\left[ #1 \right]}

\newcommand{\prodl}[2]{\prod\limits_{#1}^{#2}}
\newcommand{\sul}[2]{\sum\limits_{#1}^{#2}}
\newcommand{\pdiff}[2]{\fr{\partial #1}{\partial #2}}
\newcommand{\intg}[3]{\int\limits_{#1}^{#2}\rmd#3}
\newcommand{\ve}{\bi{v}}
\newcommand{\fref}[1]{Fig.~\ref{#1}}
\newcommand{\Fref}[1]{Fig.~\ref{#1}}
\newcommand{\sref}[1]{Sect.~\ref{#1}}
\newcommand{\Sref}[1]{Sect.~\ref{#1}}
\newcommand{\eref}[1]{(\ref{#1})}
\newcommand{\Eref}[1]{Equation (\ref{#1})}
\newcommand{\es}{e$^*$}
\newcommand{\hs}{h$^*$}

\newcommand{\earef}[2]{(\ref{#1}#2)}

\newcounter{mycounter}[equation]

\begin{document}
\title{Conductance and Thermopower of Ballistic Andreev Cavities}
\author{Thomas Engl}
\affiliation{Institut f\"ur Theoretische Physik, Universit\"at Regensburg, D-93040 Regensburg, Germany}
\author{Jack Kuipers}
\affiliation{Institut f\"ur Theoretische Physik, Universit\"at Regensburg, D-93040 Regensburg, Germany}
\author{Klaus Richter}
\affiliation{Institut f\"ur Theoretische Physik, Universit\"at Regensburg, D-93040 Regensburg, Germany}
\email{\mailto{Thomas.Engl@physik.uni-regensburg.de}, \mailto{Jack.Kuipers@physik.uni-regensburg.de}, \mailto{Klaus.Richter@physik.uni-regensburg.de}}
\date{\today}

\begin{abstract}
When coupling a superconductor to a normal conducting region the physical properties of the system are highly affected by the superconductor. We will investigate the effect of one or two superconductors on the conductance of a ballistic chaotic quantum dot to leading order in the total channel number using trajectory based semiclassics. The results show that the effect of one superconductor on the conductance is of the order of the number of channels and that the sign of the quantum correction from the Drude conductance depends on the particular ratios of the numbers of channels of the superconducting and normal conducting leads. In the case of two superconductors with the same chemical potential we additionally study how the conductance and the sign of quantum corrections are affected by their phase difference. As far as random matrix theory results exist these are reproduced by our calculations. Furthermore in the case that the chemical potential of the superconductors is the same as that of one of the two normal leads the conductance shows, under certain conditions, similar effects as a normal metal-superconductor junction. The semiclassical framework is also able to treat the thermopower of chaotic Andreev billiards consisting of one chaotic dot, two normal leads and two superconducting islands and shows it to be antisymmetric in the phase difference of the superconductors.
\end{abstract}

\maketitle

\section{Introduction}
\label{intro}
Transport problems have always attracted a lot of attention in condensed matter physics. While the Landauer-B\"uttiker formalism which connects the electrical current to the quantum transmission probabilities of a conductor is of key importance for transport through nanosystems, similar formulae have also been derived for hybrid structures consisting of normal conducting (N) regions connected to superconductors (S) \cite{btktheory,generalized_landauer,islandconductance} in which Andreev reflection \cite{ref:andreevrefl} plays a crucial role.

Andreev reflection \cite{ref:andreevrefl} can occur whenever a normal metal region is coupled to a superconductor. If an electron hits the normal metal-superconductor (N-S) interface with an energy closely above the Fermi energy an additional electron-hole pair can be created, and the two electrons enter the superconductor forming a Cooper pair. The hole however has to compensate the momentum of the original electron, therefore it retraces the electron path. Moreover the hole picks up a phase equal to the phase of the macroscopic superconducting wave function.

The early theoretical and experimental investigations of transport properties focused on the current through the interface of normal metal-superconductor, normal metal-insulator-superconductor (N-I-S) and S-N-S junctions \cite{transportreview}. For these the BTK-theory \cite{btktheory} applies, which is based on the Landauer type equation
\begin{equation}
 I=\fr{2e}{h}\Omega\intg{-\infty}{\infty}{\epsilon}\left[1-R_0+R_A\right]\cbr{f(\en-eV)-f(\en)},
 \label{btk_eq}
\end{equation}
where $I$ is the current through the N-S interface with an applied voltage $V$ and $\Omega$ a measure of the area of the junction. In \eref{btk_eq} $R_0$ is the probability for normal reflection, $R_A$ is the probability for Andreev reflection,  and $f$ is the Fermi function. The BTK theory predicts for N-S junctions with sufficiently large barrier strengths at the N-S interface that the differential conductance $\rmd I/\rmd V$ vanishes for voltages smaller than the superconducting gap $\Delta/e$. In this regime the conductance is doubled compared to the conductance of the same normal conducting region with a normal conducting lead instead of the superconducting one: an indication of the proximity effect \cite{proximity_exp,proximity_theo}. When increasing the voltage the differential conductance has a peak at $eV\approx\Delta$ and finally approaches the conductance of the normal conducting region without the superconductor. However, the total value of the current for high voltages exceeds that of a metallic junction by the so-called excess current. The early experiments on N-I-S junctions were in agreement with BTK-theory. However later experiments \cite{zba1,zba2} found an enhancement of the differential conductance at $V=0$ later known as the zero bias anomaly.

\begin{figure}
 \subfigure[\label{fig:andreev_dot}]{\includegraphics[width=0.45\columnwidth]{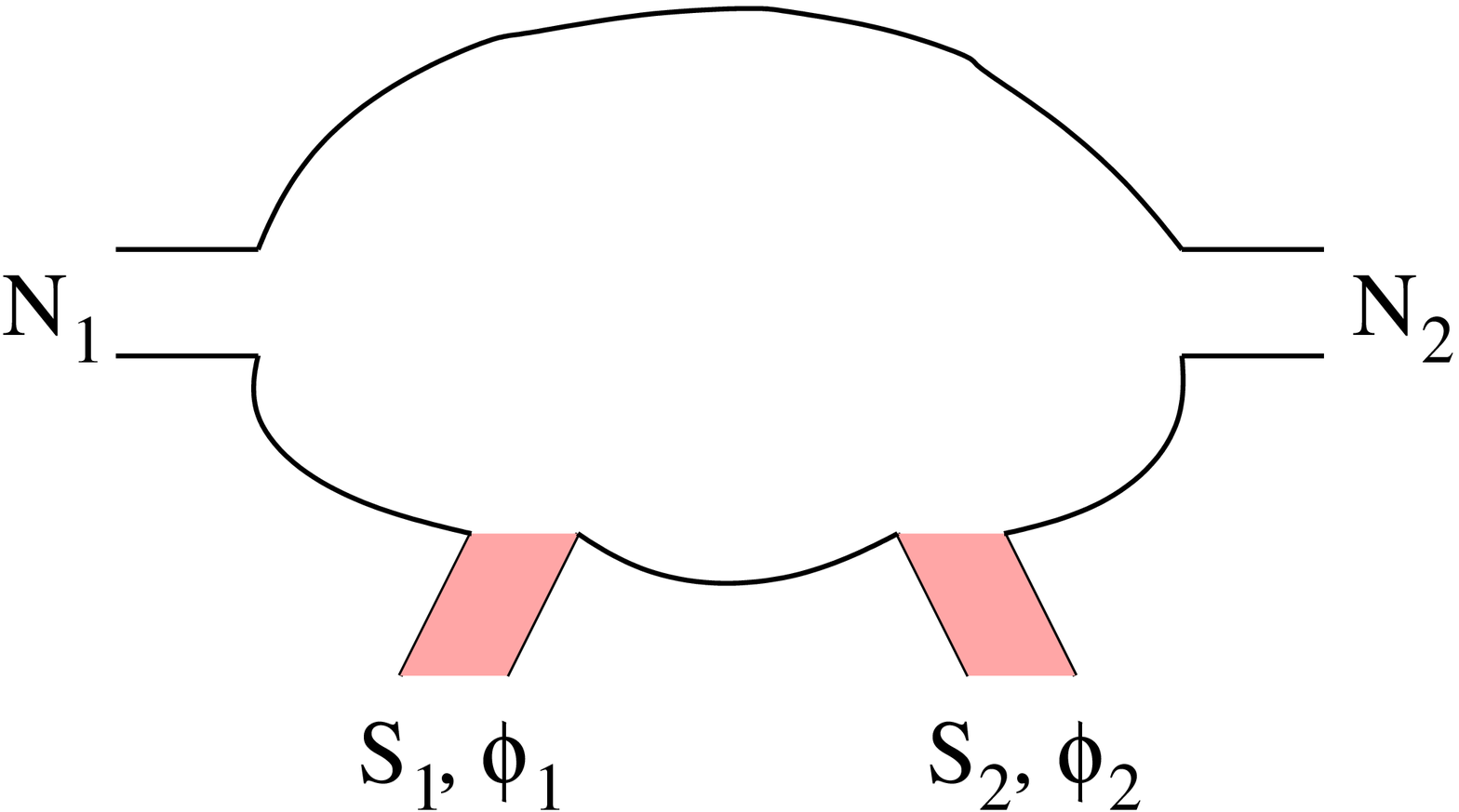}}
 \subfigure[\label{fig:parallelogram}]{\includegraphics[width=0.45\columnwidth]{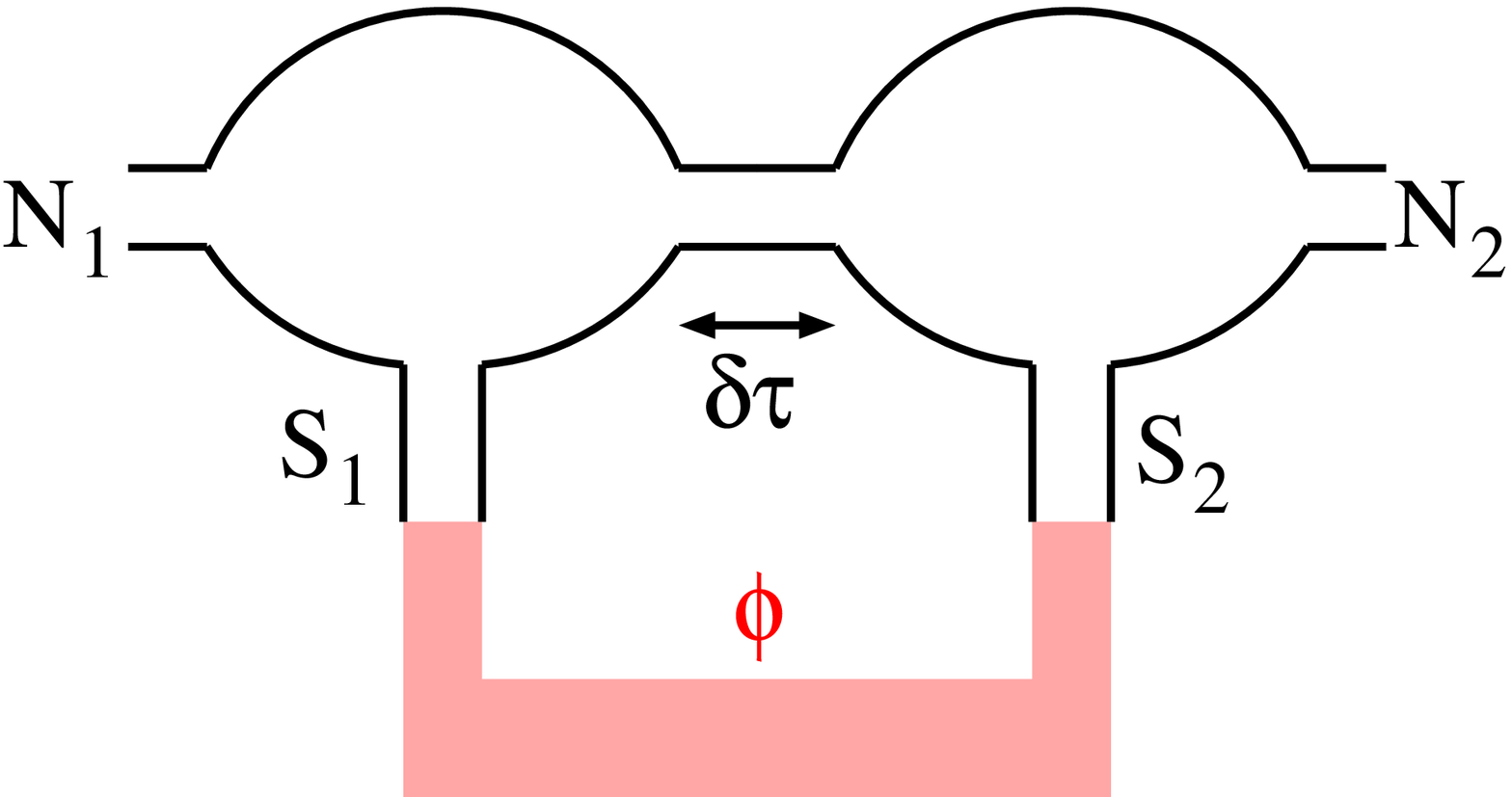}}
\\
 \subfigure[\label{island_setup}]{\includegraphics[width=0.45\columnwidth]{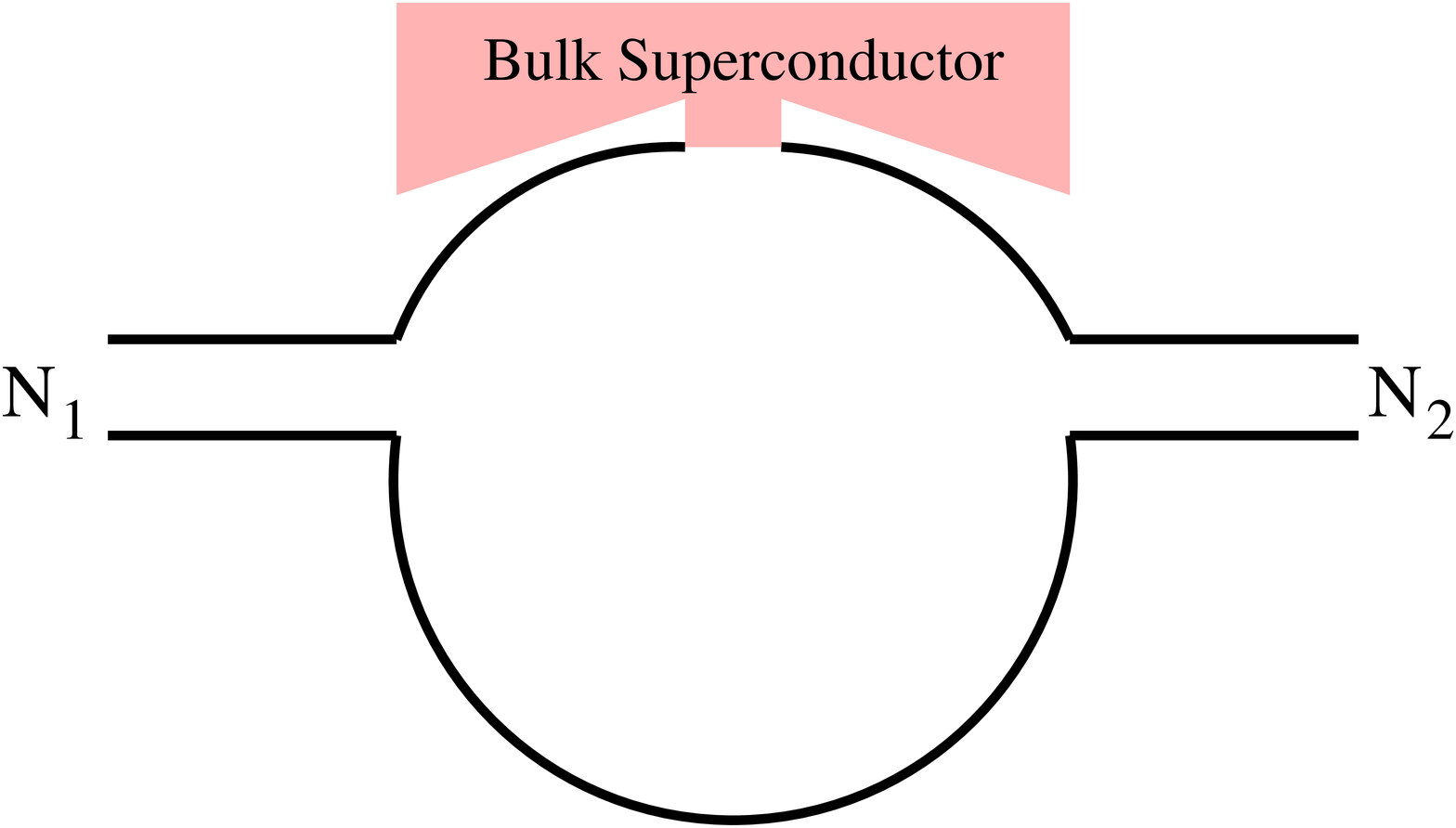}}
 \subfigure[\label{lead_setup}]{\includegraphics[width=0.45\columnwidth]{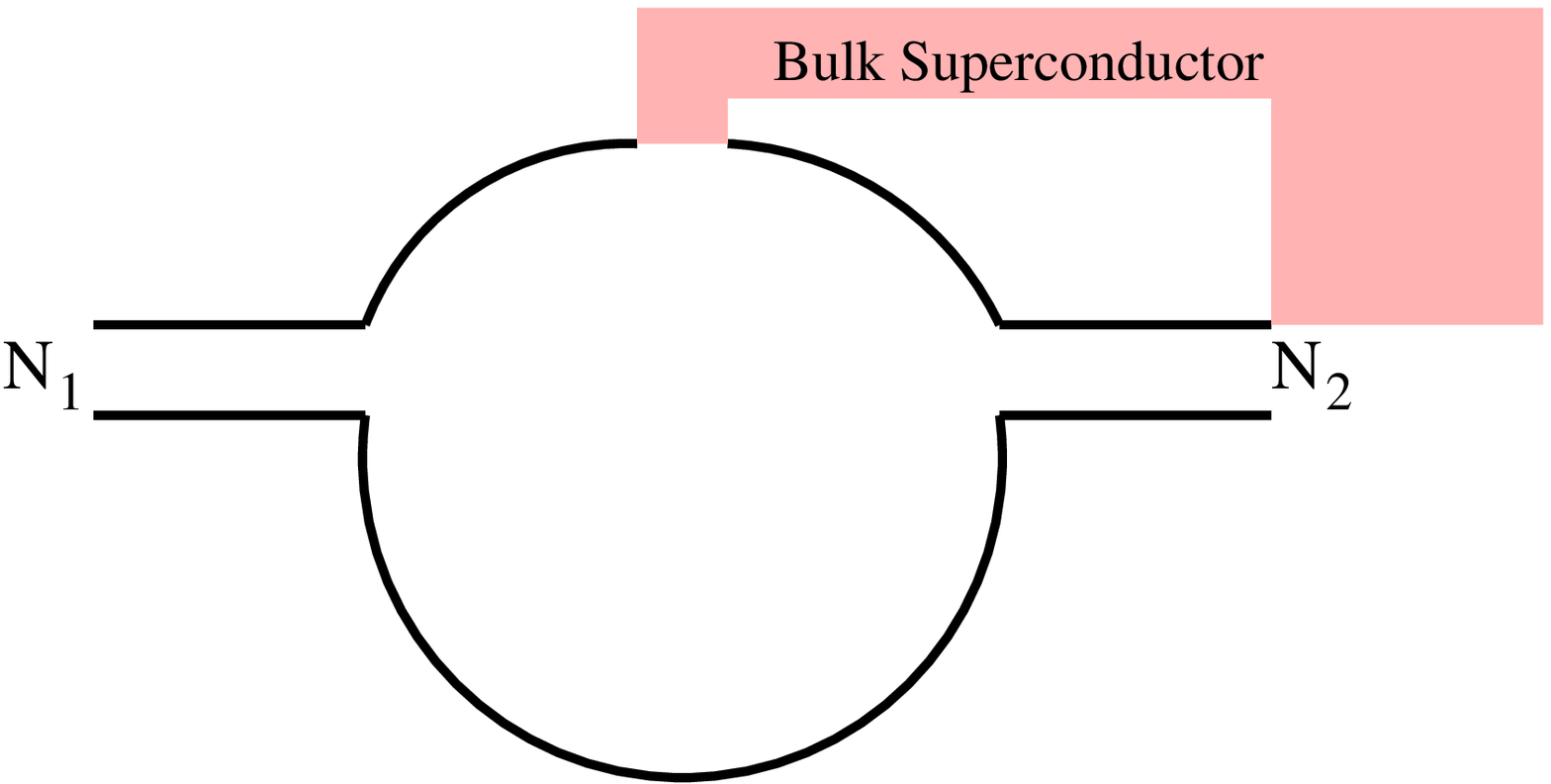}}
\\
\subfigure[\label{fig:thermo_setups_sym}]{\includegraphics[width=0.3\columnwidth]{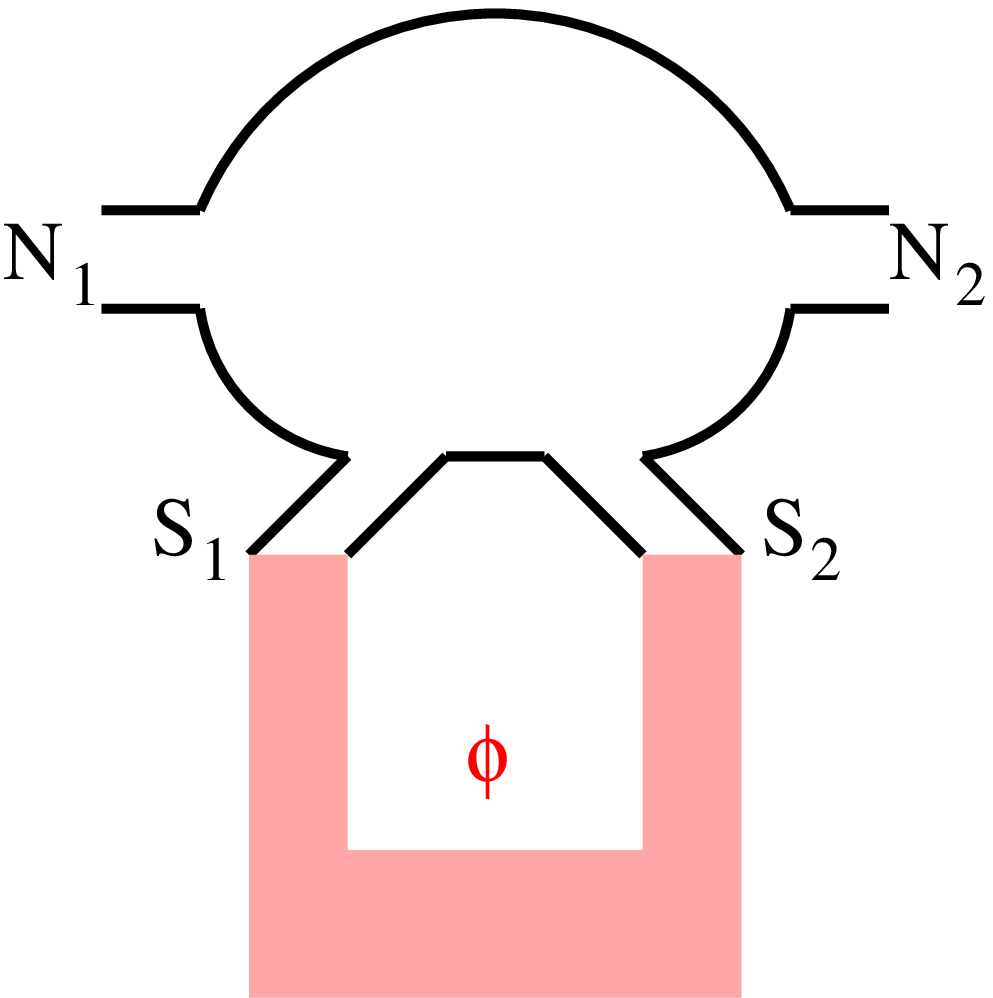}}
\rule{0.133\columnwidth}{0pt}
\subfigure[\label{fig:thermo_setups_asym}]{\includegraphics[width=0.3\columnwidth]{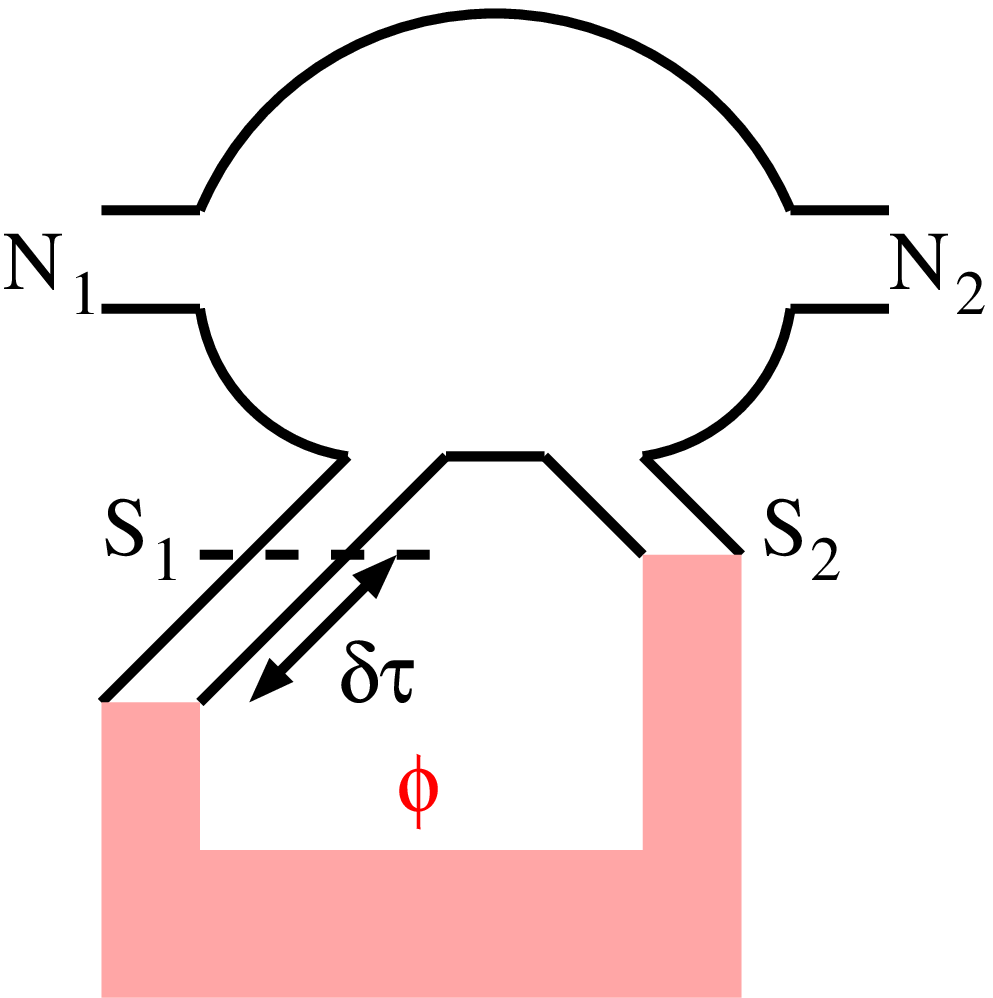}}
 \caption{\label{fig:andreev_setups}Schema of various Andreev billiards considered here: (a) Andreev interferometer with two superconducters, (b) double dot setup, (c) chaotic quantum dot coupled to one superconducting island. (d) the case of a superconducting lead with the same chemical potential as the right lead, (e) the so-called ``symmetric house'', (f) the ``asymmetric house'' where at lead $1$ a neck is additionally inserted compared to (e).}
\end{figure}

Recently Whitney and Jacquod \cite{nsntrans} considered a somewhat different type of setup. They considered a ballistic normal conducting region with a boundary giving rise to classically chaotic dynamics. Andreev reflection and interference between quasiparticles with slightly different paths lead to a hard gap in the density of states of such chaotic ballistic conductors coupled to a superconductor \cite{melsenetal97,semiclassical_dos,my_dos}. In Ref.~\onlinecite{nsntrans} such a chaotic Andreev quantum dotis coupled to two normal conducting and one superconducting lead and its transport characteristics was studied. Using a trajectory based semiclassical method they calculated the average conductance between the two normal leads of such chaotic Andreev billiards up to second order in the ratio $N_\mathrm{S}/N_\mathrm{N}$ where $N_\mathrm{S}$ is the total number of superconducting channels and $N_\mathrm{N}=N_1+N_2$ the sum of the number of channels in the normal leads. If the superconducting chemical potential is the same as that of one of the two normal conducting leads (abbreviated to `superconducting lead' and depicted in \Fref{lead_setup}) they found that the correction to the classical conductance be huge (of order of $N=N_\mathrm{N}+N_\mathrm{S}$) compared to usual weak localization effects, in particular it was shown that the quantum correction may become negative or positive depending on the ratio $N_1/N_2$. A similar change in the sign of the quantum correction to the conductance may be caused by a change in the transparencies of the leads \cite{ref:imperfect_leads}.

Using the same approach Whitney and Jacquod showed furthermore that to leading order in $N_\mathrm{S}/N_\mathrm{N}$ the thermopower of a chaotically shaped normal metal quantum dot with two normal leads and two superconducting islands (called a ``symmetric house'' and depicted in \fref{fig:thermo_setups_sym}) is antisymmetric in the phase difference of the superconductors. They also argued that the thermopower vanishes if the two superconductors carry the same amount of channels as long as no symmetry breaking neck is inserted at one of the two superconductors (\textit{c.f.}~\fref{fig:thermo_setups_asym}).

Here we combine the trajectory based semiclassical approaches of Refs.~\onlinecite{nsntrans},~\onlinecite{semiclassical_dos},~\onlinecite{my_dos} and~\onlinecite{countingstatistic} and provide a comprehensive calculation of the conductance and the thermopower of Andreev billiards. In Refs.~\onlinecite{semiclassical_dos} and~\onlinecite{my_dos} a method was developed for the systematic evaluation of multiple sums over electron and hole type orbits arising in a semiclassical approach to the proximity effect on the density of states of Andreev billiards. Here we further extend this recent approach to the conductance. To this end a diagonal backbone is introduced which is given by a path and its complex conjugated. The quantum correction in leading order in $1/N$ is then obtained by attaching an even number of so called trees (or complex conjugated trees) as those used in Ref.~\onlinecite{countingstatistic}. In this diagrammatical language, in Ref.~\onlinecite{nsntrans} the authors restricted themselves to at most two trees consisting of just one path pair. Therefore their results are valid only for small $N_\mathrm{S}/N_\mathrm{N}$ and the validity of their results for larger $N_\mathrm{S}$ is not known. Unlike the results in Ref.~\onlinecite{ref:conductance_distribution}, where the authers considered the distribution of the conductance of chaotic quantum dots with one open channel per lead, our results will be valid for large numbers of channels in the normal leads.
\renewcommand{\thesubfigure}{}
\begin{figure}
\hfill \subfigure{\includegraphics[width=0.45\columnwidth]{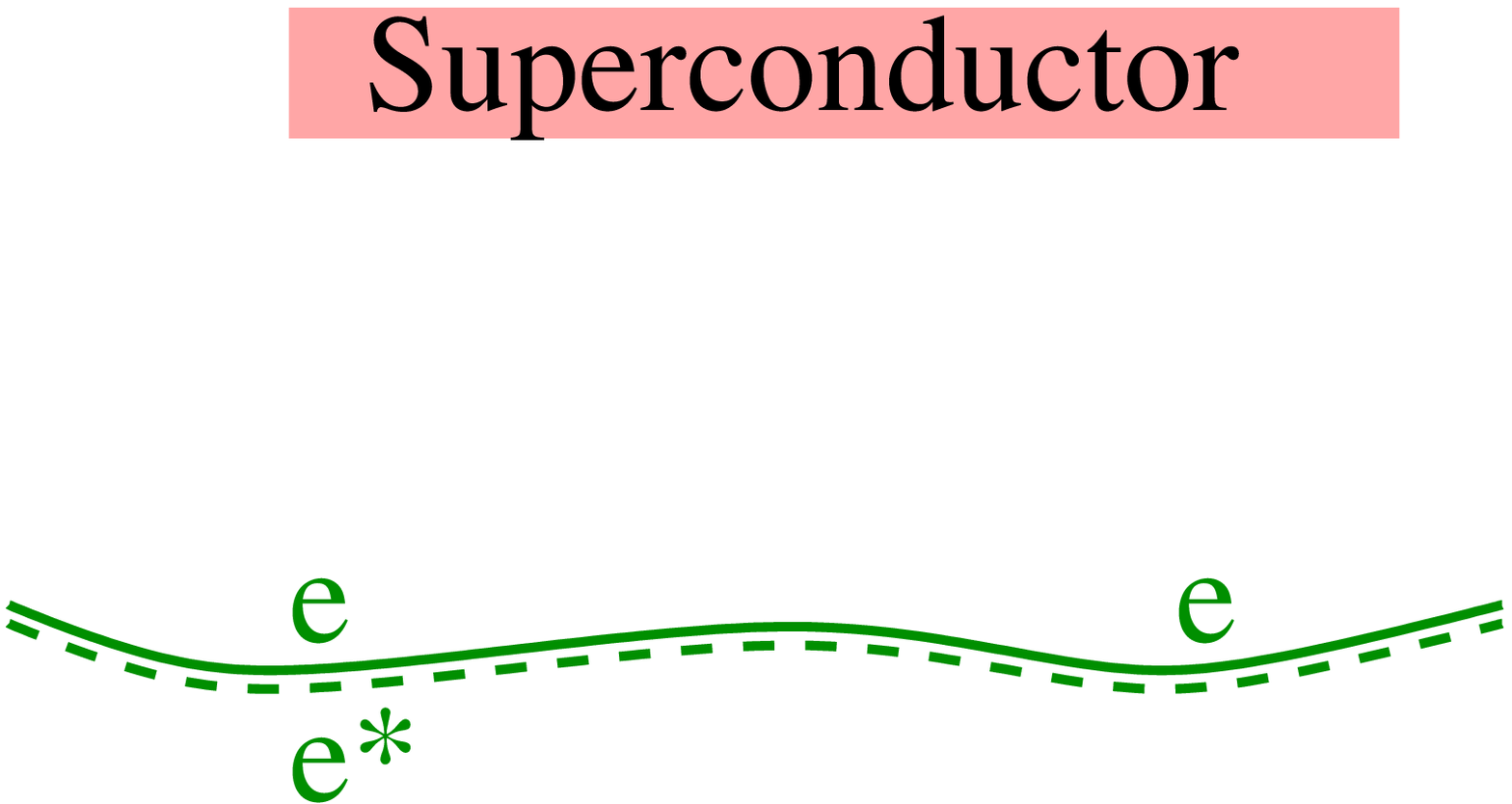}}\hfill
\subfigure{\includegraphics[width=0.45\columnwidth]{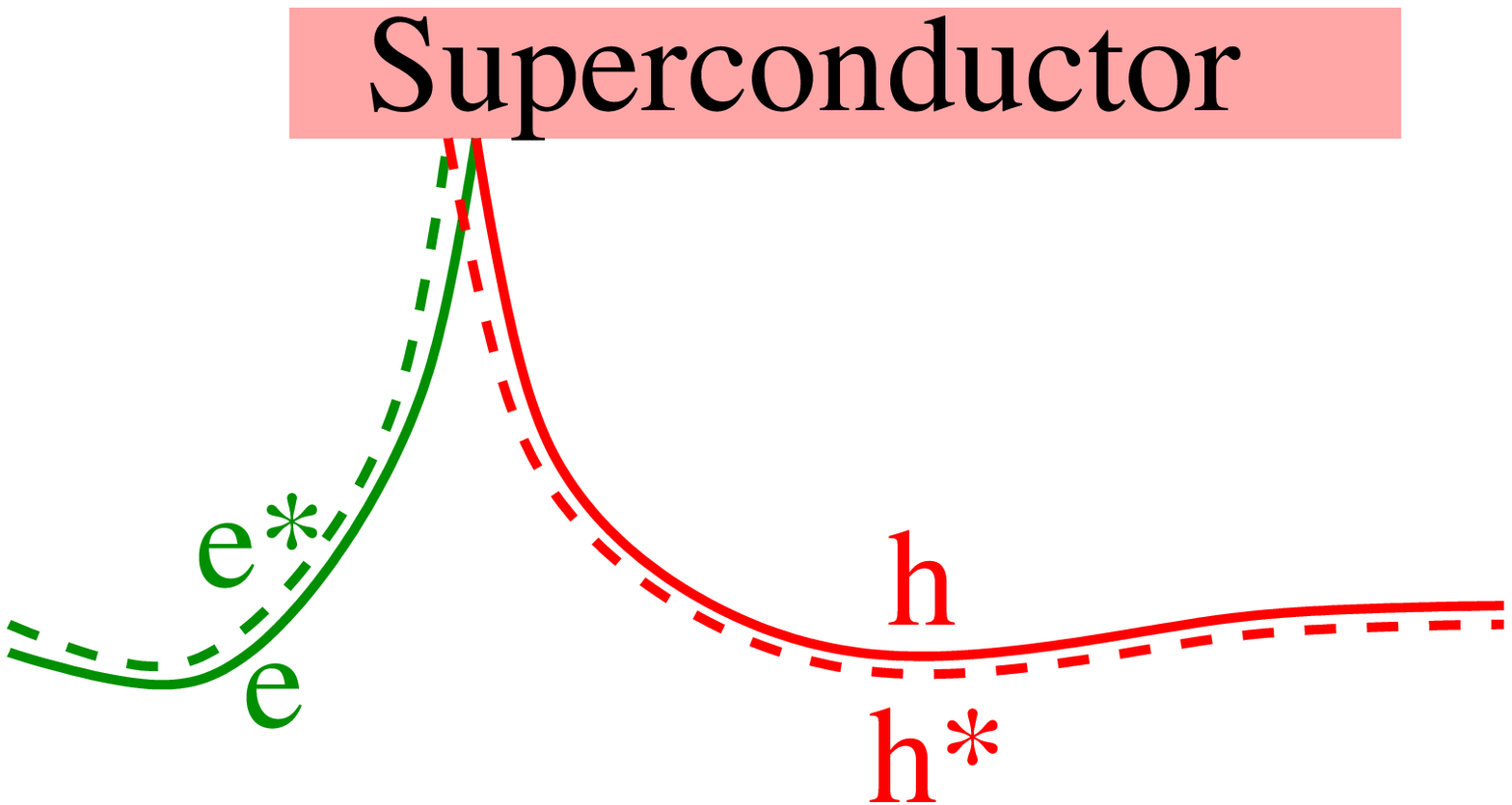}}\hfill\\
\hfill\subfigure{\includegraphics[width=0.45\columnwidth]{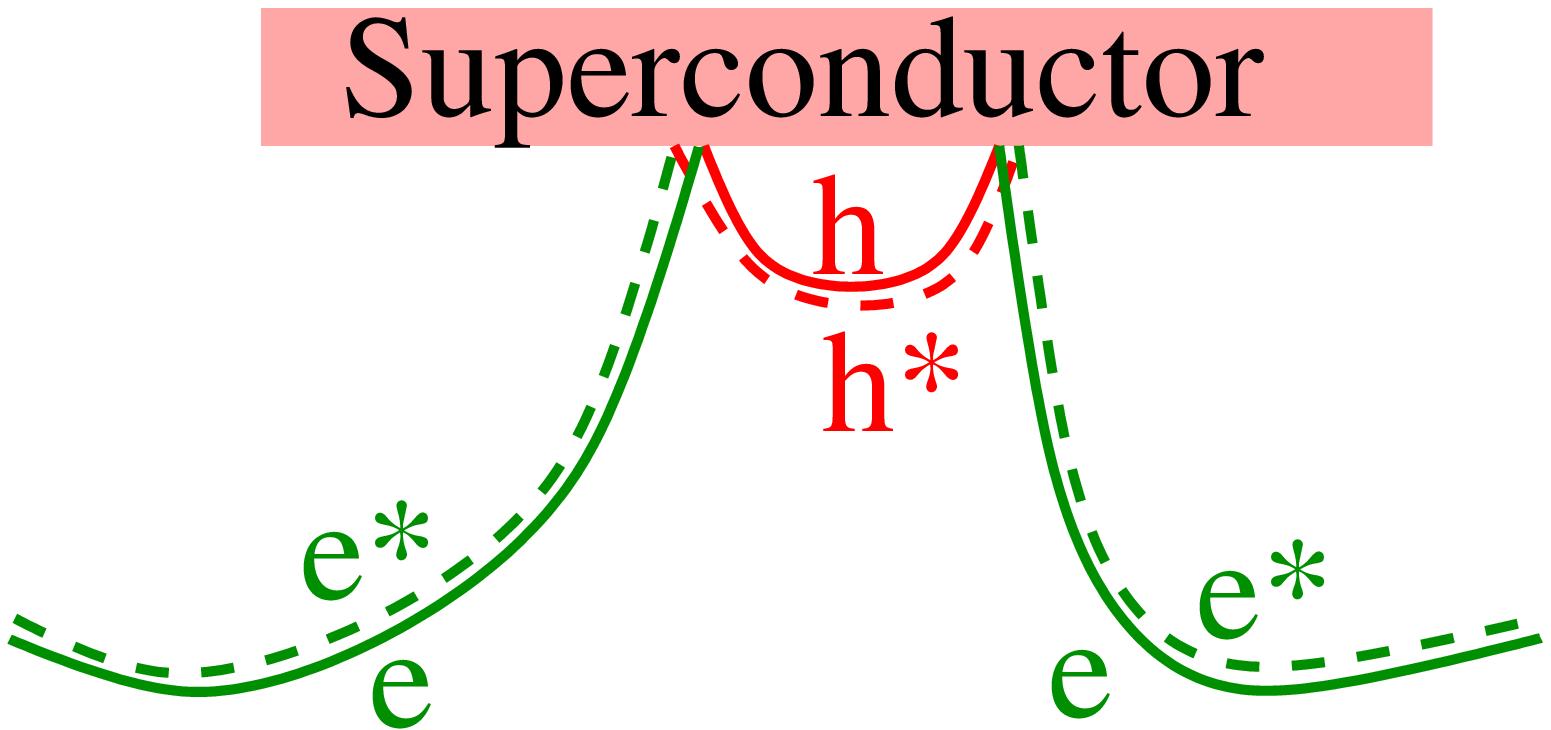}}\hfill
\subfigure{\includegraphics[width=0.45\columnwidth]{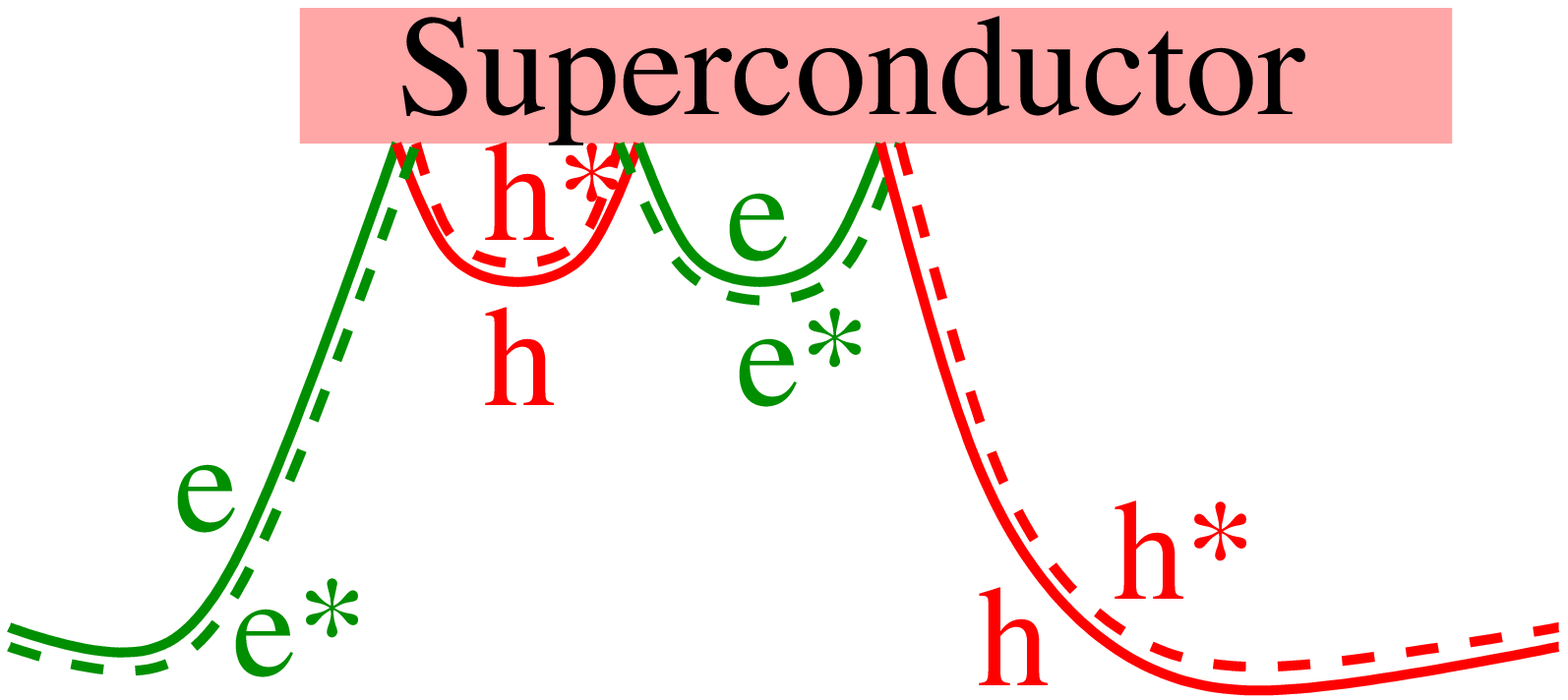}}\hfill
\caption{\label{fig:diagonal_contr_cond}The diagonal diagrams contributing to the conductance up to third order in the number $N_\mathrm{S}$ of channels of the superconductor. Here, e and h denote electron type and hole type quasiparticles and the asterisk denotes that the path enters the calculations with the complex conjugated factors.}
\end{figure}
\renewcommand{\thesubfigure}{(\alph{subfigure})}

We will derive the conductance of the two setups in Ref.~\onlinecite{nsntrans} - namely the setup with a superconducting island (see \fref{island_setup}) and with a superconducting lead (see \fref{lead_setup}) - to all orders in $N_\mathrm{S}/N_\mathrm{N}$.
To this end we start in Sects. \ref{sec:cond_diagrams} - \ref{sec:cond_transmission_probs} by considering the semiclassical diagrams and their contribution to the transmission probabilities and thus to the conductance to leading order in $1/N$. In \Sref{island_sec} we apply this approach to the setup with a superconducting island. We show that our semiclassical result for the conductance coincides with previous random matrix theory results \cite{rmttrans} existing for zero magnetic field and temperature (though still with a phase difference $\phi=\phi_1-\phi_2$ between the superconductors). We furthermore consider the magnetic field and temperature dependence of the conductance of setup \fref{island_setup}.
\renewcommand{\thesubfigure}{}
\begin{figure*}
 \subfigure[ee3I]{\raisebox{-\height}{\includegraphics[width=0.18\textwidth]{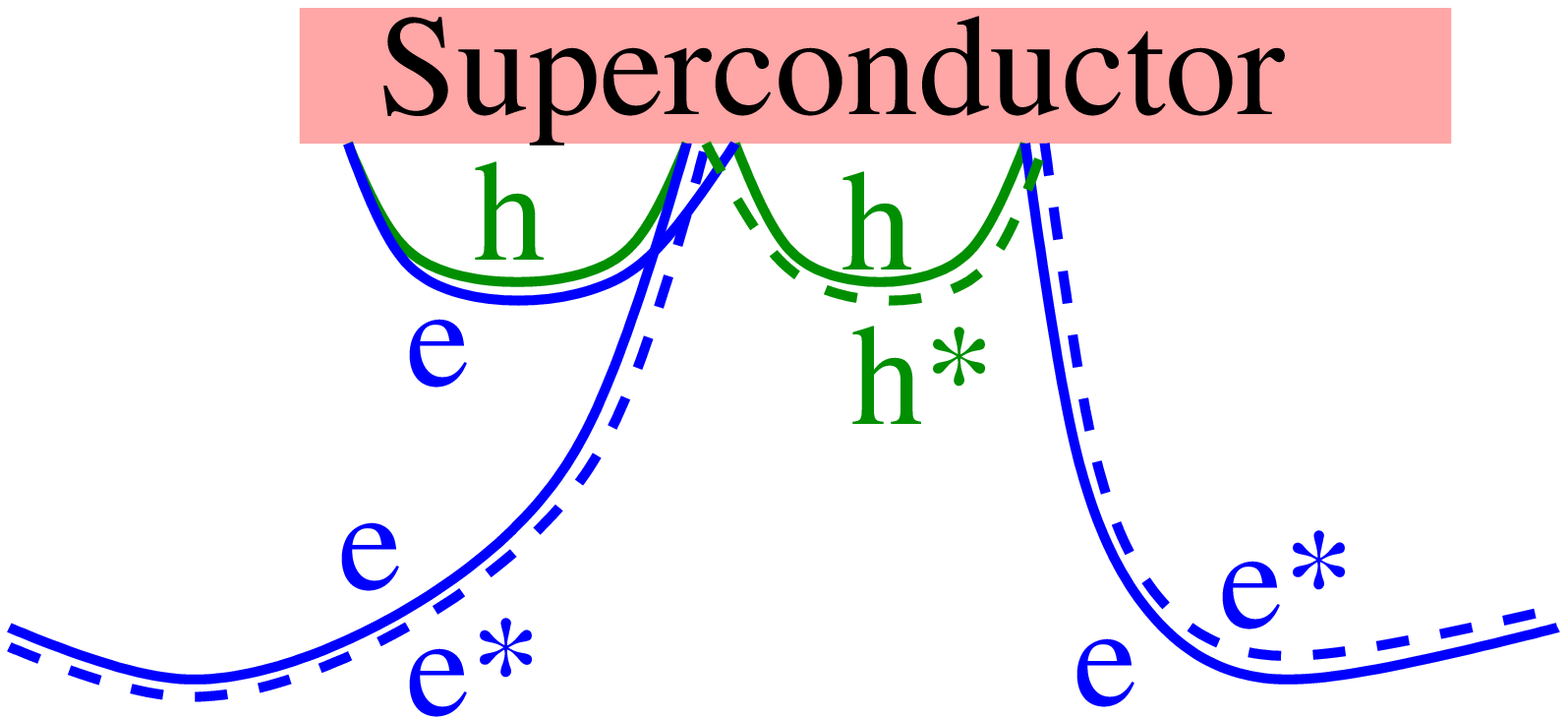} }\raisebox{-\height}{\rule{0pt}{52pt}}}
 \subfigure[ee3II]{\raisebox{-\height}{\includegraphics[width=0.18\textwidth]{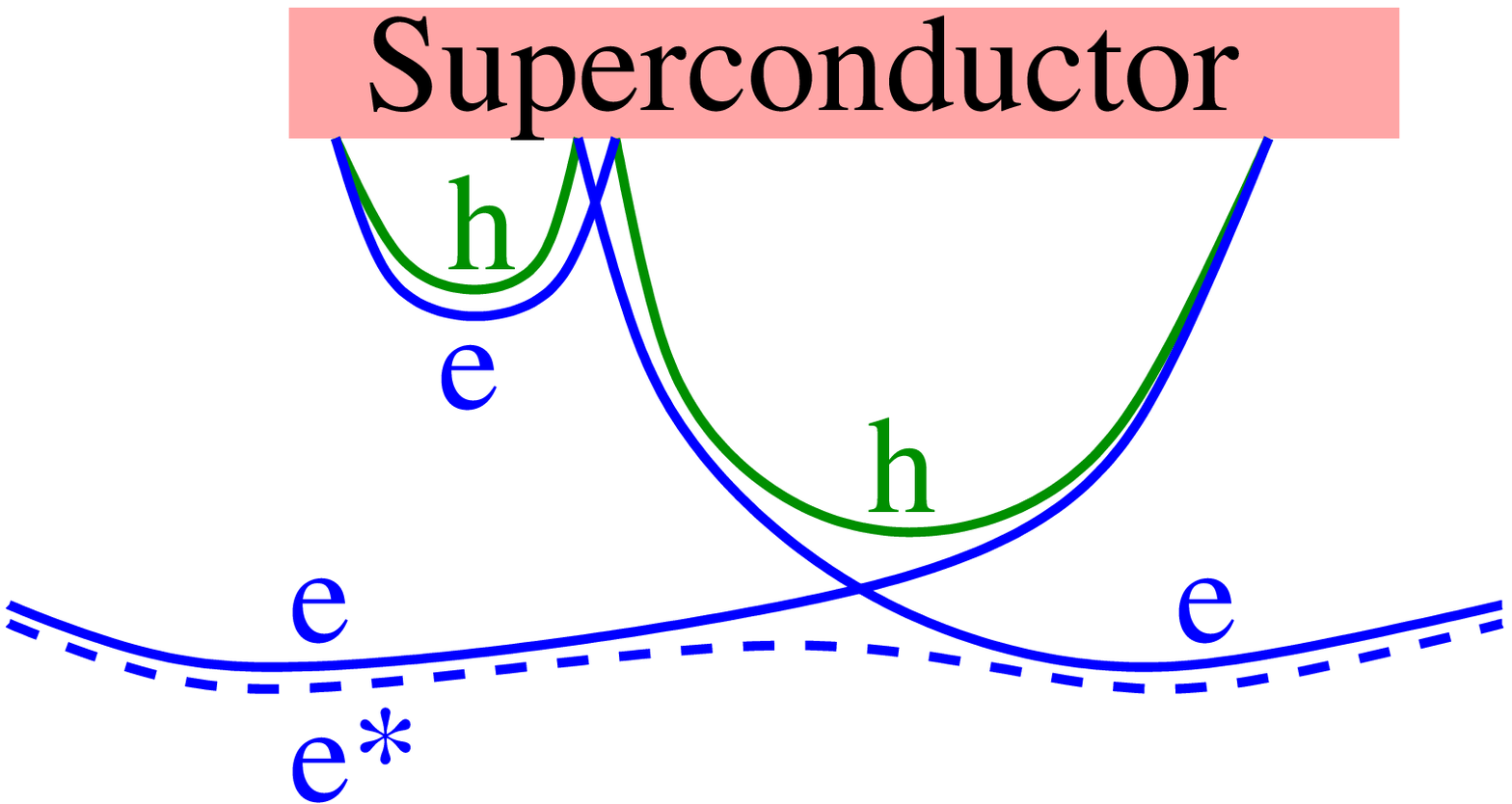} }\raisebox{-\height}{\rule{0pt}{52pt}}}
 \subfigure[ee3III]{\raisebox{-\height}{\includegraphics[width=0.18\textwidth]{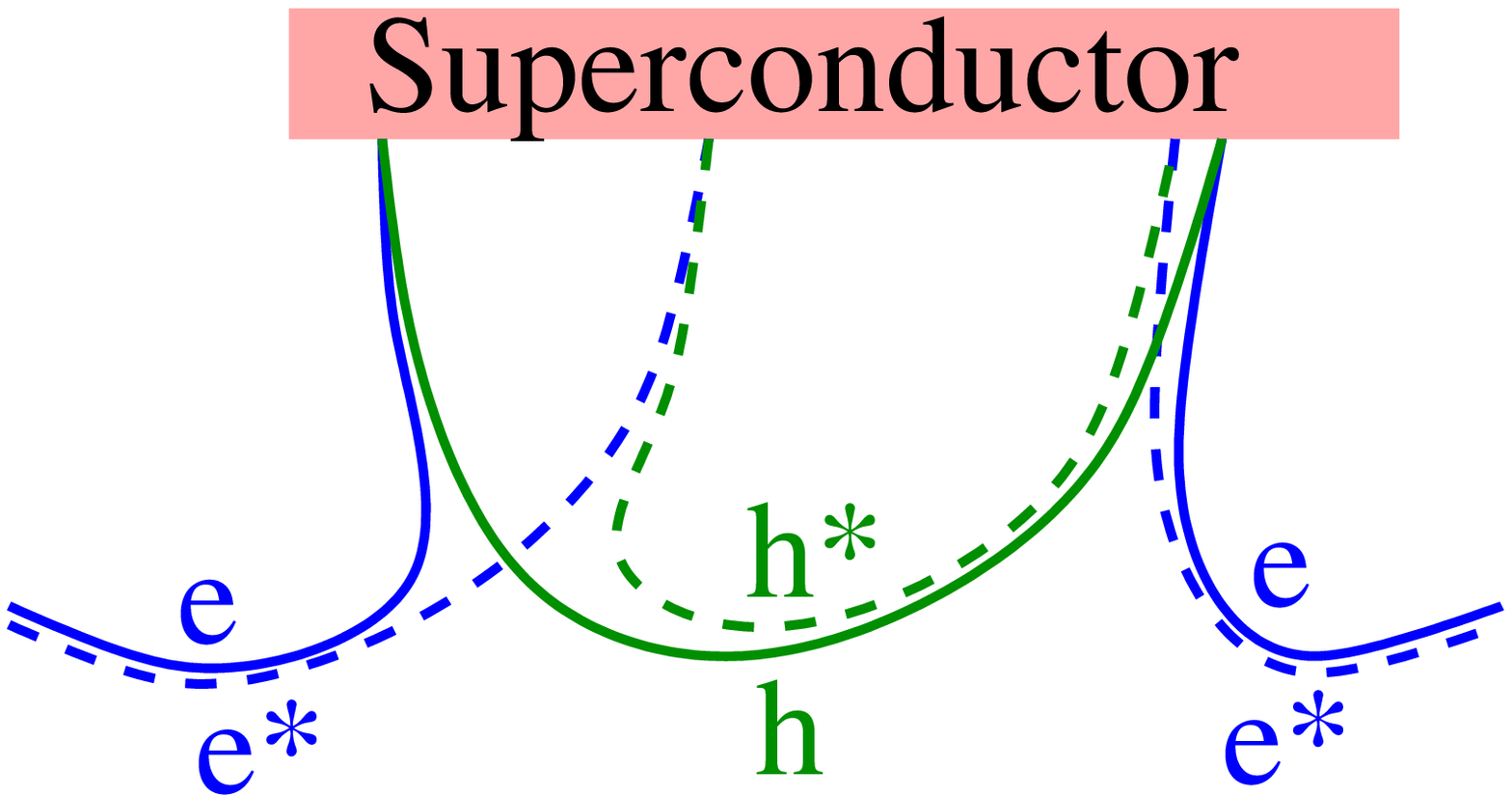} }\raisebox{-\height}{\rule{0pt}{52pt}}}
 \subfigure[he3I]{\raisebox{-\height}{\includegraphics[width=0.18\textwidth]{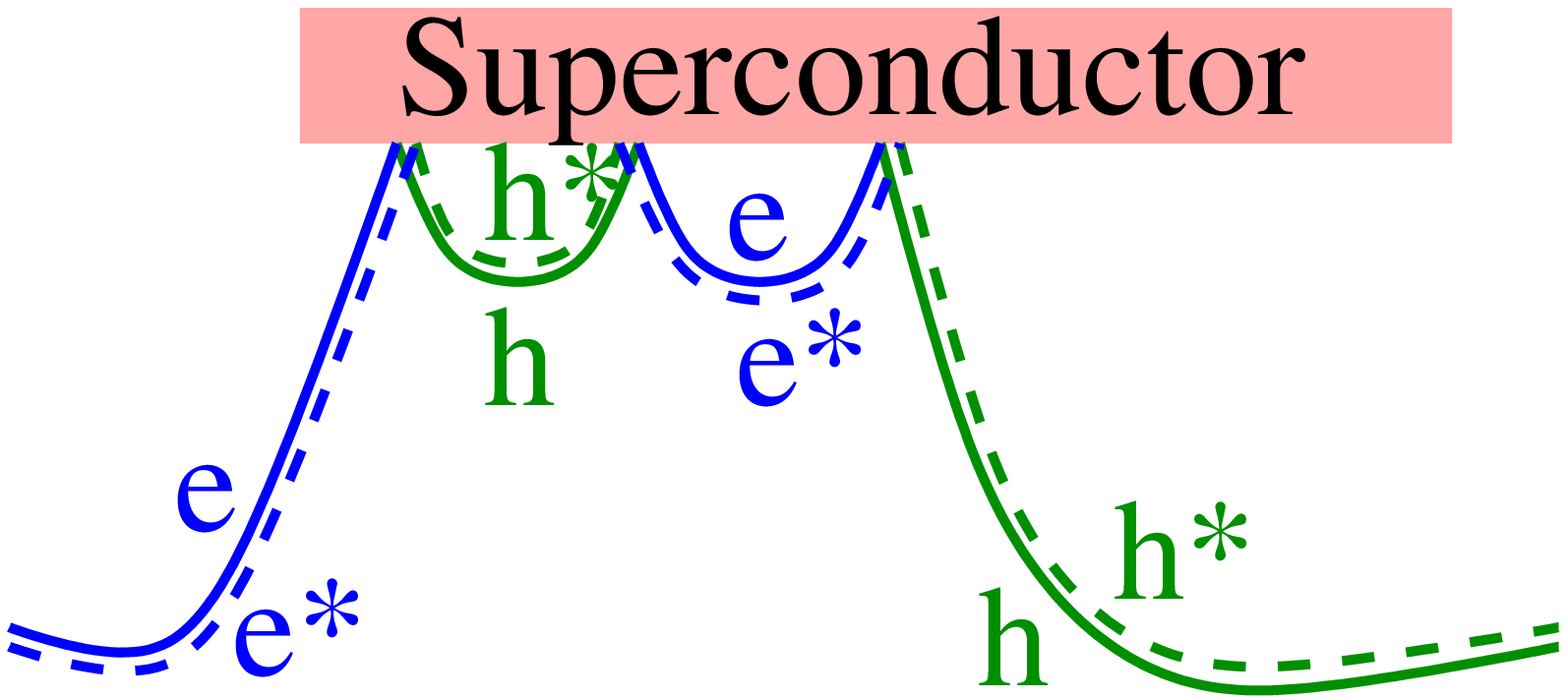} }\raisebox{-\height}{\rule{0pt}{52pt}}}
 \subfigure[he3II]{\raisebox{-\height}{\includegraphics[width=0.18\textwidth]{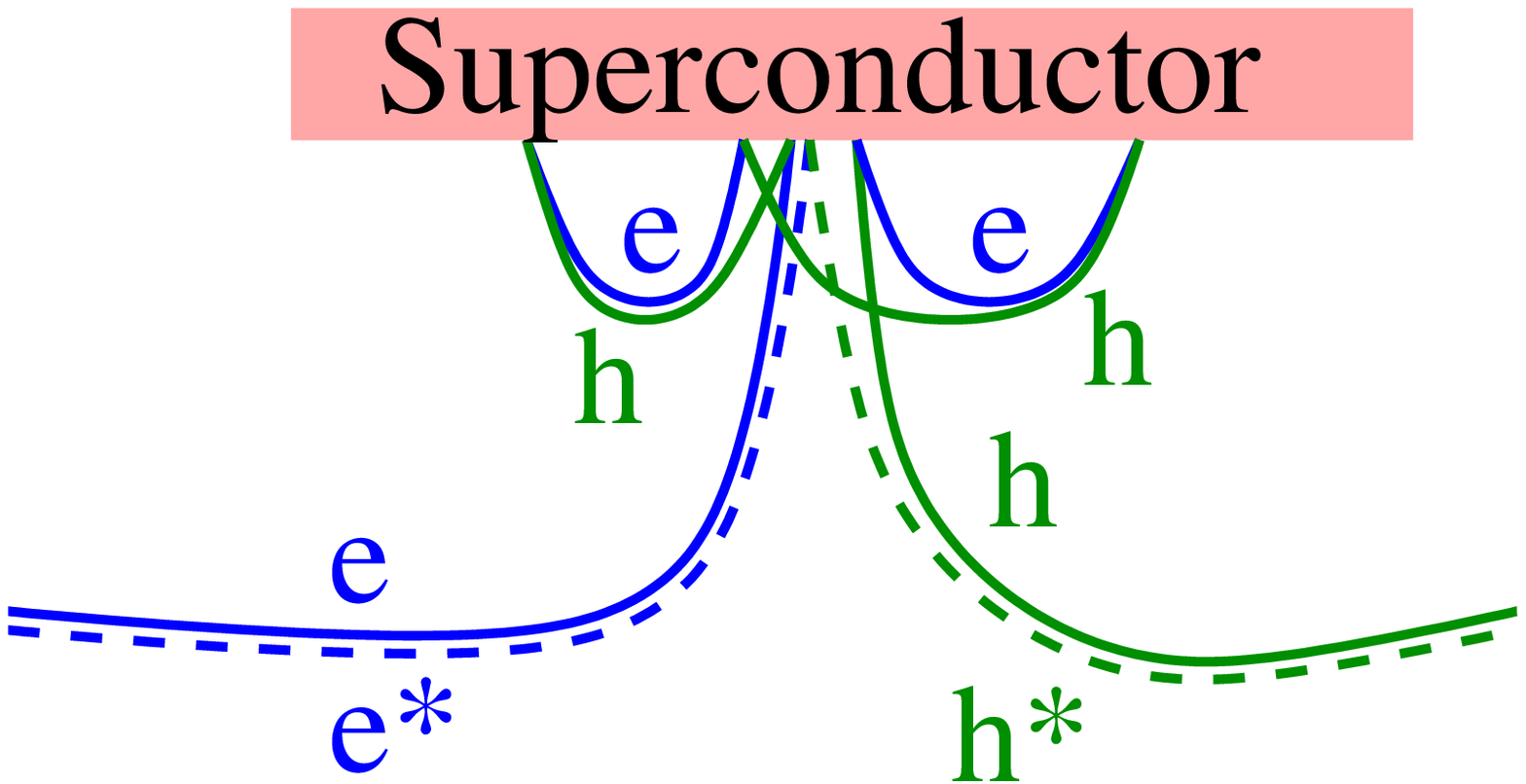} }\raisebox{-\height}{\rule{0pt}{52pt}}}\\
 \subfigure[he3III]{\raisebox{-\height}{\includegraphics[width=0.18\textwidth]{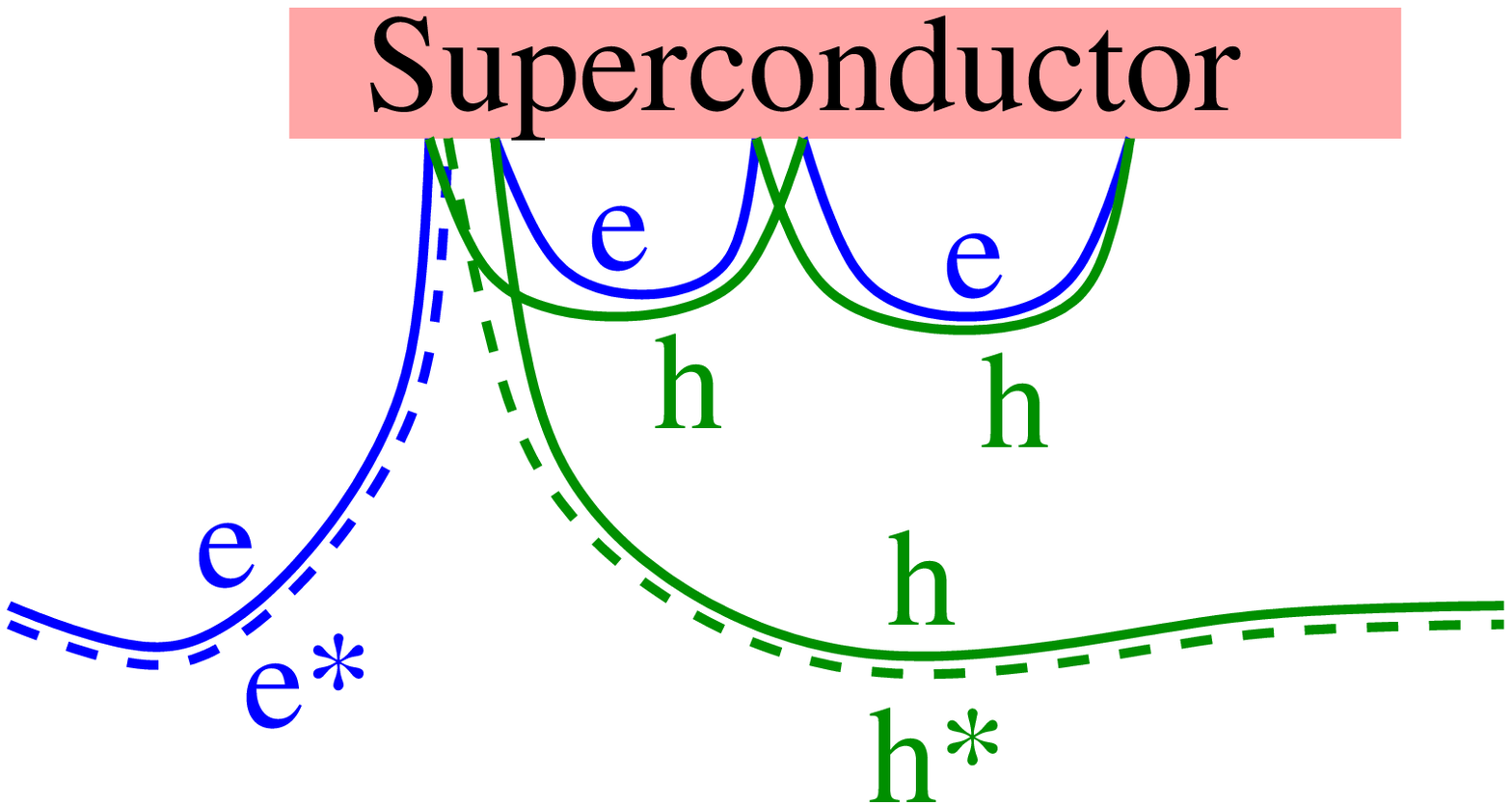} }\raisebox{-\height}{\rule{0pt}{52pt}}}
 \subfigure[he3IV]{\raisebox{-\height}{\includegraphics[width=0.18\textwidth]{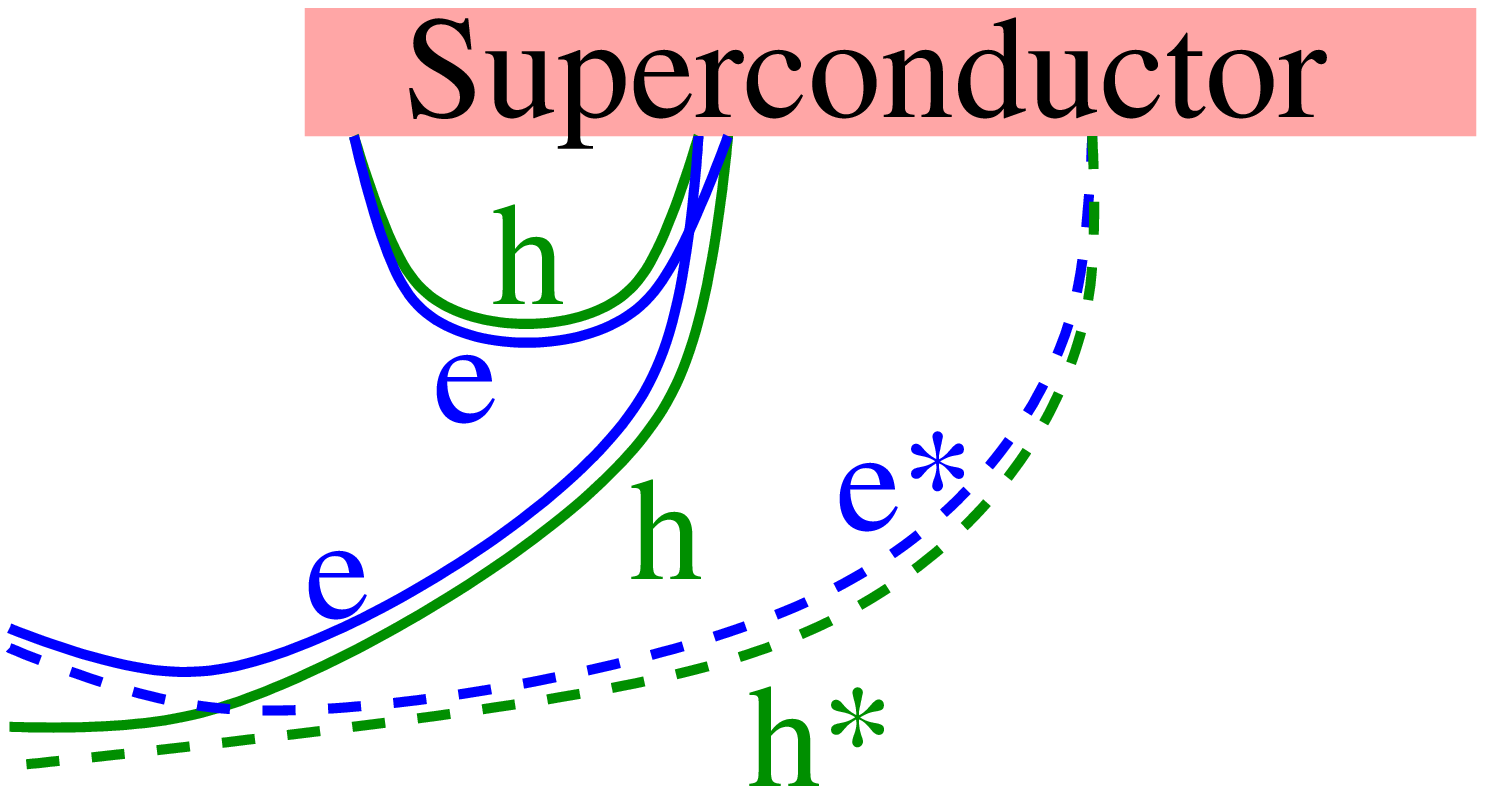} }\raisebox{-\height}{\rule{0pt}{52pt}}}
 \subfigure[{he3V}]{\raisebox{-\height}{\includegraphics[width=0.18\textwidth]{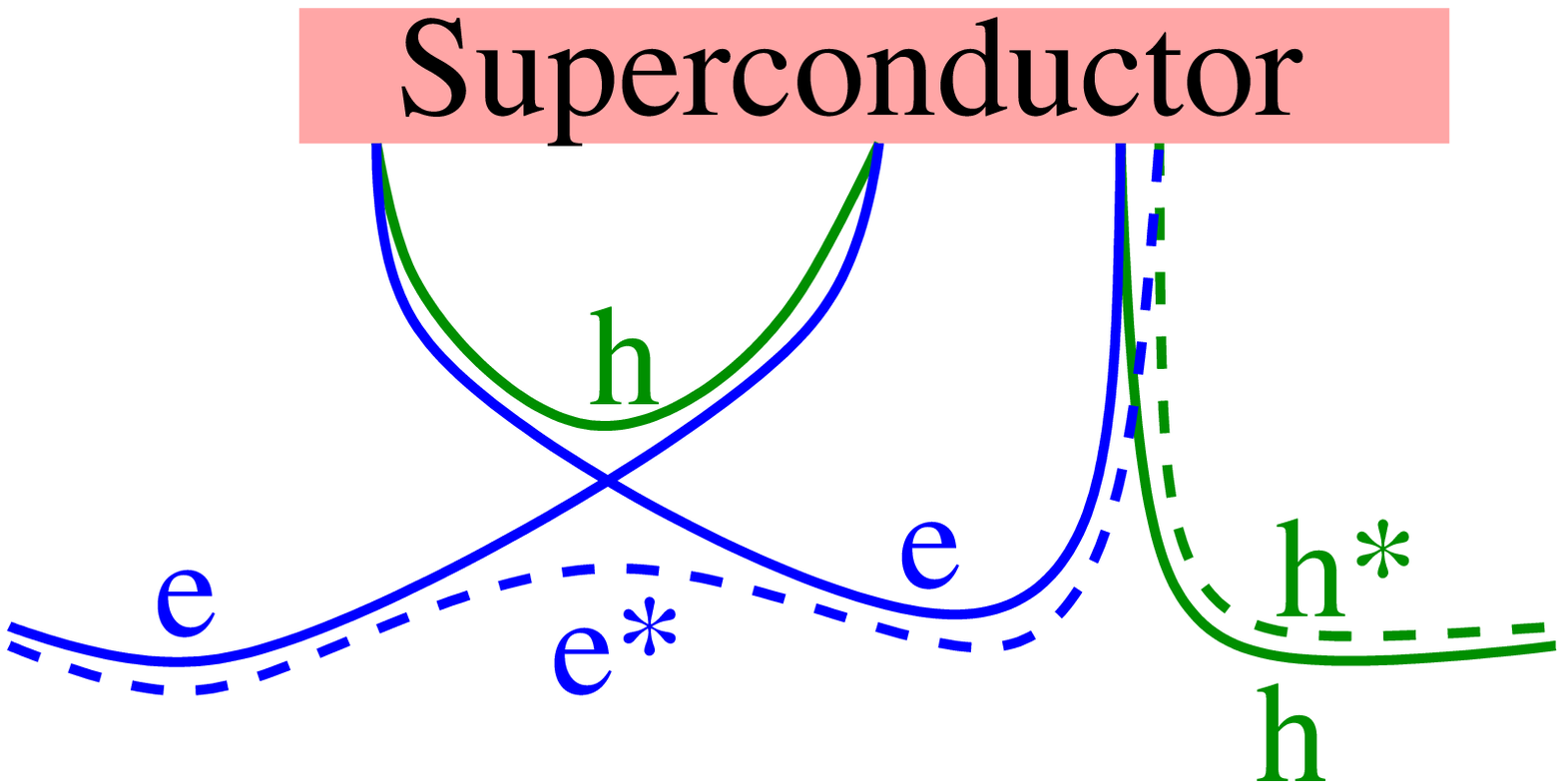} }\raisebox{-\height}{\rule{0pt}{52pt}}}
 \subfigure[he3VI]{\raisebox{-\height}{\includegraphics[width=0.18\textwidth]{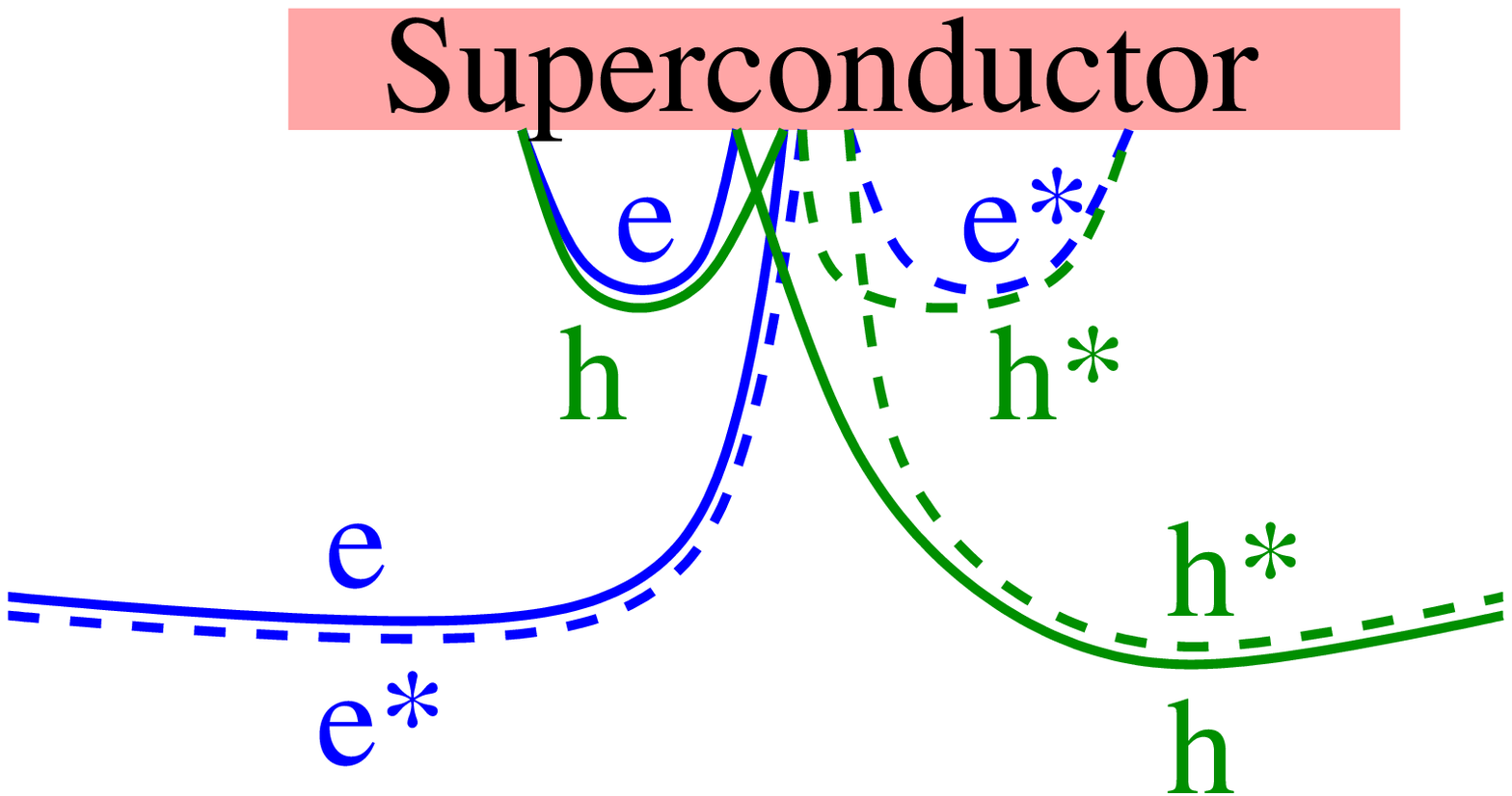} }\raisebox{-\height}{\rule{0pt}{52pt}}}
 \subfigure[he3VII]{\raisebox{-\height}{\includegraphics[width=0.18\textwidth]{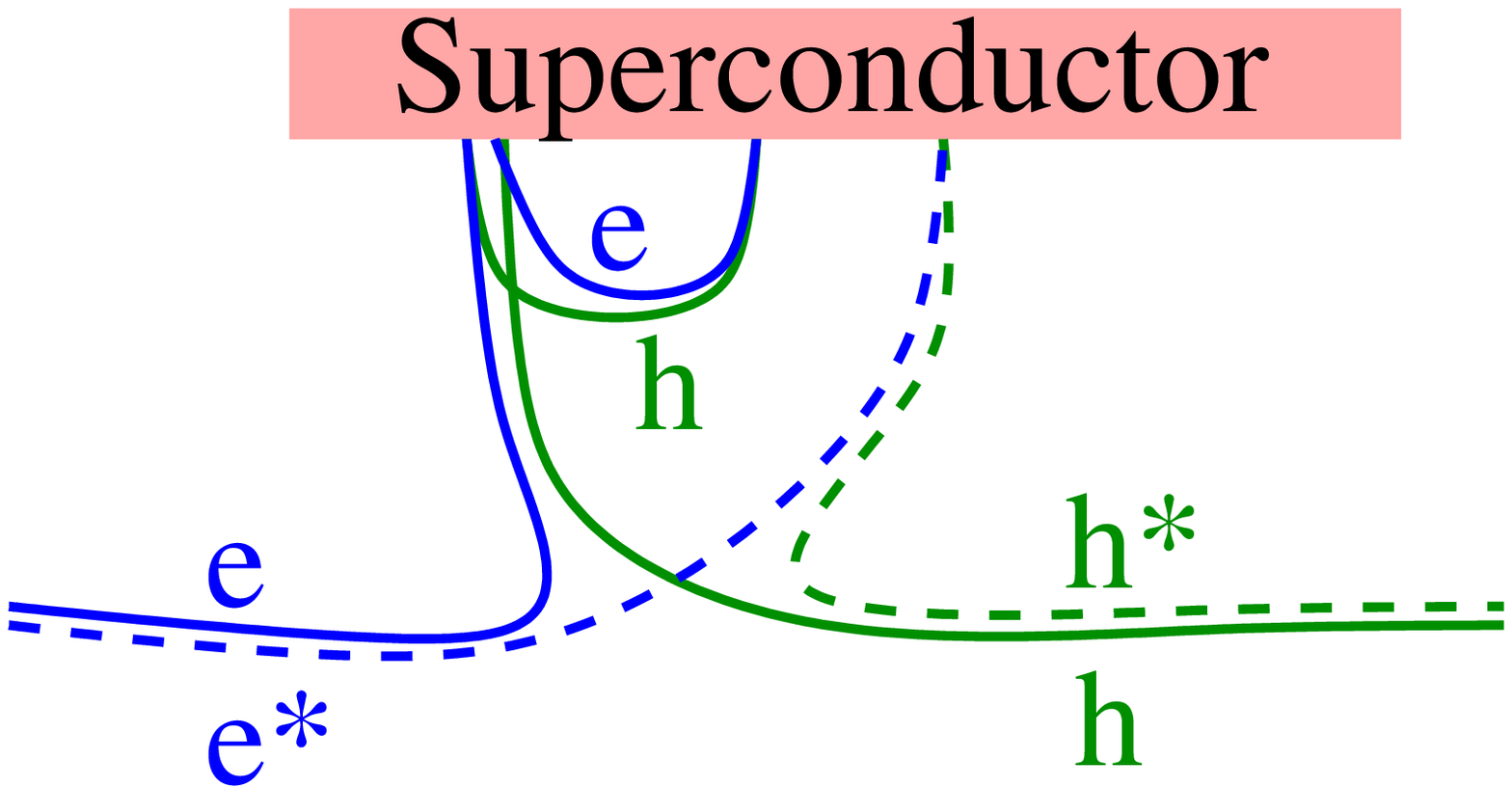} }\raisebox{-\height}{\rule{0pt}{52pt}}}
 \caption{\label{3rd_order}Pairs of paths contributing to the third-order term in $x=N_\mathrm{S}/N_\mathrm{N}$ of the transmission. Electron (hole) paths are green (red). The solid (dashed) lines belong to $\gamma$ ($\gamma^\prime$). A trajectory pair entering from the left and exiting to the right can connect the two normal conducting leads while a trajectory pair entering and exiting at the same side can only contribute if the incoming and outgoing channel belong both to the same lead.}
\end{figure*}
\renewcommand{\thesubfigure}{(\alph{subfigure})}

For the other setup of an Andreev billiard coupled to one or two separate superconducting leads (\fref{lead_setup}) we will show as a main result in \sref{lead_sec} that the quantum correction to the classical value of the conductance changes its sign not only with the ratio of the number of channels in the two normal conducting leads $N_1/N_2$ but also by tuning the ratio $x=N_\mathrm{S}/N_\mathrm{N}$. This sign change wa not anticipated in Ref.~\onlinecite{nsntrans}, since it requires an analysis to higher orders in $x$. This conductance correction is also shown to oscillate with the phase difference $\phi$ between the two superconducting leads with period $2\pi$. Finally, we study the dependence of the conductance on an applied magnetic field and temperature. The effects we observe for some combinations of the ratios $x$ and $N_1/N_2$ turn out to be fairly similar to those found in the structures containing only one normal conducting lead.
%

In \sref{sec:parallelogram_conductance} we show how the methods derived before can be extended to calculate the transmission coefficients of two dots connected to each other by a neck wehre each dot has one further normal and one superconducting lead (see \fref{fig:parallelogram}). The conductance of this setup is shown to also be symmetric in the phase difference. The sign of the quantum correction depends on the ratios $x$ and $n=N_\mathrm{n}/\rbr{N_1+N_2}$, where $N_\mathrm{n}$ is the channel number of the neck.

In \sref{sec:thermopower} we finally apply our calculations to the thermopower of the setup shown in \fref{fig:thermo_setups_sym} with both equal and different numbers of channels as well as to the setup shown in \fref{fig:thermo_setups_asym}. We find that for the symmetric house with different channel numbers and for the antisymmetric house the thermopower is antisymmetric in the phase difference.

\section{Contributing diagrams}
\label{sec:cond_diagrams}
We will evaluate the quantum transmission between two normal conducting leads coupled to a classically chaotic, ballistic quantum dot which is additionally connected to superconducting leads such as depicted in \fref{fig:andreev_setups}.
\renewcommand{\thesubfigure}{}
\begin{figure*}
 \subfigure[ee3I]{\raisebox{-23pt}{\includegraphics[width=0.18\textwidth]{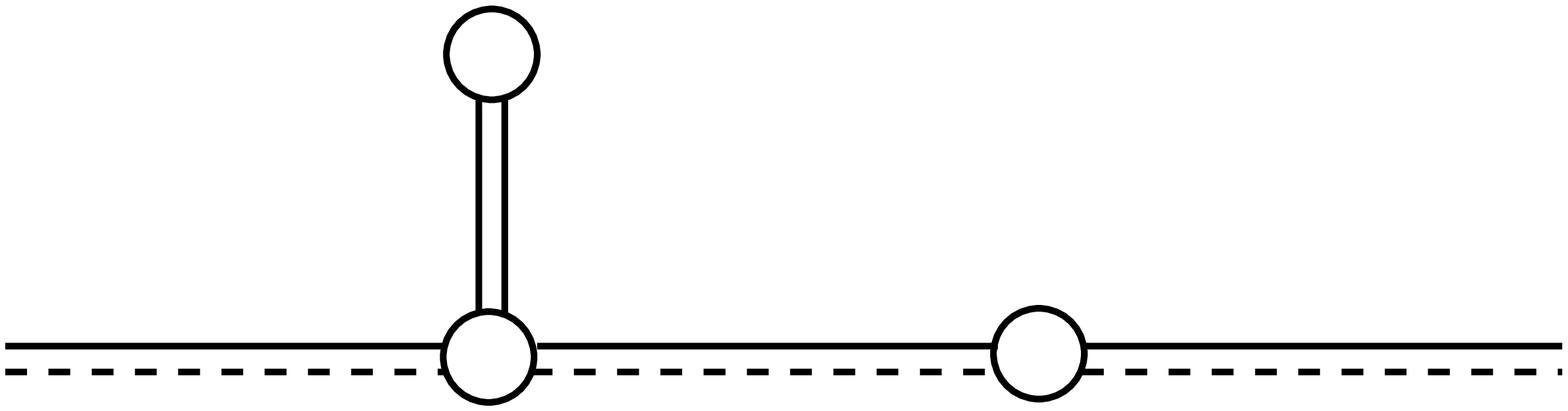}}\raisebox{-\height}{\rule{0pt}{45pt}}}\raisebox{-\height}{\rule{1pt}{0pt}}
 \subfigure[ee3II]{\raisebox{-23.5pt}{\includegraphics[width=0.18\textwidth]{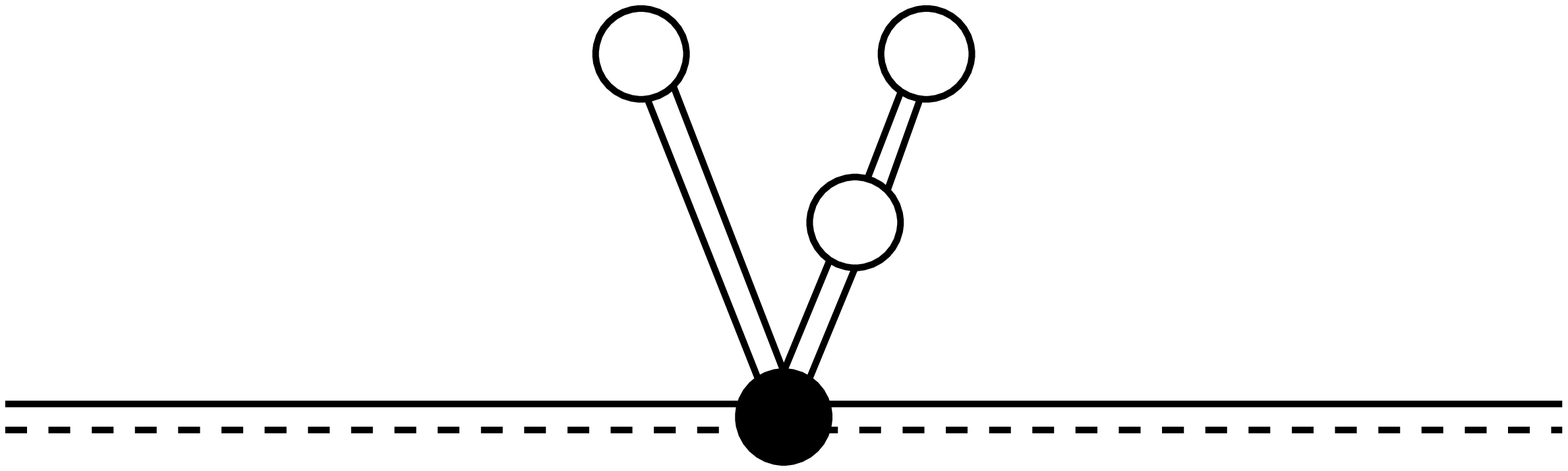} }\raisebox{-\height}{\rule{0pt}{45pt}}}
 \subfigure[ee3III]{\raisebox{-\height}{\includegraphics[width=0.18\textwidth]{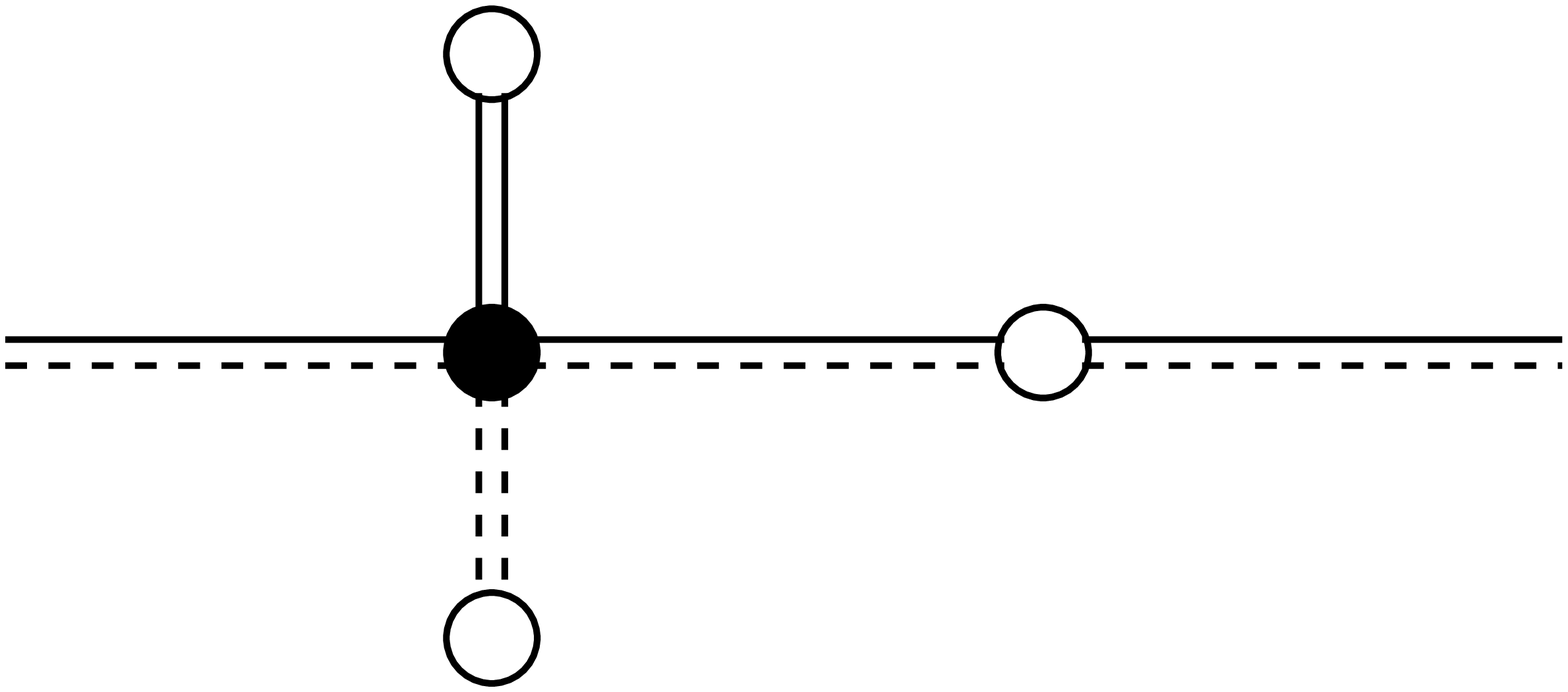} }\raisebox{-\height}{\rule{0pt}{45pt}}}
 \subfigure[he3I]{\raisebox{-23.5pt}{\includegraphics[width=0.18\textwidth]{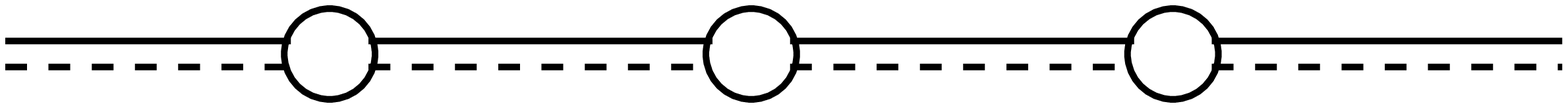} }\raisebox{-\height}{\rule{0pt}{45pt}}}
 \subfigure[he3II]{\raisebox{-\height}{\includegraphics[width=0.18\textwidth]{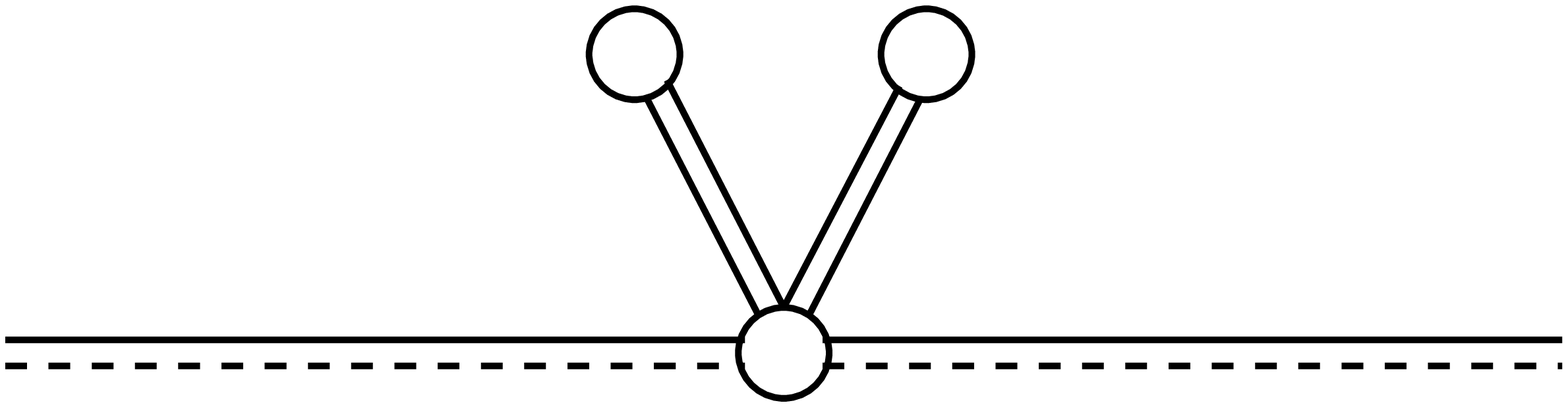} }\raisebox{-\height}{\rule{0pt}{45pt}}}\\
 \subfigure[he3III]{\raisebox{-\height}{\includegraphics[width=0.18\textwidth]{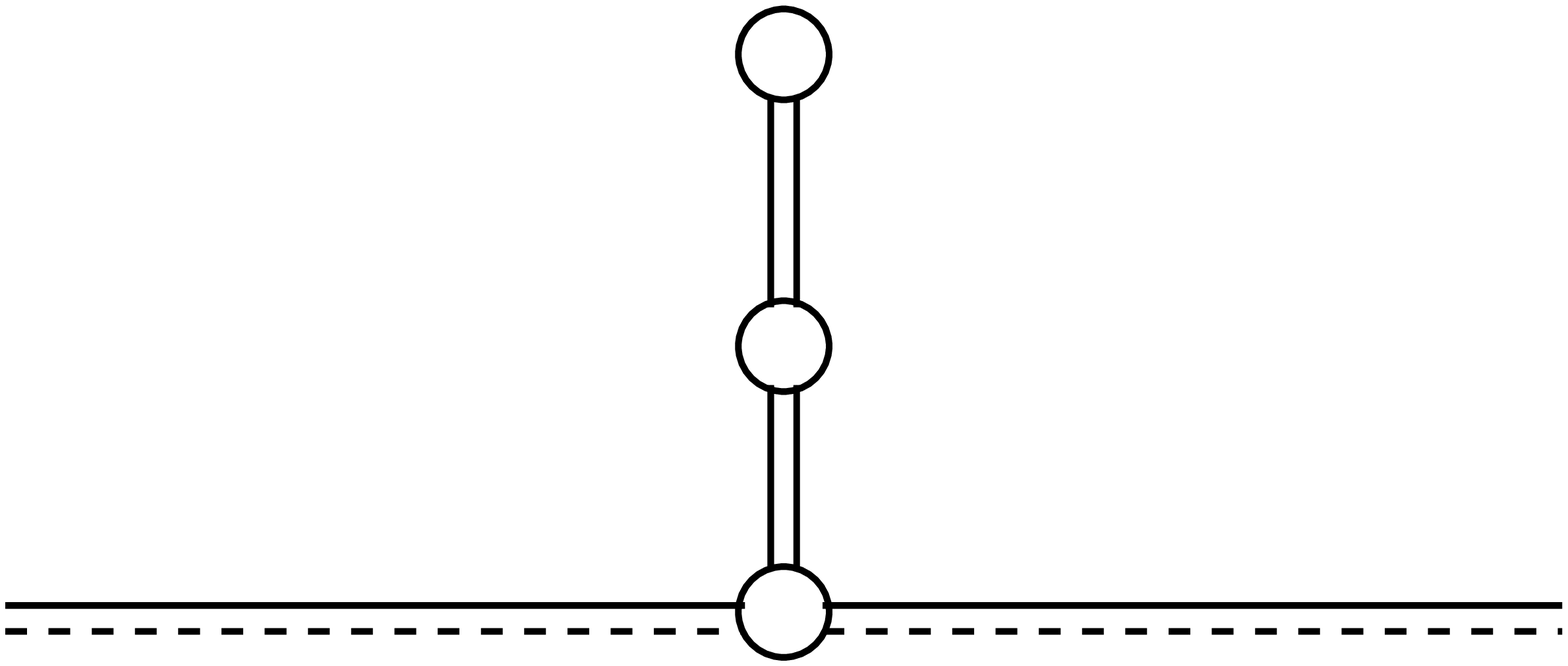} }\raisebox{-\height}{\rule{0pt}{60pt}}}\raisebox{-\height}{\rule{1pt}{0pt}}
 \subfigure[he3IV]{\raisebox{-56.5pt}{\rule{0.085\textwidth}{0pt}\includegraphics[height=60pt]{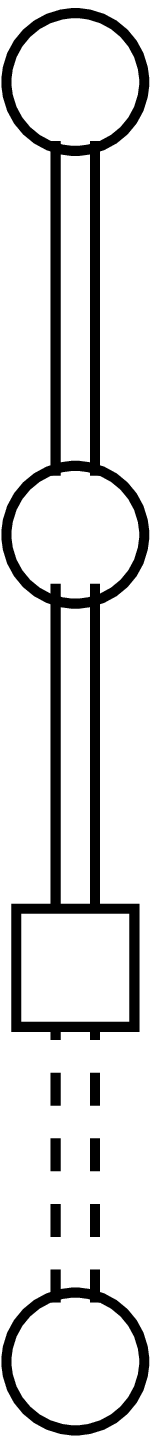}\rule{0.085\textwidth}{0pt} }\raisebox{-\height}{\rule{0pt}{60pt}}}
 \subfigure[he3V]{\raisebox{-39pt}{\includegraphics[width=0.18\textwidth]{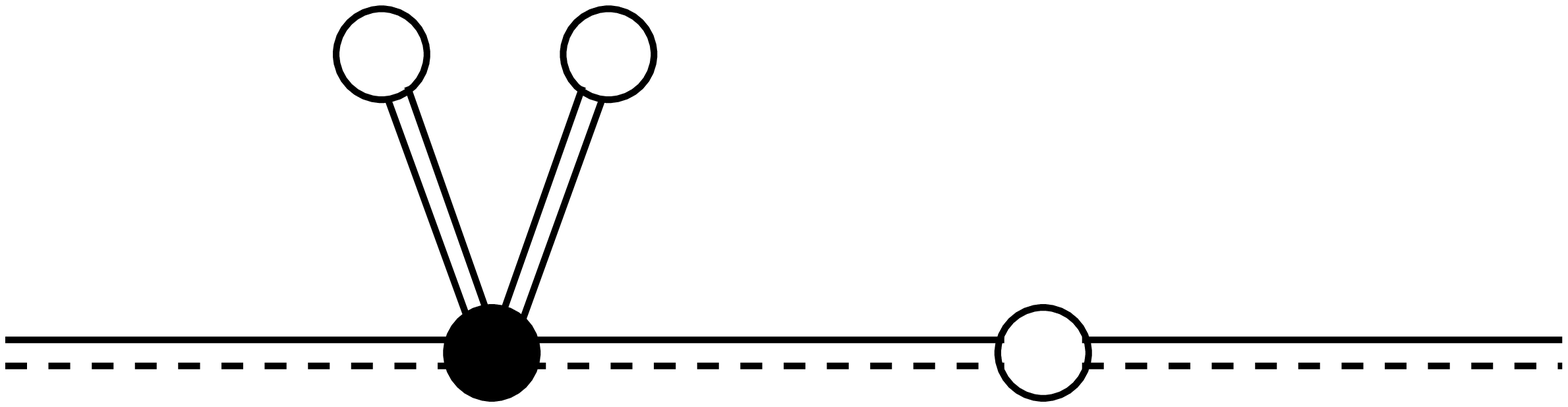}}\raisebox{-\height}{\rule{0pt}{60pt}}}\raisebox{-\height}{\rule{1pt}{0pt}}
 \subfigure[he3VI]{\raisebox{-55.5pt}{\includegraphics[width=0.18\textwidth]{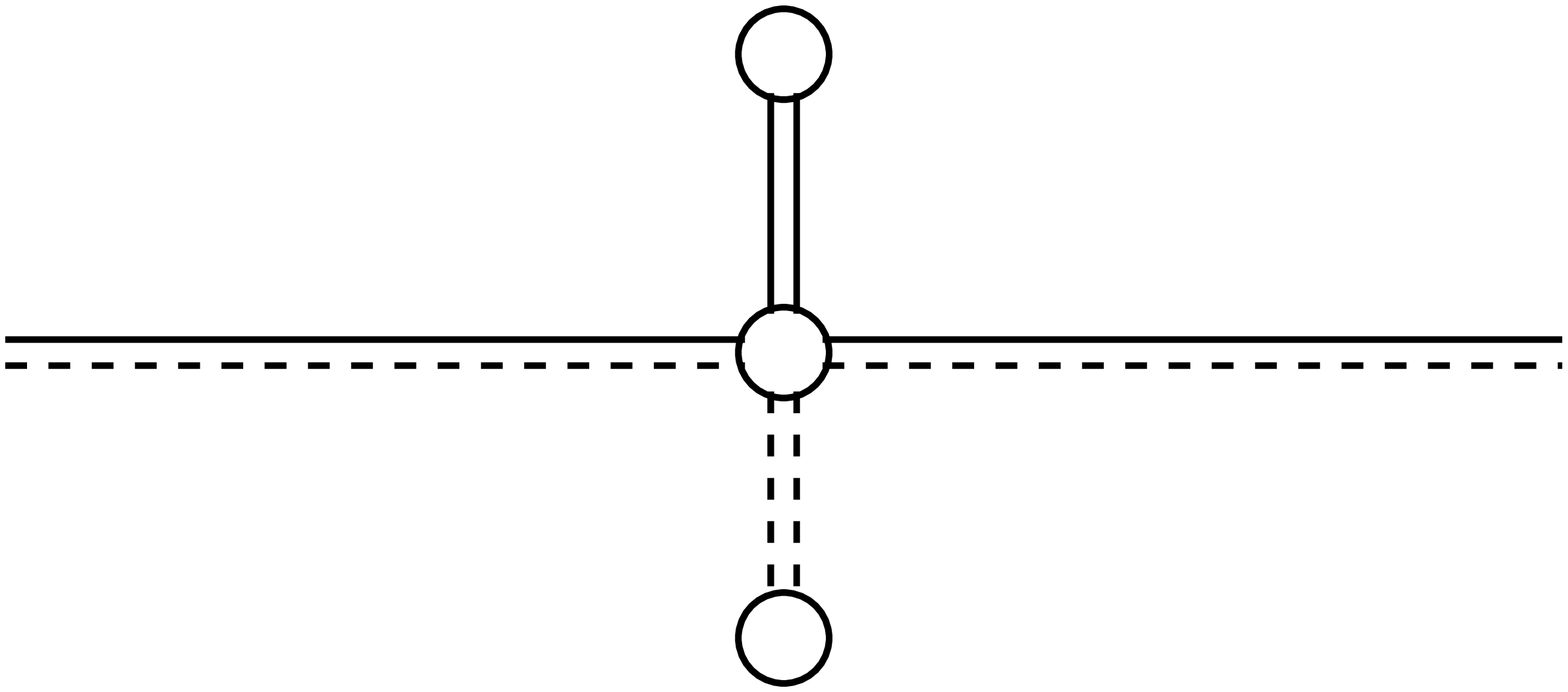} }\raisebox{-\height}{\rule{0pt}{60pt}}}
 \subfigure[he3VII]{\raisebox{-56.5pt}{\includegraphics[width=0.18\textwidth]{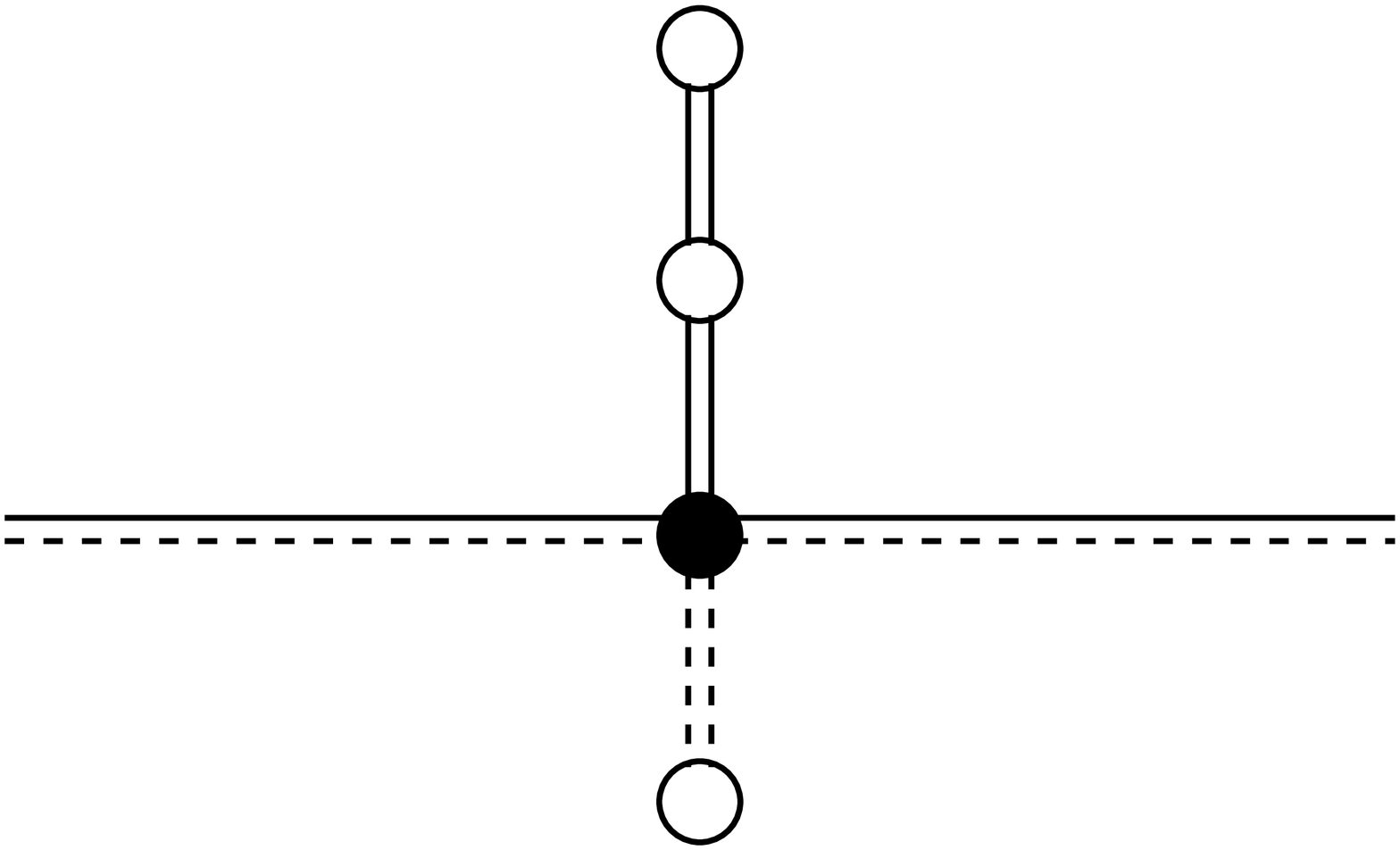} }\raisebox{-\height}{\rule{0pt}{60pt}}}
 \caption{\label{3rd_order_diag}Diagrams corresponding to the trajectory pairs shown in \fref{3rd_order}. The full circles denote encounters while the empty circles denote Andreev reflections. Note that an encounter touching the superconductor is marked as Andreev reflection. An encounter touching a normal conducting lead is shown as an empty box. The solid (dashed) line represents $\zeta$ ($\zeta^\prime$).}
\end{figure*}
\renewcommand{\thesubfigure}{(\alph{subfigure})}
In a trajectory-based semiclassical approach the transmission probabilities may be written as \cite{ref:semiclassical_scat_alternative,ref:trans_semicl,semiclassical_mesoscopic}
\begin{equation}
 T_{ij}=\fr{1}{T_\mathrm{H}}\sul{b=1}{N_i}\sul{a=1}{N_j}\sul{\zeta,\zeta^\prime}{}\sqrt{A_{\zeta}A^\text{\raisebox{0.5mm}{*}}_{\zeta^\prime}}\rme^{\rmi\rbr{S_{\zeta}-S_{\zeta^\prime}}/\hbar},
 \label{trans_semicl}
\end{equation}
where the $^*$ denotes complex conjugation. Here, $a$ and $b$ label the channels in lead $i\in\{1,2\}$ and $j\in\{1,2\}$, respectively. $\zeta$ and $\zeta^\prime$ are classical trajectories starting at channel $a$ and ending at channel $b$. The amplitudes $A_{\zeta}$ contain the stability of the trajectory $\zeta$ and $S_{\zeta}$ is its classical action. Moreover, $T_\mathrm{H}$ is the Heisenberg time the time dual to the mean level spacing.

We are interested in the conductance averaged over the shape of the quantum dot or an energy range small compared to the Fermi energy but large enough to smooth out fluctuations. Moreover we will take the semiclassical limit $\hbar\rightarrow0$. Therefore the energy dependent action difference in \eref{trans_semicl} causes fluctuations cancelled on average unless it is of order of $\hbar$. Thus we have to pair the trajectories in such a way that their action difference becomes sufficiently small. The easiest way to do this is to require that $\zeta=\zeta^\prime$ which is known as the diagonal approximation \cite{ref:diagonal_approximation}. In \Fref{fig:diagonal_contr_cond} the trajectory pairs contributing to the diagonal approximation are drawn schematically for up to three Andreev reflections. The contributions of the diagonal pairs to the conductance provide the classical conductance \cite{nsntrans}
\begin{equation}
 g_{cl}=\fr{N_1N_2}{N_1+N_2}
 \end{equation}
 if the superconductors are isolated and
 \begin{equation}
 g_{cl}=\fr{N_1\rbr{N_2+2N_\mathrm{S}}}{N_1+N_2+2N_\mathrm{S}}
\end{equation}
 in the case of the superconducting leads. However as shown semiclassically in Refs.~\onlinecite{semiclassical_dos} for the density of states and~\onlinecite{nsntrans} for the conductance of Andreev quantum dots one has to go beyond the diagonal approximation to fully account for quantum effects. The non-diagonal trajectory pairs contributing to the conductance of normal junctions in the limit $\hbar\to0$ have been first considered in Ref.~\onlinecite{semiclassic} and generalized to higher orders in $1/N$ in Ref.~\onlinecite{ref:diagramatic}: There are small regions in which an arbitrary even number - say $2l$ - of trajectory stretches come close to each other. $l$ of these trajectory stretches `cross' each other in this region while the remaining $l$ ones avoid crossing. The difference between a trajectory $\zeta$ and its partner $\zeta^\prime$ then leads to a small action difference as long as these stretches are close enough to each other. Such a region with $l$ crossing trajectory stretches and $l$ trajectory stretches `avoiding crossing' will be referred to as $l$-encounter. Between these $l$-encounters two different trajectory stretches retrace each other forming a path pair with vanishing action difference which will also be referred to as a link.
 
In what follows we will identify the relevant trajectory pairs contributing to the conductance beyond the diagonal approximation in leading order in the inverse channel number $1/N$. The diagrams with two Andreev reflections may be found in Ref.~\onlinecite{nsntrans}. However we want to go beyond second order in $x=N_\mathrm{S}/N_\mathrm{N}$. The trajectories contributing in third order in $x$, \textit{i.e.}~trajectories with three Andreev reflections, are shown in \fref{3rd_order}.
The first task is to find a structure in the diagrams contributing at leading order in the channel number. To facilitate this we can redraw our semiclassical diagrams in a skeleton form and represent encounters and path pairs by nodes and lines. For example the diagrams contributing to third order in $x$, shown in \fref{3rd_order}, can be redrawn as in \fref{3rd_order_diag}.

\begin{figure}
\includegraphics[width=\columnwidth]{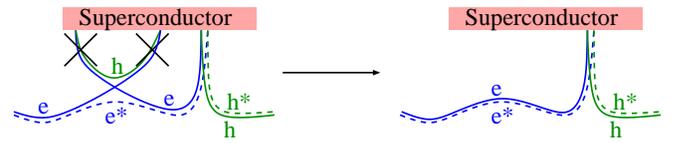}
\caption {\label{fig:cut_offdiagonal}If the e-h path pairs are cut off a diagonal type diagram remains.}
\end{figure}

We first consider how to read of the channel number dependence from a given diagram, \textit{i.e.}~we use the diagrammatic rules used in Ref.~\onlinecite{ref:diagramatic} disregarding an energy and magnetic field dependence and any signs for the moment. A path pair hitting lead $j$ contributes a factor of channel number $N_j$. The path pair, or link, itself however contributes a factor $1/N$ while each encounter contributes a factor $N$. From the trajectory pairs shown in \fref{3rd_order} we see that if we cut off all e-h and \es-\hs~pairs we again get a diagonal like contribution as depicted in \fref{fig:cut_offdiagonal}.
For example if we cut the e-h pair at the very left of ee3I we get the diagonal contribution to the second order in $x$ in \fref{fig:diagonal_contr_cond} since there are two Andreev reflections and if we cut the `off-diagonal' parts in say ee3III we get a diagonal contribution at first order in $x$.

\begin{figure}
 \subfigure[\label{fig:neglected_contr_a}]{\includegraphics[width=0.3\columnwidth]{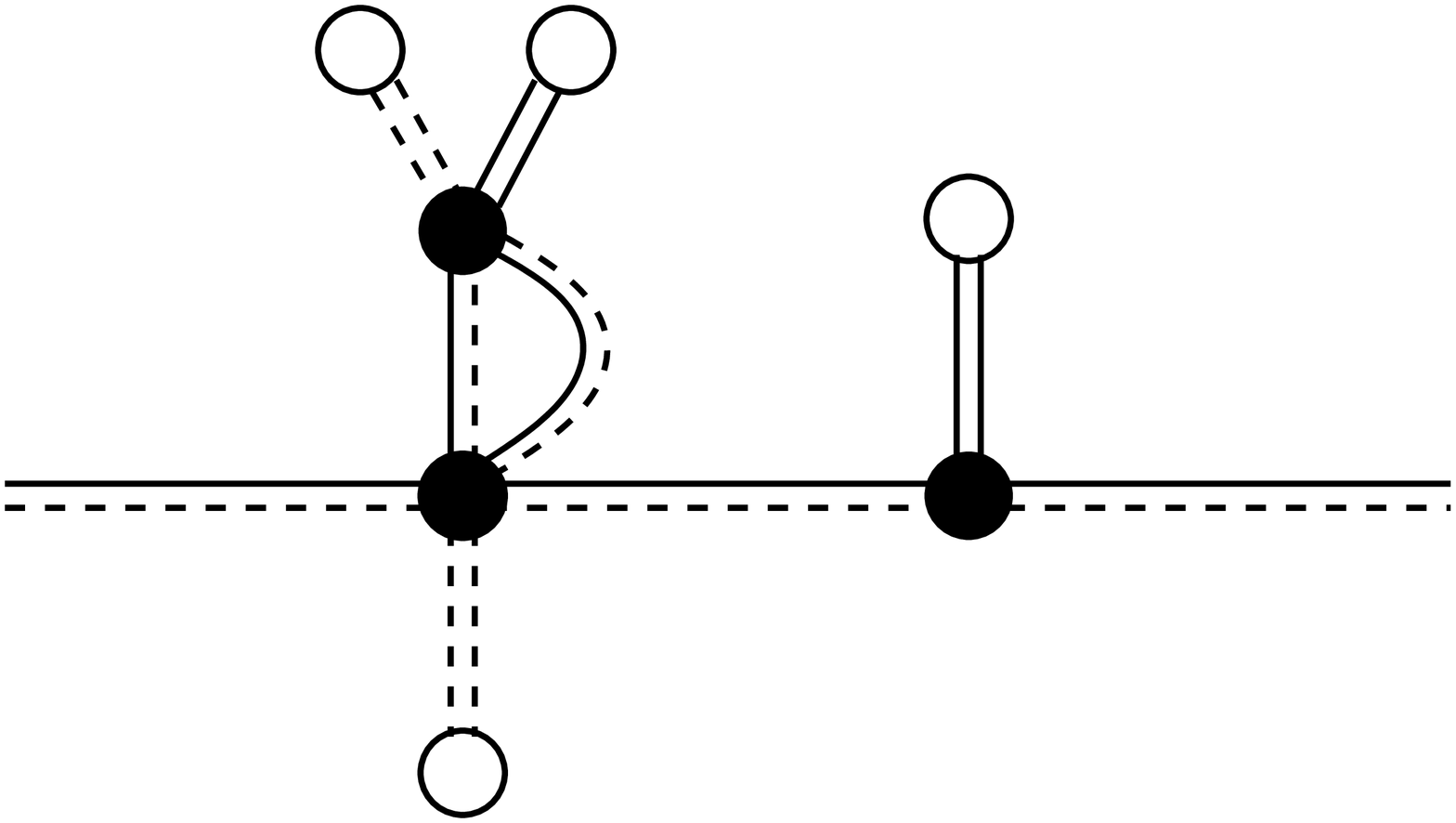}}
 \subfigure[\label{fig:neglected_contr_b}]{\includegraphics[width=0.3\columnwidth]{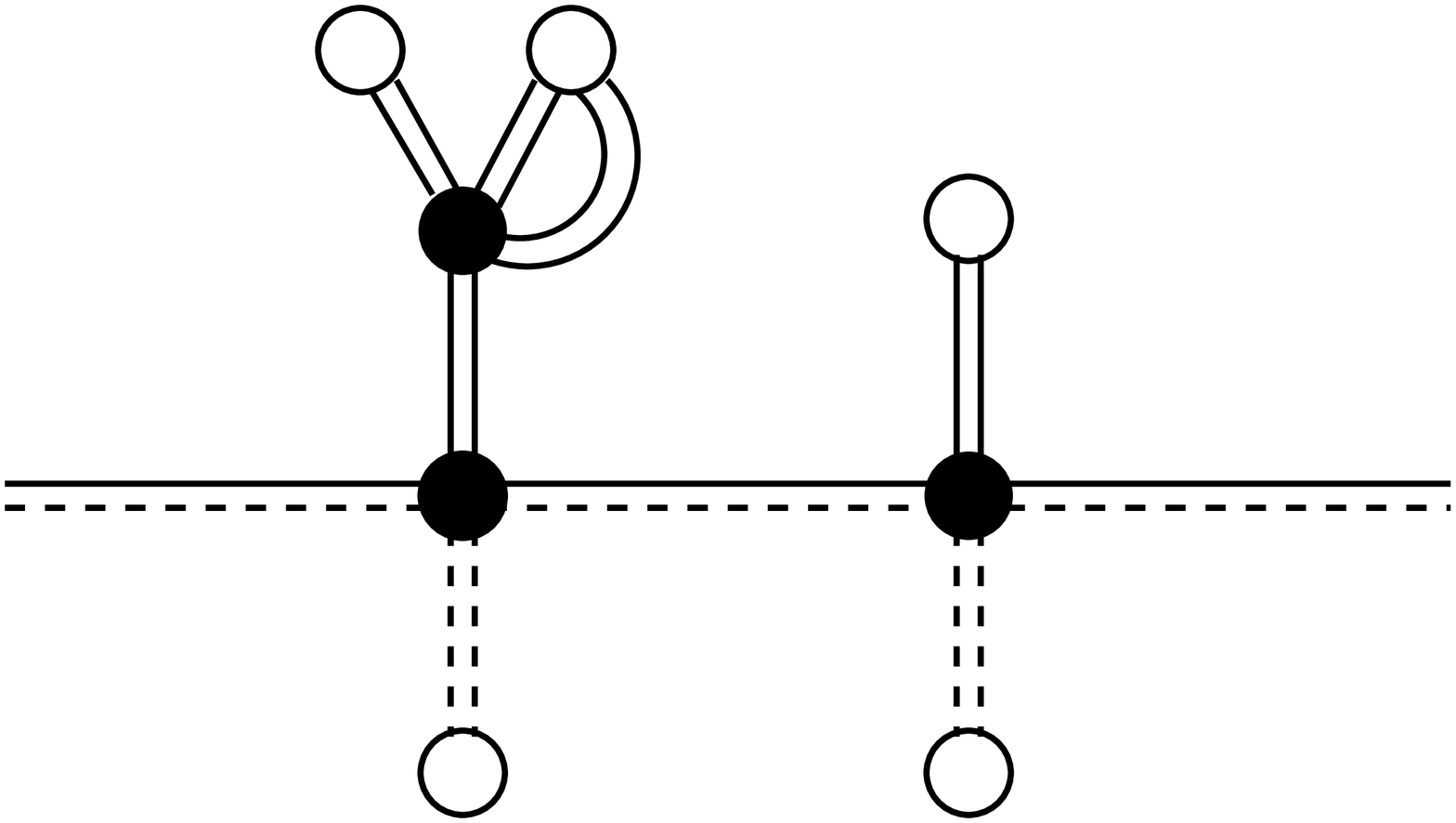}}
 \subfigure[\label{fig:neglected_contr_c}]{\raisebox{11pt}{\includegraphics[width=0.3\columnwidth]{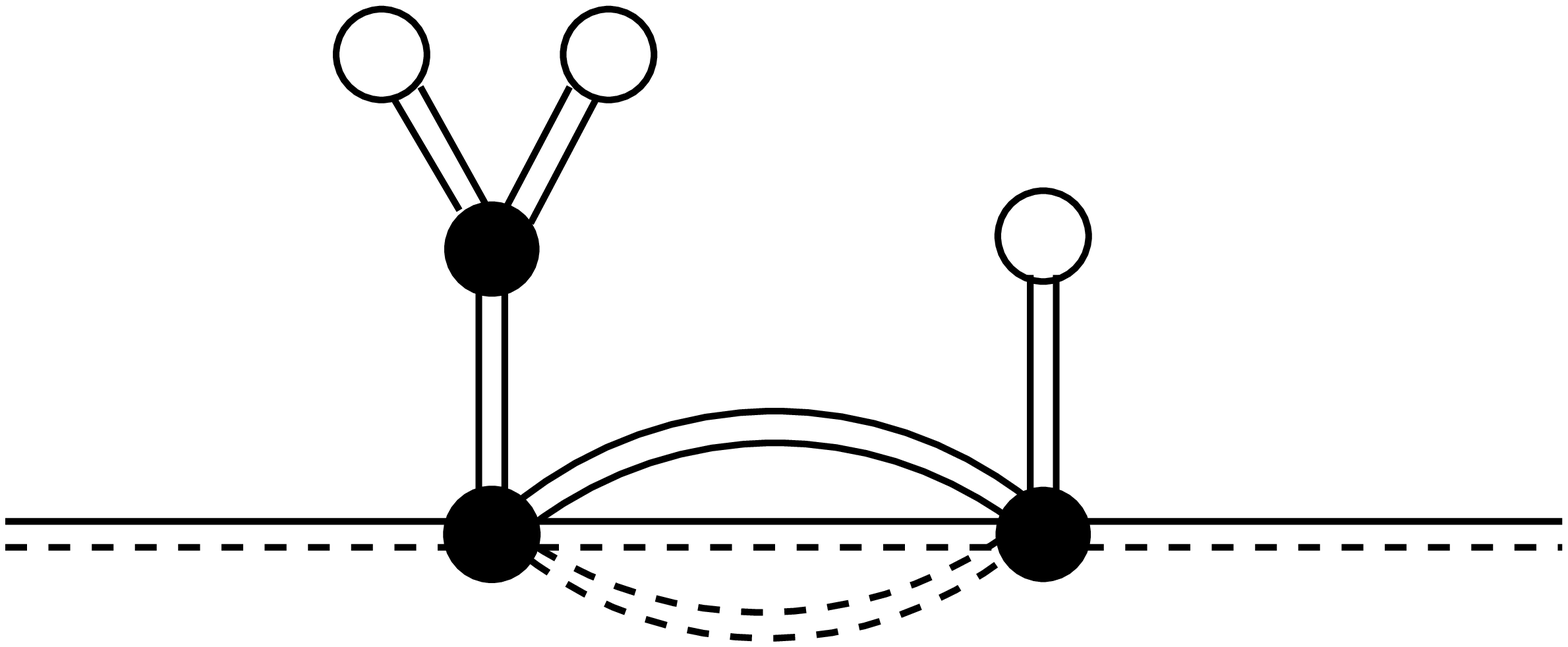}}}
 \caption{\label{fig:neglected_contr} Diagrams we neglect in leading order in $1/N$ due to the formation of loops: (a) A non-diagonal $\zeta$-$\zeta^\prime$ pair causes the formation of a loop. (b) A loop formed by an off-diagonal e-h path pair. (c) A loop formed due to the lack of a diagonal-type $\zeta$-$\zeta^\prime$ path pair.}
\end{figure}

Staying at leading order in $1/N$ implies that the `off-diagonal' path pairs cannot consist of one $\zeta$- and one $\zeta^\prime$-stretch since each of those path pairs has to be traversed by $\zeta$ and $\zeta^\prime$ in the same direction. Thus in order to come back from the off-diagonal part starting with an $\zeta$-$\zeta^\prime$ pair we have to connect this off-diagonal part to the diagonal `backbone' by a second $\zeta$-$\zeta^\prime$ path pair thus forming a loop as indicated in \fref{fig:neglected_contr_a}.
This loop however adds a link, thus giving a factor $1/N$ compared to the same diagram without the loop, and therefore decreases the number of Andreev reflections by at least one and therefore the contribution to the conductance is suppressed by a factor of the order $1/N_{\mathrm{S}}$ such that it would contribute to sub-leading order in the inverse channel number. Therefore the `off-diagonal' parts may only consist of e-h or \es-\hs~pairs. In the same way we may neglect loops formed by e-h or \es-\hs~path pairs as the one in \fref{fig:neglected_contr_b}.

In terms of graphs, the `off-diagonal' parts again become rooted plane trees as in Refs.~\onlinecite{my_dos},~\onlinecite{countingstatistic}. These trees start with a link (root) which connecting an encounter to the diagonal like backbone. From this encounter several further links emerge all ending again at an encounter or at a superconducting channel. In contrast to the trees in Refs.~\onlinecite{my_dos},~\onlinecite{countingstatistic}, the trees here - we will call them `side-trees' - start at the `diagonal encounter' such that their root does not touch a channel but the diagonal backbone instead. Note that we draw the diagrams such that the non-complex conjugated side trees are at the upper side of the diagonal backbone while the complex conjugated ones are on the lower side of the diagonal backbone.

The fact that the path pairs along the backbone are composed only of $\zeta$-$\zeta^\prime$ pairs is again due to neglecting loops: The two trajectories $\zeta$ and $\zeta^\prime$ must both start at lead $j$ and end at lead $i$. Therefore the path pairs hitting the normal leads have to be $\zeta$-$\zeta^\prime$ pairs and so, if there is a `diagonal' encounter entered by a $\zeta$-$\zeta^\prime$ pair and left only by e-h and \es-\hs~pairs, there must be a corresponding encounter entered only by e-h and \es-\hs~pairs and left by a $\zeta$-$\zeta^\prime$ pair. Therefore we again would get a loop essentially formed by one e-h and one \es-\hs~pair as shown in \fref{fig:neglected_contr_c}.

All told the diagrams have to consist of a diagonal type `backbone' consisting of $\zeta$-$\zeta^\prime$ path pairs and encounters (which may also touch the superconductor) and $\zeta$- and $\zeta^\prime$-side trees emerging from these diagonal encounters.
%
Note that when pairing a $\zeta$ with a $\zeta^\prime$ stretch these stretches have to be traversed by the same kind of quasiparticle, \textit{i.e.}~it has to be an e-\es~or a h-\hs~pair. This is related to the fact that each encounter has an even number of entering and exiting path pairs.

However, there is still one possibility left we have not mentioned yet but that needs a special treatment. If the `diagonal' part consists of only two path pairs and one $2$-encounter with one $\zeta$-side tree (a side tree formed by $\zeta$) and one $\zeta^\prime$-side tree this encounter can be moved into one of the normal conducting leads, say lead $i$. An example for an $2$-encounter touching the incoming lead is the trajectory labelled by he3IV in figures \ref{3rd_order} and \ref{3rd_order_diag} which arises from the trajectory labelled by he3VII by moving the encounter into the lead. However this is only possible if the trajectory connects lead $i$ to itself and thus if the electron is coherently back scattered. In this case we have only one side tree and one complex conjugated side tree but no `diagonal' part.


Since we know the structure of the trajectory pairs contributing at leading order in the inverse channel number we can start evaluating them. Because the contributions of the encounters and the stretches are multiplicative \cite{ref:diagramatic} we may factorise the contribution of a given diagram into the contributions of side trees starting at the first encounter with an $\alpha$-type quasiparticle, $P^\alpha(\en,x)$, the first encounter itself and the diagram remaining when cutting the diagram after the first encounter as in \fref{fig:split_diagram}.
We will first evaluate the contribution arising from the summation over a possible side tree.

\section{Side tree contributions}
\label{cond_sidetrees}

We restrict ourself to sufficiently low temperatures such that only energies $\en E_{\mathrm{T}}$ (measured with respect to the Fermi energy and in units of the Thouless energy $E_\mathrm{T}=\hbar/2\tau_\mathrm{D}$ where $\tau_\mathrm{D}$ is the mean dwell time) much smaller than the superconducting gap $\Delta$ have to be taken into account $\en E_\mathrm{T}\ll\Delta$. This allows us to approximate $\exp[-\rmi\arccos(\en/\Delta)]\approx-\rmi$ such that the scattering matrix of Andreev reflection becomes independent of the energy \cite{ref:density}. Thus the diagrammatic rules for the $\zeta$-side trees read \cite{ref:diagramatic,my_dos}
\begin{itemize}
 \item An e-h path pair contributes $\cbr{N\rbr{1+\rmi\en+b^2}}^{-1}$
 \item An $l$-encounter contributes $-N\rbr{1+\rmi l\en+l^2b^2}$
 \item An e-h path pair hitting the superconductor S$_j$ contributes $N_{\mathrm{S}_j}$.
 \item An $l$-encounter touching the superconductor S$_j$ contributes $N_{\mathrm{S}_j}$.
 \item Each Andreev reflection at the superconducting lead $j$ converting an electron into a hole contributes $-\rmi\rme^{-\rmi\phi_j}$.
 \item Each Andreev reflection at the superconducting lead $j$ converting a hole into an electron contributes $-\rmi\rme^{\rmi\phi_j}$.
\end{itemize}
with $b\propto B/\hbar$ where $B$ is the magnetic field applied to the system. The proportionality factor depends on the actual system \cite{ref:diagramatic}.
These diagrammatic rules have to be complex conjugated for a $\zeta^\prime$-side tree and imply that when exchanging electrons and holes we just have to replace $\phi\leftrightarrow-\phi$. Thus a side tree starting with a hole gives the same contribution of a side tree starting with an electron but with negative phase. Therefore we only need to evaluate side trees starting with electrons.

The evaluation of the side trees then follows essentially those in Refs.~\onlinecite{countingstatistic}, \onlinecite{my_dos} and \onlinecite{ref:mot}. However here the root of the tree does not hit any channel and therefore can not touch the superconductor which simplifies the calculation. Moreover from the rules above for a path pair hitting a channel in the superconductor S$_1$ we get a factor $-\rmi\rme^{-\rmi\phi/2}N_{\mathrm{S}_1}$ if an electron hits the channel and $-\rmi\rme^{\rmi\phi/2}N_{\mathrm{S}_1}$ if a hole hits the channel, rather than just a factor of the numbers of channels, and equivalently for a path pair hitting S$_2$.

Similar to Ref.~\onlinecite{my_dos} as long as the phase difference $\phi$ is zero and no encounter touches the superconductor the contribution of a side tree with characteristic $\ve$ - which is the vector which $l$-th entry is the number of $l$-encounters of the tree - but without the contribution of the path pairs hitting one of the superconductors is
\begin{align}
\rbr{1+\rmi\en+b^2}^{-1}&\prodl{\alpha=1}{V(\ve)} \fr{\rbr{1+\rmi l_\alpha\en+l_\alpha^2b^2}}{\rbr{1+\rmi\en+b^2}^{2l_\alpha-1}} \nonumber \\
 & =\rbr{1+\rmi\en+b^2}^{-n}\prodl{\alpha=1}{V(\ve)}\fr{\rbr{1+\rmi l_\alpha\en+l_\alpha^2b^2}}{\rbr{1+\rmi\en+b^2}^{l_\alpha}}.
 \label{eq:treecontr_notouch}
\end{align}
Here the encounters have been labelled by $\alpha$ and we used that the side tree has to satisfy \cite{countingstatistic} $n=L(\ve)-V(\ve)+1$, where $n$ is the number of links touching the superconductor, $V(\ve)=\sul{l\geq2}{}v_l$ is the total number of encounters of the tree and $L(\ve)=\sul{l\geq2}{}lv_l$. This is because every $l$-encounter creates $2l-1$ additional path pairs and each path pair has to end either in an encounter or at the superconductor.

We then enumerate the number of $l$-encounters by $x_l$ and the number of $l$-encounters touching the superconductor S$_i$ at an odd numbered channel by $z_{o,l}^{(i)}$. An $l$-encounter touching the superconductor means that the $l$ incoming trajectory stretches hit the superconductor at one and the same channel. We look at the generating function $f(\bi{x},\bi{z}_o^{(1)},\bi{z}_o^{(2)})$ which counts the number of possible side trees and their encounter types and derive a recursion relation for it by cutting the side tree at its top encounter. If the top encounter is traversed by $2l$ stretches and does not touch the superconductor the tree then has the contribution of the top encounter times that of all $2l-1$ subtrees giving $x_lf^l\hat{f}^{l-1}$, where $\hat{f}$ is the same as $f$ but with $\phi$ replaced by $-\phi$ accounting for the fact that each even numbered subtree starts with a hole instead of an electron. If the top encounter however is an encounter traversed by $2l$ stretches touching S$_i$ its contribution is $z_{o,l}^{(i)}\hat{f}^{l-1}$. In total we therefore have
\addtocounter{equation}{1}
\begin{align}
f= & -\rmi\fr{N_{\mathrm{S}_1}}{N}\rme^{-\rmi\phi/2}-\rmi\fr{N_{\mathrm{S}_2}}{N}\rme^{\rmi\phi/2} \nonumber\\
 & +\sul{l=2}{\infty}\cbr{x_lf^l\hat{f}^{l-1}+\rbr{z_{o,l}^{(1)}+z_{o,l}^{(2)}}\hat{f}^{l-1}}
\label{eq:generating_func_general_el}
\tag{\theequation\alph{mycounter}}
\addtocounter{mycounter}{1}\\
\hat{f}= & -\rmi\fr{N_{\mathrm{S}_1}}{N}\rme^{\rmi\phi/2}-\rmi\fr{N_{\mathrm{S}_2}}{N}\rme^{-\rmi\phi/2} \nonumber\\
 & +\sul{l=2}{\infty}\cbr{x_l\hat{f}^lf^{l-1}+\rbr{\hat{z}_{o,l}^{(1)}+\hat{z}_{o,l}^{(2)}}f^{l-1}}
\label{eq:generating_func_general_ho}
\tag{\theequation\alph{mycounter}}
\addtocounter{mycounter}{1}
\end{align}
where the first two terms account for empty side trees which consist of one link and one Andreev reflection at S$_1$ or S$_2$ and $\hat{z}_{o,l}^{(i)}$ is the same as $z_{o,l}^{(i)}$ but with $\phi$ replaced by $-\phi$.

\begin{figure}
 \includegraphics[width=\columnwidth]{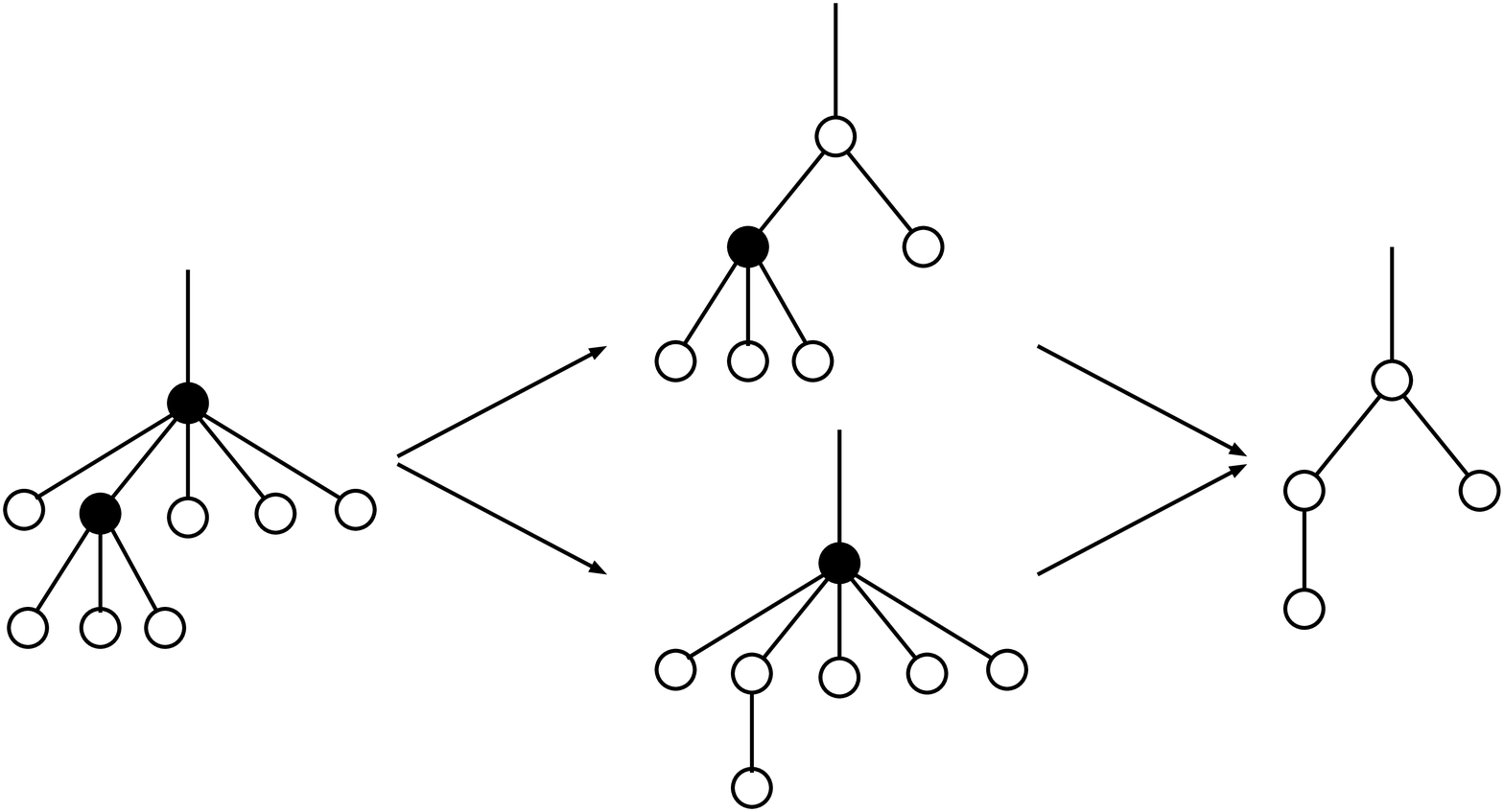}
 \caption{\label{fig:sidetree_slide}If the odd numbered subtrees have zero characteristic and hit the same superconductor the top encounter may be slid into the superconductor.}
\end{figure}

Due to the fact that the links of the side trees are traversed by one electron at energy $+\en\hbar/2\tau_\mathrm{D}$ and one hole at energy $-\en\hbar/2\tau_\mathrm{D}$ in opposite directions an $l$-encounter consists of $l$ electron-stretches traversing the encounter in the same direction and $l$ hole-stretches traversing the encounter in the opposite direction. Thus we have $x_l=-\rbr{1+\rmi l\en+l^2b^2}/\rbr{1+\rmi\en+b^2}^l\tilde{r}^{l-1}$ in line with \eref{eq:treecontr_notouch}. The powers of $\tilde{r}$ are included in order to keep track of the order of the trees.
Now consider an $l$-encounter touching S$_1$. According to the diagrammatic rules after extracting the factor $(1+\rmi\en+b^2)^{-n}$ as in \eref{eq:treecontr_notouch} the contribution of the encounter and the link connecting the top encounter to the backbone is $N_{\mathrm{S}_1}/N$. However we have to include the phase factors contributed by the Andreev reflections. To evaluate this phase factor we look at the $l$-encounter touching the superconductor as arising from an $l$-encounter inside the dot by sliding it into the superconductor as indicated in \fref{fig:sidetree_slide}.
This is only possible if the odd numbered subtrees have zero characteristic and hit the same superconductor. By sliding the encounter into the superconductor the total number of Andreev reflections do not change such that the phase factor provided by the encounter touching the superconductor is the same as the phase factor provided by the odd numbered side trees we start from. For a side tree starting with an electron the odd numbered subtrees with zero characteristic provide one Andreev reflection converting an electron into a hole. Since from an $l$-encounter $l$ odd numbered subtrees emerge the phase factor of an $l$-encounter touching S$_i$ is $-\rmi\rme^{-\rmi l\phi_i}$.
\addtocounter{equation}{1}
\setcounter{mycounter}{0}
\begin{align}
z_{o,l}^{(1)}=&(-\rmi)^l\fr{N_{\mathrm{S_1}}}{N}\rme^{-\rmi l\phi/2}\tilde{r}^{l-1},
\label{eq:first_encounter_1touch}
\tag{\theequation\alph{mycounter}}
\addtocounter{mycounter}{1}\\
z_{o,l}^{(2)}=&(-\rmi)^l\fr{N_{\mathrm{S_2}}}{N}\rme^{\rmi l\phi/2}\tilde{r}^{l-1}
\label{eq:first_encounter_2touch}
\tag{\theequation\alph{mycounter}}
\addtocounter{mycounter}{1}
\end{align}

The total power of a tree with $2n-1$ Andreev reflections is again $\sum_l(l-1)v_l=L-V=n-1$. Thus in order to get the required prefactor of $(1+\rmi\en+b^2)^{-n}$ we can make the change of variables
\begin{align}
f&=g(1+\rmi\en+b^2), & & \tilde{r}=\fr{r}{1+\rmi\en+b^2}.
\label{eq:variable_change_side_tree}
\end{align}

After making this change of variables and performing the summations in \earef{eq:generating_func_general_el}{,b} using geometric series we get in view of Eqs.~(\ref{eq:first_encounter_1touch},b)
\begin{align}
 \fr{\rbr{1+\rmi\en+b^2}g}{1-rg\hat{g}}+\fr{\rbr{2b^2+\rmi\en}r\hat{g}g^2}{\rbr{1-rg\hat{g}}^2}+\fr{b^2\rbr{1+rg\hat{g}}r\hat{g}g^2}{\rbr{1-rg\hat{g}}^3} \nonumber \\
+\fr{\rmi x(1+y)}{2(1+x)\rbr{\rme^{\rmi\phi/2}+\rmi r\hat{g}}}+\fr{\rmi x(1-y)}{2(1+x)\rbr{\rme^{-\rmi\phi/2}+\rmi r\hat{g}}}&=0
 \label{geq}
\end{align}
and the same equation with $\hat{g}$ and $g$ exchanged and $\phi$ replaced by $-\phi$. Here we used $N_\mathrm{S}/N_\mathrm{N}=x$ and introduced the difference of the numbers of channels of the two superconductors $y=(N_{\mathrm{S}_1}-N_{\mathrm{S}_2})/N_\mathrm{S}$ such that $y=0$ corresponds to the case of equal numbers of channels and $y=\pm 1$ to the case of just one superconductor.

In the case that the two superconductors provide the same number of channels (y=0) those two equations are the same implying $\hat{g}=g$ and \eref{geq} is equivalent to an algebraic equation of $7$th order in $g$. This increase in the order of the equation with respect to the same case for the density of states\cite{my_dos} is due to the fact that in the case of the density of states we had no normal leads.

The contribution $P^e$ of the side trees starting with an electron is then obtained by giving all trees the same weight by setting $r=1$ in $g$. The contribution of the side trees starting with a hole are then given by replacing $\phi$ by $-\phi$ in $g$ or setting $r=1$ in $\hat{g}$.
After setting $r=1$ and eliminating say $P^h$ the contribution of the side trees starting with an electron $P^e$ is given by a rather lengthy equation of in general 11th order which factorises in the case $y=0$ such that $P^e=P^h=P$ is given by
 \begin{align}
-{P}^{7}+\left(2\rmi\beta+\rmi\beta x\right){P}^{6} \nonumber \\
+\left(-{b}^{2}x+3+\rmi\epsilon\,x-{b}^{2}+\rmi\epsilon\right){P}^{5} \nonumber \\
+\left(-\rmi\beta x+2\rmi{b}^{2}\beta+2\epsilon\beta+2\rmi{b}^{2}\beta x+2\epsilon\beta x-4\rmi\beta
\right){P}^{4} \nonumber \\
+\left(-2\rmi\epsilon-3-2\rmi\epsilon x\right){P}^{3} \nonumber \\
+\left(2\rmi{b}^{2}\beta+2\rmi{b}^{2}\beta x-2\epsilon\beta x-2\epsilon \beta+2\rmi\beta-\rmi\beta x\right){P}^{2} \nonumber \\
+\left(\rmi\epsilon x+{b}^{2}+{b}^{2}x+1+\rmi\epsilon\right)P+\rmi\beta x&=0.
\label{alg_g_samechannels_mag_en}
\end{align}
If no magnetic field is applied ($b=0$) the equation may be factorised, and one has to solve an equation whose order is lowered by $2$.

If the Andreev interferometer consists of two superconductors with the same numbers of channels (y=0) the side tree contributions only depend on $\beta=\cos(\phi/2)$ rather than on $\phi$ itself. Therefore in this case the side tree contributions are symmetric in $\phi$ and the contribution of a side tree starting with a hole is the same as that of a side tree starting with an electron. In the most simple case of the absence of a magnetic field, zero temperature (\textit{i.e.} $\en=0$) and zero phase difference \eref{alg_g_samechannels_mag_en} reduces to a second order equation:
\begin{equation}
 -P^2+\rmi P+\rmi Px-x
 \label{simplestside treeeq}
\end{equation}
yielding
\begin{equation}
 P(0,x)=\fr{\rmi}{2}\rbr{1+x-\sqrt{1+6x+x^2}}.
 \label{simplestside treesol}
\end{equation}
Note that we take the solution satisfying $P(0,0)=0$ since when there is no superconductor the correction of leading order in the channel number has to be zero.

\section{Transmission coefficients}
\label{sec:cond_transmission_probs}
We will now demonstrate how to calculate the transmission coefficients $T_{ij}^{\alpha\beta}$ for transmission from lead $j$ to lead $i$ while converting an $\alpha$-type quasiparticle in a $\beta$-type one, using $T_{ij}^{ee}$ as an example, as the evaluation of the other transmission coefficients will be similar.
\begin{figure}
 \subfigure[\label{fig:split_diagram_a}]{\includegraphics[width=\columnwidth]{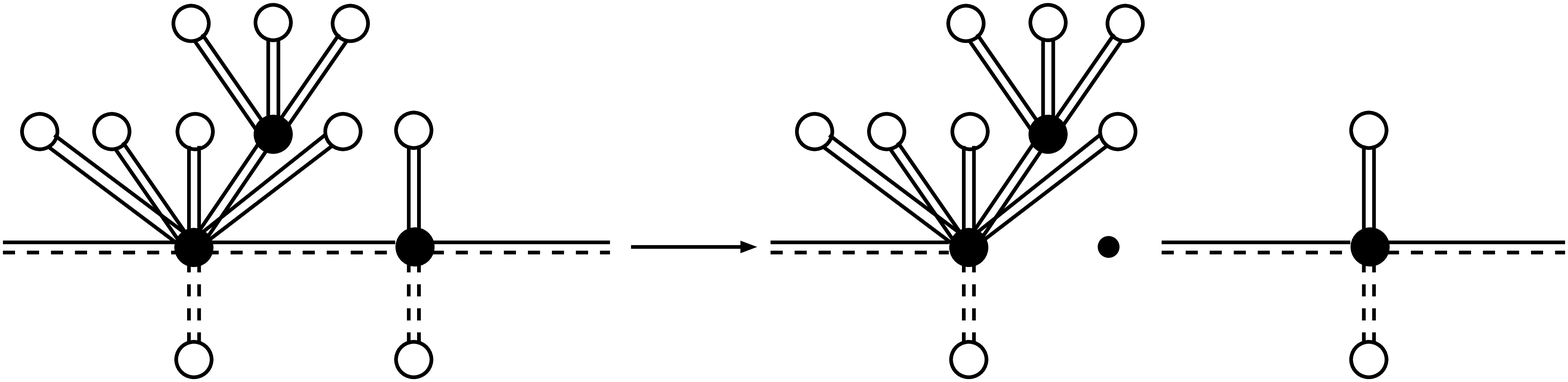}}
\\
 \subfigure[\label{fig:split_diagram_b}]{\includegraphics[width=\columnwidth]{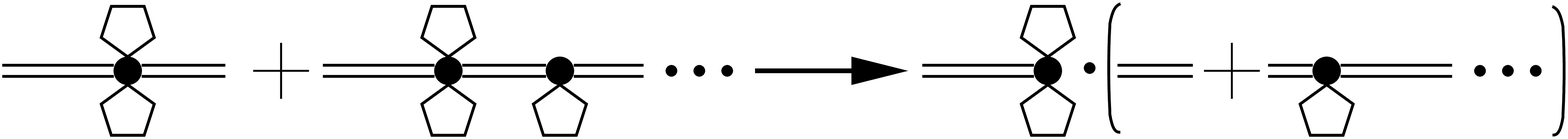}}
 \caption{\label{fig:split_diagram}(a) A diagram contributing to $X_4$ is split right after the first $4$-encounter and decomposes into two separate diagrams where the second one contributes to $T_{ij}^{eh}$. (b) To sum over all diagrams starting with an $l$-encounter we can remove a factor corresponding to the first encounter (and its side trees) and a sum again over the transmission diagrams.}
\end{figure}

We first order the sum over all diagrams contributing in leading order in the channel number with respect to the first encounter. Then the first summand is of course the diagram corresponding to the upper left trajectory in \fref{fig:diagonal_contr_cond}. Next there are all diagrams whose first encounter is a $2$-encounter followed by all diagrams whose first encounter is a $3$-encounter etc. Note that we also allow for the first encounter to touch the superconductor or (if the first encounter is a $2$-encounter and $i=j$) the normal lead. We denote the contribution of the sum over all diagrams having an $l$-encounter as their first encounter and contributing to $T_{ij}^{ee}$ in leading order in the number of channels by $X_l$. We may include the diagonal diagram without any encounter by setting $X_1=N_iN_j/N$. The transmission coefficients are then given by $T_{ij}^{ee}=\sum_{l\geq 1}X_l$. Now we fix $l\geq 2$ and split all diagrams contributing to $X_l$ right after the first encounter into one part consisting of the first path pair and the first encounter together with its side trees and the remaining part such as indicated in \fref{fig:split_diagram_b}. Note that the diagonal type path pair leaving the first encounter is completely included in the second part. Since the diagrammatic rules are multiplicative the contribution of a diagram is given by the product of the two parts and hence they all have a common factor which is given by the first diagonal type link, the first encounter and the side trees emerging from it. To sum over all diagrams starting with an $l$-encounter we pull out this factor and are left with a sum over the transmission diagrams as depicted in \fref{fig:split_diagram_b}. This sum runs over all possible diagrams contributing to $T_{ij}^{ee}$ if the first encounter is left by an electron and to $T_{ij}^{eh}$ if it is left by a hole. However in order to be able to fully identify the sum over the second parts as the transmission we have to reassign the contributed number of channels $N_j$ contributed by the first path pair leaving lead $j$ to the second part. Due to the two possibilities of which transmission coefficient the remaining diagrams contribute to we split $X_l$ into two parts \[X_l=A_l^eT_{ij}^{ee}+B_l^eT_{ij}^{eh},\] where $A_l^e$ is the contribution of the first e-\es~pair and the $l$-encounter the path pair enters together with all side trees and with the entering and exiting quasiparticle being the same. $B_l^e$ is the same but with the entering and exiting quasiparticle being different.

The transmission coefficients may therefore be written as
\addtocounter{equation}{1}
\begin{align}
T_{ij}^{ee}&=\fr{N_iN_j}{N}+\sul{l=2}{\infty}A_l^eT_{ij}^{ee}+\sul{l=2}{\infty}B_l^eT_{ij}^{eh}
\label{islrecA},
\tag{\theequation a}\\
 T_{ij}^{eh}&=\sul{l=2}{\infty}A_l^hT_{ij}^{eh}+\sul{l=2}{\infty}B_l^hT_{ij}^{ee}.
\label{islrecB}
\tag{\theequation b}
\end{align}
$A_l^h$ and $B_l^h$ are the same as $A_l^e$ and $B_l^e$, respectively, but with electrons and holes exchanged. \Eref{islrecB} is obtained in the same way as \eref{islrecA} but with the additional condition that there is no diagram without any Andreev reflection contributing to it since converting an electron to a hole requires at least one Andreev reflection and therefore one encounter. The formulae for $T_{ij}^{hh}$ and $T_{ij}^{he}$ are the same but with $e$ and $h$ exchanged.

\begin{figure}
 \includegraphics[width=\columnwidth]{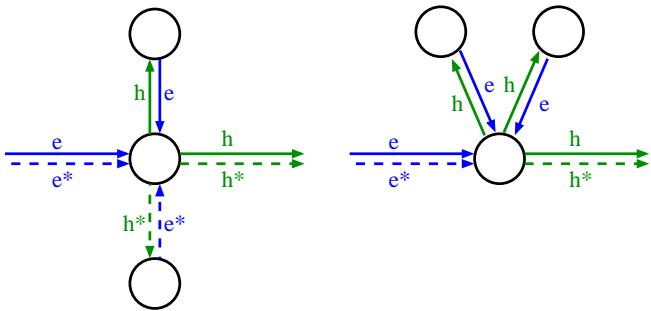}
 \caption{\label{andreevenc}Simple examples for encounters touching a superconductor. The electron paths are shown green while the hole paths are shown red. The solid lines belong to $\zeta$ while the dashed ones belong to $\zeta^\prime$. If the quasiparticles entering an encounter touching the superconductor following an diagonal-type path pair the diagonal-type path pairs leaving it are traversed by holes and vice versa.}
\end{figure}

The next task is to find out what causes the encounter which is entered by an electron to be left by an electron or a hole. The trajectories in \fref{3rd_order} and their corresponding diagrams in \fref{3rd_order_diag} indicate that, as long as the first encounter does not touch the superconductor, an encounter entered by an electron is left by an electron if the number of side trees on each side of the diagonal backbone emerging from this encounter is even (such as in the diagrams ee3II and he3V) and by a hole if it is odd (such as in the diagrams ee3III and he3VII). If the first encounter however touches the superconductor the encounter is always left by a hole if it was entered by an electron. This is also indicated in \fref{andreevenc}

We will now show that this indeed holds for all `diagonal encounters' entered by a diagonal type e-\es~pair by starting with considering encounters not touching the superconductor. Since an $l$-encounter connects $2l$ links to each other, each diagonal $l$-encounter, where $2$ of the links belong to the backbone, provides in total $(2l-2)$ side trees implying that if the number of $\zeta$-side trees is even the number of $\zeta^\prime$-side trees is even too, or they are both odd. Furthermore each side tree provides an odd number of Andreev reflection and therefore a conversion of an electron into a hole or vice versa, since each of its $l$-encounters is left by $(2l-1)$ additional path pairs and each path pair increases the number of Andreev reflections by one (this is closely related to the fact that we consider diagrams contributing at leading order in the number of channels). Thus, as long as the first encounter does not touch the superconductor, the entering electron leaves the encounter as an electron if the number of side trees $\tilde{p}$ built by $\zeta$ is even and as a hole if the number of side trees built by $\zeta$ is odd.

\begin{figure*}
\subfigure[\label{fig:sliding_a}]{\includegraphics[width=0.38\textwidth]{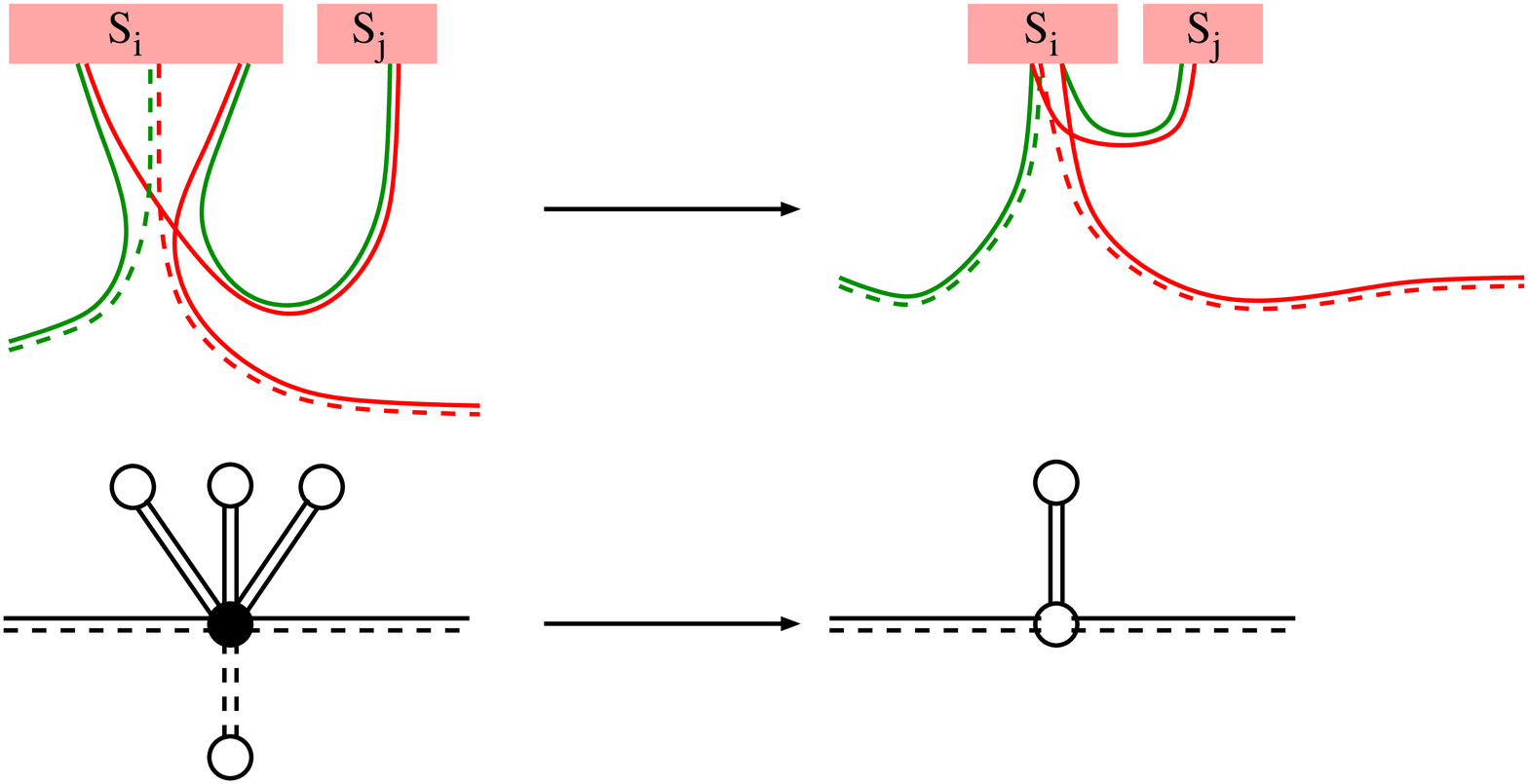}}
\rule{0.01\textwidth}{0pt}
\subfigure[\label{fig:sliding_b}]{\includegraphics[width=0.58\textwidth]{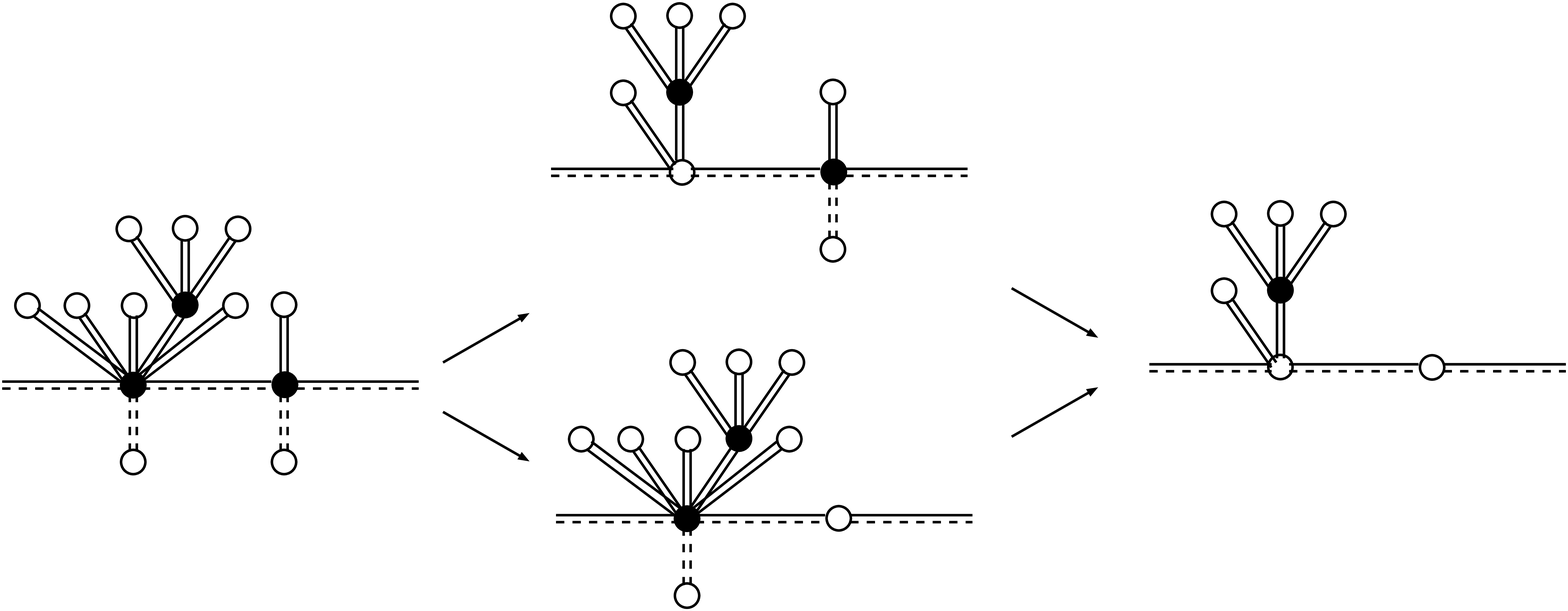}}
\caption{\label{fig:sliding} (a) An $3$-encounter may touch the superconductor S$_i$ if the odd numbered side trees have zero characteristic and hit the same superconductor. The number of Andreev reflections stays the same. If the encounter touches the superconductor an entering electron is converted into a hole. (c) A more complicated diagram with two diagonal encounters that may touch the superconductor. Note that additionally the fourth side tree may also touch the superconductor but this does not affect the diagonal encounter but is instead included in the side tree recursion.}
\end{figure*}

However if the first diagonal $l$-encounter touches the superconductor the first side tree starts with a hole instead of an electron and is therefore left by an electron. Since the electron leaving the first side tree hits the superconductor the second side tree again starts with a hole. If one proceeds inductively one finds that every side tree starts with a hole and is left by an electron which after that undergoes again an Andreev reflection. Therefore if the first encounter entered by an electron touches the superconductor it is always left by a hole and we can view it as arising from an $l$-encounter with an odd number $\tilde{p}$ of $\zeta$-side trees slid into the superconductor as indicated in \fref{fig:sliding} and therefore contributes to $B_l^e$.
An $l$-encounter may then touch the superconductor if the number of $\zeta$-side trees $\tilde{p}$ is odd and the odd numbered $\zeta$-side trees, which are the side trees traversed by $\zeta$ after an odd number of traversals of the encounter, as well as the odd numbered $\zeta^\prime$-side trees have zero characteristic (\textit{i.e.}~consist of just one link and one Andreev reflection). Moreover the links of the odd numbered side trees have to hit the same superconductor such that the channels can coincide. When sliding such an encounter into the superconductor the channels at which the odd numbered side trees hit the superconductor coincide and the links vanish. Therefore beside the diagonal-type path pairs from such a diagonal $l$-encounter touching the superconductor $p=(\tilde{p}-1)/2$ even numbered $\zeta$-side trees starting with a hole and $[(2l-2-\tilde{p})-1]/2=l-2-p$ even numbered $\zeta^\prime$-side trees, which also start with a hole emerge.

Thus if we denote the contribution of the first $\alpha$-$\alpha^*$ pair and of the first $l$-encounter inside the dot with $\tilde{p}$ $\zeta$-side trees by $x_{l,\tilde{p}}^\alpha$ and the contribution of the Andreev reflections provided by the first $l$-encounter touching the superconductor S$_j$ created by sliding an $l$-encounter with originally $\tilde{p}$ $\zeta$-side trees into the superconductor S$_j$ by $z_{l,\tilde{p},j}^\alpha$, we find
\begin{widetext}
\addtocounter{equation}{1}
\begin{align}
A_l^e&=\sul{p=0}{l-1}x_{l,2p}^e\rbr{P^e}^{p}\rbr{P^h}^p\cbr{\rbr{P^e}^*}^{l-1-p}\cbr{\rbr{P^h}^*}^{l-1-p},
\tag{\theequation a}
\label{eq:coeff_definition_Ae}\\
B_l^e&=\sul{p=0}{l-2}\cbr{x_{l,2p+1}^e\rbr{P^e}^{p+1}\rbr{P^h}^p\cbr{\rbr{P^e}^*}^{l-1-p}\rbr{\rbr{P^h}^*}^{l-2-p}+\sul{j}{}z_{l,2p+1,j}^e\rbr{P^h}^p\cbr{\rbr{P^h}^*}^{l-2-p}},
\label{eq:coeff_definition_Be}
\tag{\theequation b}\\
A_l^h&=\sul{p=0}{l-1}x_{l,2p}^h\rbr{P^h}^{p}\rbr{P^e}^p\cbr{\rbr{P^h}^*}^{l-1-p}\cbr{\rbr{P^e}^*}^{l-1-p},
\tag{\theequation c}
\label{eq:coeff_definition_Ah}\\
B_l^h&=\sul{p=0}{l-2}\cbr{x_{l,2p+1}^h\rbr{P^h}^{p+1}\rbr{P^e}^p\cbr{\rbr{P^h}^*}^{l-1-p}\rbr{\rbr{P^e}^*}^{l-2-p}+\sul{j}{}z_{l,2p+1,j}^h\rbr{P^e}^p\cbr{\rbr{P^e}^*}^{l-2-p}},
\label{eq:coeff_definition_Bh}
\tag{\theequation d}
\end{align}
\end{widetext}
where we have used that $\tilde{p}$ has to be even for $A_l^\alpha$ and thus replaced $\tilde{p}=2p$ and odd for $B_l^\alpha$ with $\tilde{p}=2p+1$.

The next and final step is to find the contribution of the encounters. For that we would like to recall the diagrammatic rule for an $l$-encounter traversed by trajectories with energies $\pm\en$ and in presence of a magnetic field $b$ from Ref.~\onlinecite{ref:diagramatic}:
\begin{itemize}
\item An $l$-encounter inside the dot contributes a factor $-N\rbr{1+\eta\rmi\en+\mu^2b^2}$.
\end{itemize}
Here $\eta$ is the difference between the number of traversals of e-stretches and the number of traversals of \es-stretches and $\mu$ is the difference between the number of $\zeta$-stretches traversed in a certain direction and the number of $\zeta^\prime$-stretches traversed in the same direction. Since every electron path of the side tree is retraced by a hole every second stretch connected to a $\zeta$-side tree is an e-stretch and they are all traversed in the same direction we choose arbitrarily as `positive'. Therefore if the number of $\zeta$-side trees is even the number of e-stretches traversed in positive direction is simply $\tilde{p}/2$. If $\tilde{p}$ is odd we have to account for the fact that the first $\zeta$-side tree starts with an electron and the last one also does. Thus there are $(\tilde{p}+1)/2$ e-stretches traversed in positive direction in the encounter. In the same way one finds that the number of \es-stretches are $(2l-2-\tilde{p})/2$ and $[(2l-2-\tilde{p})+1]/2$, respectively. Since the diagonal path pair is traversed by $\zeta$ and $\zeta^\prime$ in the same direction the directions of the \es-paths are also positive. Since the holes retrace the electron paths their directions in the encounters is negative. Thus in both cases one finds that $\eta=\mu=(\tilde{p}-l+1)$. So
\begin{equation}
 x_{l,\tilde{p}}=-\cbr{1+\rbr{\tilde{p}-l+1}\rmi\en+\rbr{\tilde{p}-l+1}^2b^2}.
 \label{eq:diag_encounter_indot}
\end{equation}

For the contribution $\tilde{z}_{l,p,j}^\alpha$ which arises by sliding an $l$-encounter into the superconductor \cite{countingstatistic}, as shown in \fref{fig:sliding} we remember that the number $\tilde{p}$ of $\zeta$-side trees emerging from it is odd and the odd numbered $\zeta$-side trees as well as the odd numbered $\zeta^\prime$-side trees consist of only one path pair and one Andreev reflection (\textit{i.e.}~they have zero characteristic). Moreover the Andreev reflections of the odd numbered side trees have to be all at the same superconductor. The contribution of the encounter itself and the first path pair is then $N_{\mathrm{S}_j}/N$. However we also include the factors contributed by the Andreev reflections in $z_{l,\tilde{p},j}^\alpha$, too, which are stated in \sref{cond_sidetrees}. As for the side trees, these phase factors may be determined by looking at the odd numbered side trees before sliding the encounter into the superconductor since the number of Andreev reflections of the $\zeta$- and $\zeta^\prime$-trajectory can not change when sliding the encounter into the superconductor. Consider the $p+1=(\tilde{p}+1)/2$ odd numbered side trees which have zero characteristic and hit say S$_i$: The Andreev reflections provided by these side trees convert an electron into a hole and thus the Andreev reflections each provide a factor $-\rmi\rme^{-\rmi\phi_i}$. Hence in total the Andreev reflections of the odd numbered $\zeta$-side trees provide a factor $\rbr{-\rmi}^{p+1}\rme^{-\rmi(p+1)\phi_i}$. Analogously the Andreev reflections of the odd numbered $\zeta^\prime$-side trees contribute a factor $\rmi^{l-p-1}\rme^{\rmi(l-p-1)\phi_i}$. Thus in the case of two superconductors with phases $\phi_1=-\phi_2=\phi/2$ the phase factor included in $z_{l,\tilde{p},1}^e$ is given by $(-\rmi)^{p}\rmi^{l-p-2}\rme^{-\rmi(2p-l+2)\phi/2}$. We thus have
\begin{equation}
 z_{l,\tilde{p},1}=\fr{N_{\mathrm{S}_1}}{N}\rmi^{l-\tilde{p}-1}\rme^{-\rmi\rbr{\tilde{p}-l+1}\phi/2}.
\end{equation}
For $z_{l,\tilde{p},2}^e$ we have to exchange $\phi\leftrightarrow-\phi$ and replace $N_{\mathrm{S}_1}$ by $N_{\mathrm{S}_2}$. Moreover $z_{l,\tilde{p},j}^h=z_{l,\tilde{p},j}^e\vert_{\phi\to-\phi}$. Therefore we have
\begin{widetext}
\addtocounter{equation}{1}
\begin{align}
 A_l^e=-\sul{p=0}{l-1}&\left[1+\rmi\rbr{2p-l+1}\en+\rbr{2p-l+1}^2b^2\right]\rbr{P^e}^{p}\rbr{P^h}^p\cbr{\rbr{P^e}^*}^{l-p-1}\cbr{\rbr{P^h}^*}^{l-p-1}
\label{eq:encounter_coeff_A}
\tag{\theequation a}\\
B_l^e=-\sul{p=0}{l-2}&\Bigg[\rbr{1+\rmi\rbr{2p-l+2}\en+\rbr{2p-l+2}^2b^2}\rbr{P^e}^{p+1}\rbr{P^h}^p\cbr{\rbr{P^e}^*}^{l-p-1}\cbr{\rbr{P^h}^*}^{l-p-2} \nonumber \\
 & -\fr{x(1+y)\rme^{-\rmi\rbr{2p-l+2}\phi/2}\rbr{-\rmi P^h}^p\rbr{\rmi \rbr{P^h}^*}^{l-p-2}}{2\rbr{1+x}}-\fr{x(1-y)\rme^{\rmi\rbr{2p-l+2}\phi/2}\rbr{-\rmi P^h}^p\rbr{\rmi \rbr{P^h}^*}^{l-p-2}}{2\rbr{1+x}}\Bigg].
 \label{eq:encounter_coeff_B}
 \tag{\theequation b}
\end{align}
\end{widetext}
where we again used $y=(N_{\mathrm{S}_1}-N_{\mathrm{S}_2})/N_\mathrm{S}$. The case $N_{\mathrm{S}_1}=N_{\mathrm{S}_2}$ is then obtained by setting $y=0$ while the case of just one superconducting lead corresponds to $y=\pm1$. Since exchanging electrons and holes corresponds to replacing $\phi$ by $-\phi$, $A_l^h$ and $B_l^h$ are obtained by the same formulae but with $\phi$ replaced by $-\phi$ including an exchange $P^e\leftrightarrow P^h$. The sums may be performed using geometric series and yield our main result. Along with \eref{islrecA} and \eref{islrecB} it contains all rhe diagrams, and their semiclassical contributions, generated recursively.\\
Note that if the numbers of channels of the superconducting leads are equal the symmetry of $P$ towards the phase implies that $A_l^\alpha$ and $B_l^\alpha$ are symmetric in $\phi$ yielding $A_l^e=A_l^h$ and $B_l^e=B_l^h$ and thus $T_{ij}^{ee}=T_{ij}^{hh}$ and $T_{ij}^{he}=T_{ij}^{eh}$.

%
%
Therefore we now have all the necessary utilities to calculate the conductance of Andreev billiards with two superconducting islands. When the incoming and outgoing lead are the same, $i=j$, and the first encounter is an $2$-encounter this encounter may enter the lead. In this case however the encounter simply contributes $N_i$ and the diagrams consist of one $\zeta$-side tree and one $\zeta^\prime$-side tree. The contribution of these diagrams is therefore simply
\addtocounter{equation}{1}
\begin{align}
 \delta_{ij}N_i\vert P^e\vert^2 & & & \text{if the dot is entered by an electron,}
 \label{reflectioncorr_e}
 \tag{\theequation\alph{mycounter}}
 \addtocounter{mycounter}{1}\\
 \delta_{ij}N_i\vert P^h\vert^2 & & & \text{if the dot is entered by a hole}.
 \label{reflectioncorr_h}
 \tag{\theequation\alph{mycounter}}
 \addtocounter{mycounter}{1}
\end{align}
The transmission coefficients necessary for calculating the conductance may be calculated by evaluating the side tree contribution by solving \eref{geq}, inserting this into (\ref{eq:encounter_coeff_A},b) and performing the sums and finally inserting into (\ref{islrecA},b) and solving for the transmission coefficients.

\section{Conductance with Superconducting islands}
\label{island_sec}
We now evaluate the conductance of Andreev billiards with two normal leads. We first consider a chaotic quantum dot coupled to two normal conducting leads and one or two isolated superconductors with equal number of channels as shown in \fref{island_setup}. The chemical potential of the superconducting lead is then adjusted by the dot such that the net current in the superconductor vanishes. The dimensionless conductance $g=\pi\hbar I/(\rme^2V)$ with $I$ the current and $V$ the voltage drop between the two normal leads, in this case is given at zero temperature by \cite{islandconductance}
\begin{equation}
g=T_{21}^{ee}+T_{21}^{he}+2\fr{T_{11}^{he}T_{22}^{he}-T_{21}^{he}T_{12}^{he}}{T_{11}^{he}+T_{22}^{he}+T_{21}^{he}+T_{12}^{he}}.
 \label{nsn_island_current}
\end{equation}

\subsection{Low temperature}
\subsubsection{One superconductor}
\begin{figure}
\includegraphics[width=0.7\columnwidth]{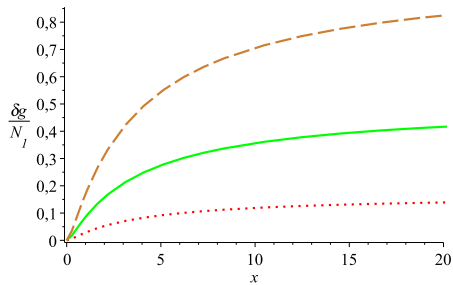}
 \caption{\label{isl_rmt}Conductance correction scaled by $N_1$ as a function of the number of channels of the isolated superconductor $x=N_\mathrm{S}/N_\mathrm{N}$ for $N_2/N_1=0.2$ (dotted line), $N_2/N_1=1$ (solid line) and $N_2/N_1=100$ (dashed line)}
\end{figure}

Using Eqs.~(\ref{reflectioncorr_e},b) in the simplest case without phase difference the random matrix result for the conductance correction $\delta g=g-g_{cl}$ \cite{rmttrans} can be reproduced:
\begin{equation}
 \delta g=\fr{N_1N_2}{N_\mathrm{N}}\cbr{x+\fr{1}{2}\rbr{1+x}^2-\fr{1}{2}\rbr{1+x}\sqrt{1+6x+x^2}}.
 \label{island_rmt}
\end{equation}
The conductance correction is shown in \fref{isl_rmt} as a function of $x=N_\mathrm{S}/N_\mathrm{N}$.
From \eref{island_rmt} it can be easily seen that the conductance in this case is symmetric in exchanging $N_1$ and $N_2$, as one would expect due to the symmetry of the setup. Moreover with this setup the superconductor always increases the conductance. In the limit of large numbers of superconducting channels we find that the conductance is doubled compared to the classical limit $x=0$ and hence approaches the conductance of an N-S interface \cite{btktheory}.

\subsubsection{Two superconductors with phase difference}
\label{subsubsec:island_phase_difference}
\begin{figure}
\subfigure[\label{island_phase_rmt_a}]{\includegraphics[width=0.45\columnwidth]{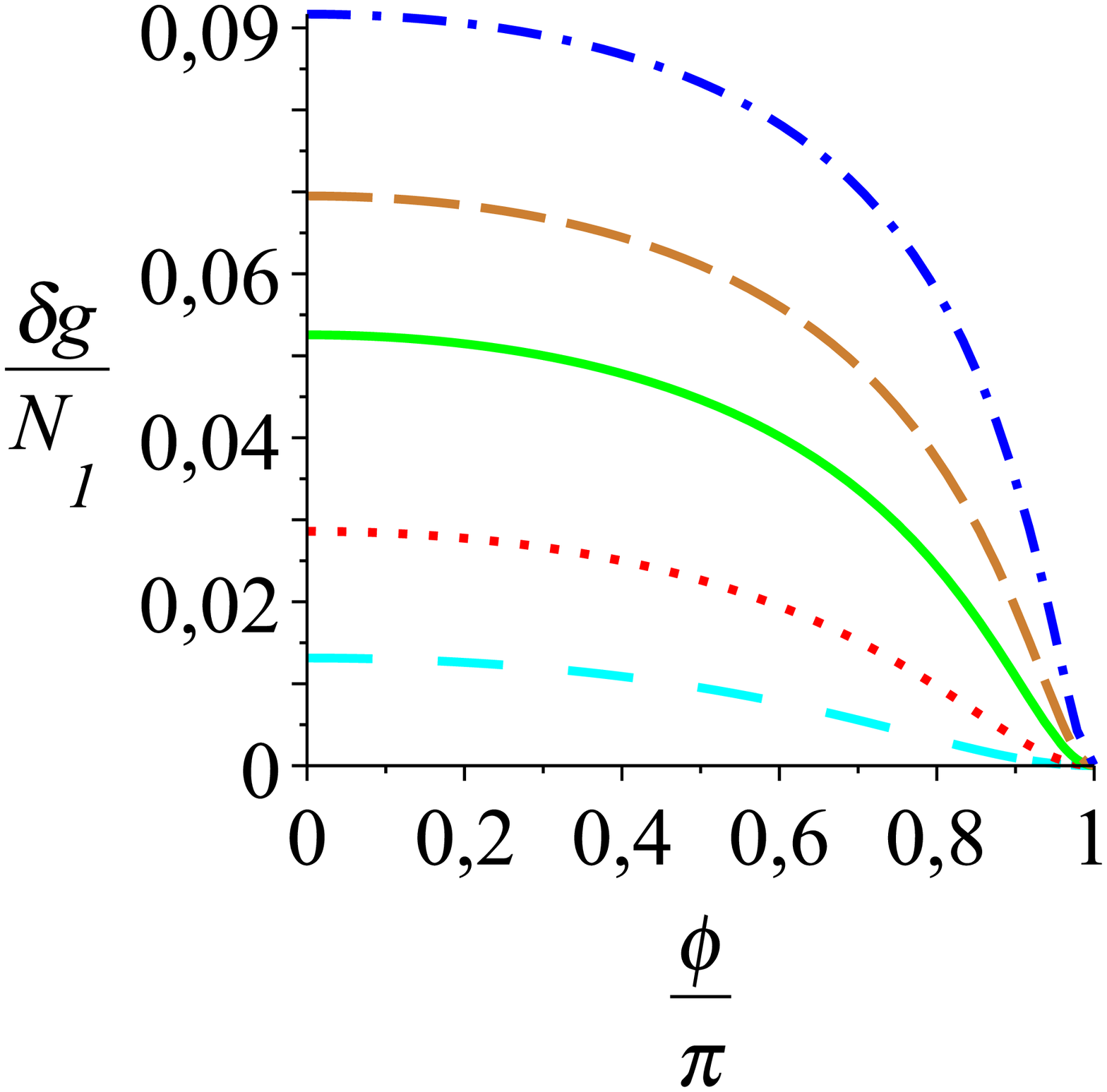}}
 \rule{0.033\columnwidth}{0pt}
\subfigure[\label{island_phase_rmt_b}]{\includegraphics[width=0.45\columnwidth]{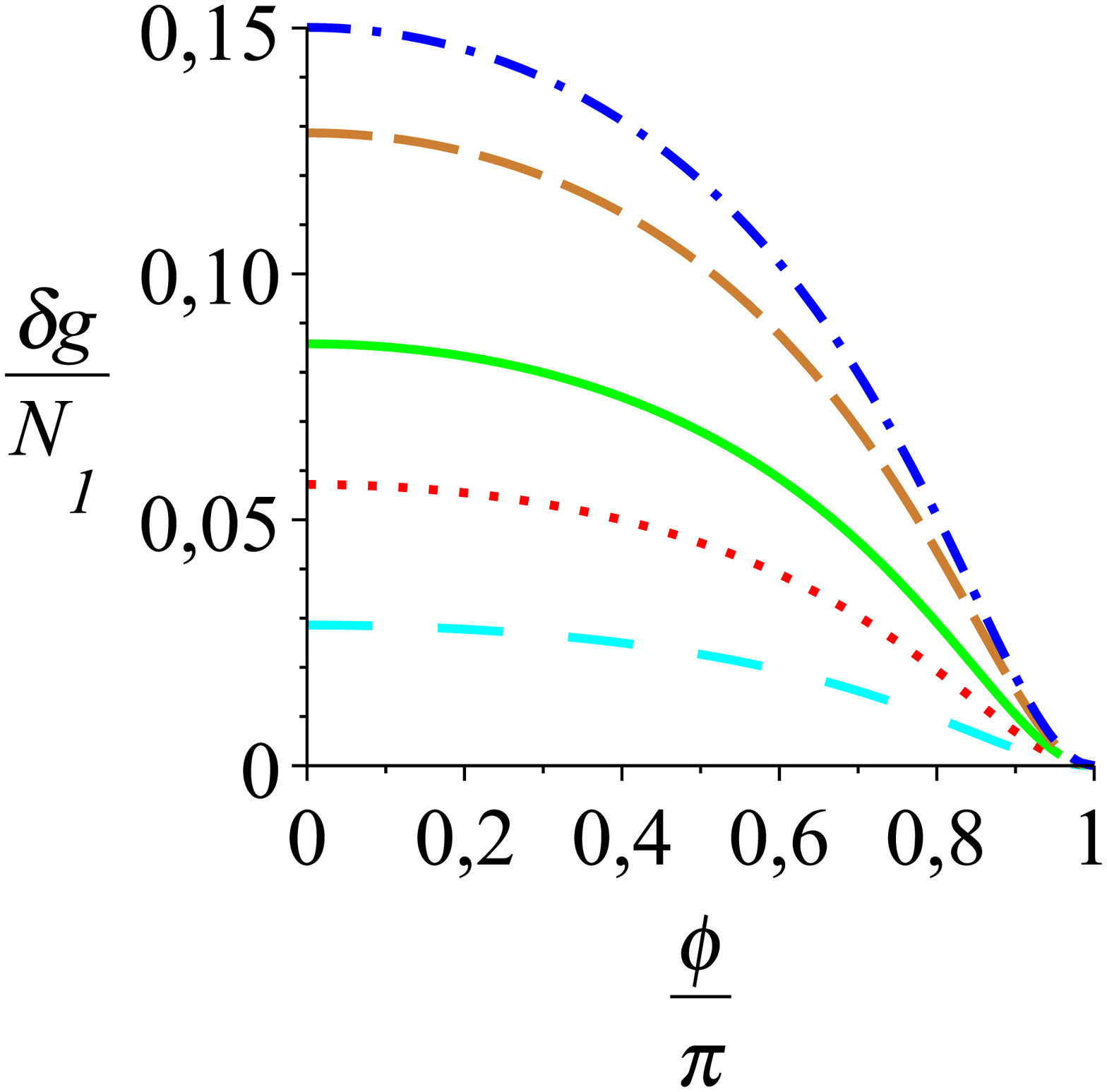}}
 \caption{\label{island_phase_rmt} The conductance correction as a function of the phase difference for $N_{\mathrm{S}_1}=N_{\mathrm{S}_2}$ (a) $N_2/N_1=0.2$ and $x=0.5$ (space dashed line), $x=1$ (dotted line), $x=2$ (solid line), $x=3$ (dashed line), $x=5$ (dashed dotted line), (b) $x=1$ and $N_2/N_1=0.2$ (space dashed line), $N_2/N_1=0.5$ (dotted line), $N_2/N_1=1$ (solid line), $N_2/N_1=3$ (dashed line) and $N_2/N_1=7$ (dashed dotted line).}
 \end{figure}

\begin{figure}
\subfigure[\label{island_phase_diff_channels_a}]{\includegraphics[width=0.45\columnwidth]{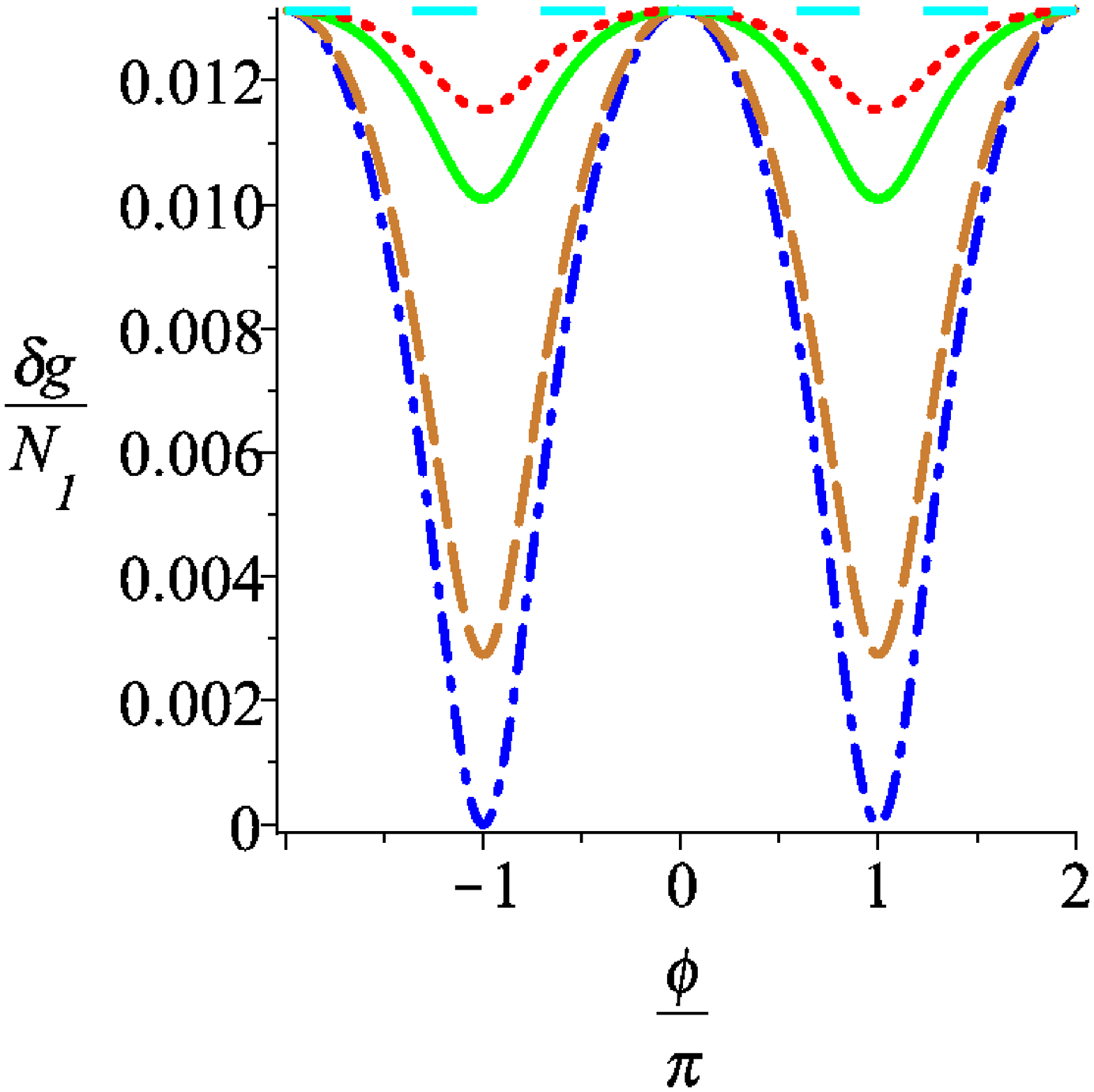}}
 \rule{0.033\columnwidth}{0pt}
\subfigure[\label{island_phase_diff_channels_b}]{\includegraphics[width=0.45\columnwidth]{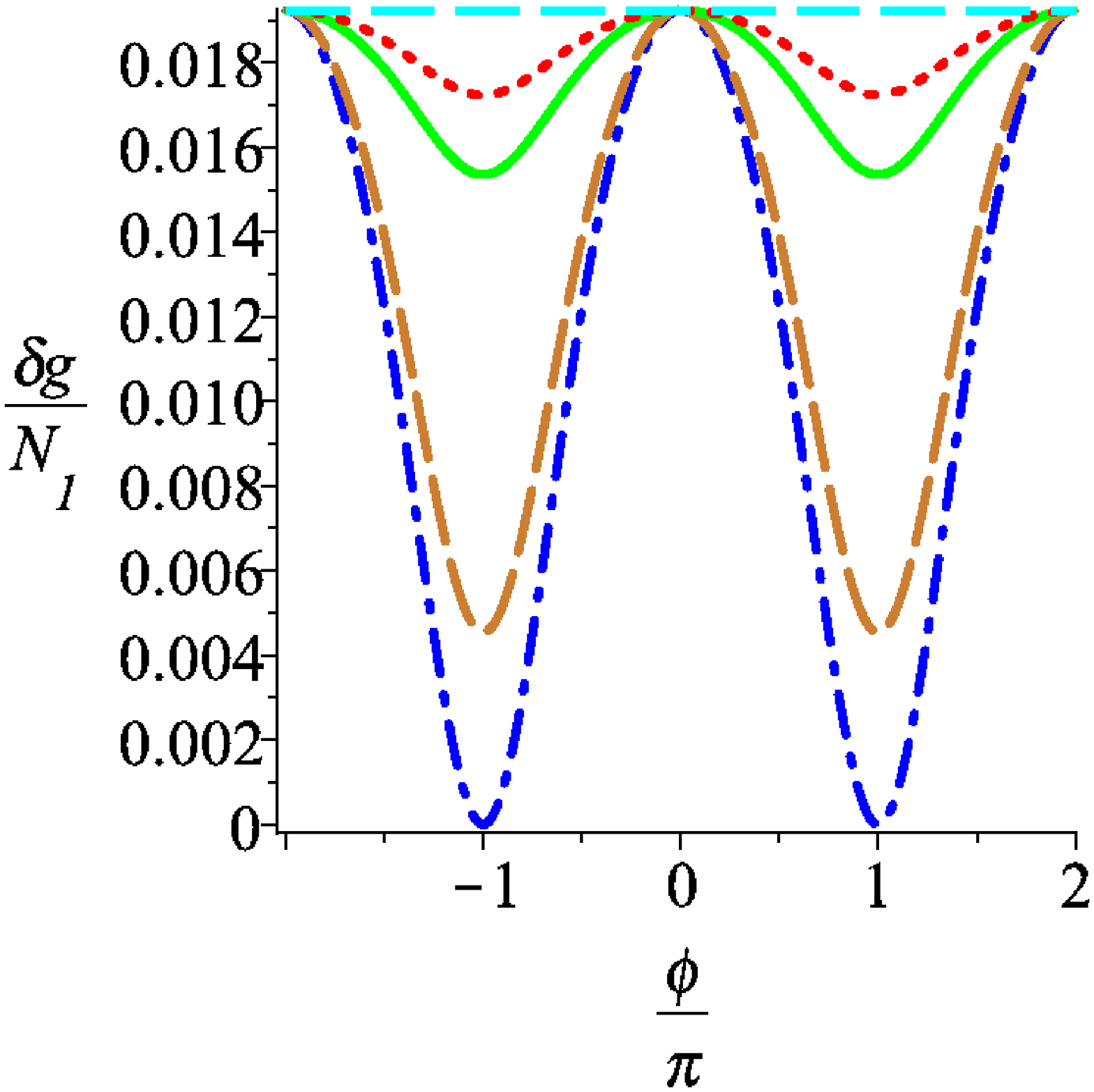}}
\caption{\label{island_phase_diff_channels} Dependence on the difference of the numbers of superconducting channels with $y=1$ (space dashed line), $y=0.95$ (dotted line), $y=0.9$ (solid line), $y=0.5$ (dashed line) and $y=0$ (dashed dotted line) for (a) $N_2/N_1=0.2$, $x=0.5$ and (b) $N_2/N_1=7$, $x=0.2$.}
\end{figure}

In Ref.~\onlinecite{rmttrans} using RMT for a finite phase difference the authors could calculate the transmission only numerically but for all $N$. Moreover they restricted themselves to the case $N_{\mathrm{S}_1}=N_{\mathrm{S}_2}$. With our semiclassical approach however we are able to calculate it at zero temperature analytically for all cases, as long as $N$ is large. Using this we could reproduce the large-$N$ limit in Ref.~\onlinecite{rmttrans} shown in \fref{island_phase_rmt} for the case $N_{\mathrm{S}_1}=N_{\mathrm{S}_2}$.
 Moreover \fref{island_phase_diff_channels} shows the dependence of the conductance on the channel number difference of the superconductors. The conductance correction vanishes for $\phi\rightarrow\pi$ since the phase accumulated at each Andreev reflection causes destructive interference. In this case the conductance is a monotonic function of the phase difference $\phi$ up to $\pi$.

The symmetry found in Figs.~\ref{island_phase_rmt} and \ref{island_phase_diff_channels} results from the fact that electrons and holes contribute symmetrically to the conductance. The $2\pi$-periodicity may also be found using \earef{eq:encounter_coeff_A}{,b}: If we increase $\phi$ by $2\pi$ the side tree contribution changes its sign. This does not affect $A_l^\alpha$ or the first part of $B_l^\alpha$. If $l$ is odd in the last two terms in $B_l^\alpha$ changing the sign of the side tree contribution results in a change of the sign of these two parts. However increasing the phase by $2\pi$ also yields an exchange of the sign of the phase factors cancelling the change of sign of the side tree contributions. If $l$ is even we again have an even number of side tree contributions and the phase factors also do not change their sign.

The crossover from two superconductors to just one superconductor is smooth and monotonic as shown in \fref{island_phase_diff_channels}. We found that the bigger the difference in the numbers of channels the faster the amplitude changes.

\subsubsection{Magnetic field}
\begin{figure}
 \includegraphics[width=0.7\columnwidth]{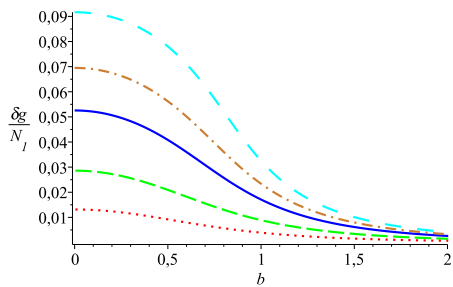}
 \caption{\label{isl_mag} The conductance correction for the setup with one superconducting island as a function of the magnetic field dependence for $N_2/N_1=0.2$ and $x=0.5$ (space dotted line), $x=1$ (dotted line), $x=2$ (solid line), $x=3$ (dashed line) and $x=5$ (dashed dotted line)}
\end{figure}

Whitney and Jacquod found in Ref.~\onlinecite{nsntrans} that for small $x$ the conductance of an Andreev quantum dot with an isolated superconductor decays at $T=0$ with increasing magnetic field as $(1+b^2)^{-2}$. For higher orders in $x$ the $(1+b^2)^{-2}$-decay mixes up with terms decaying as $(1+b^2)^{n}$ with $n\geq2$. This leads to the behaviour shown in \fref{isl_mag}. It can be seen that the conductance correction decays very quickly. Since the magnetic field enters the transmission coefficients and therefore the conductance quadratically the conductance is symmetric in reversing the magnetic field.

\subsection{Temperature dependence}
\begin{figure}
 \includegraphics[width=0.7\columnwidth]{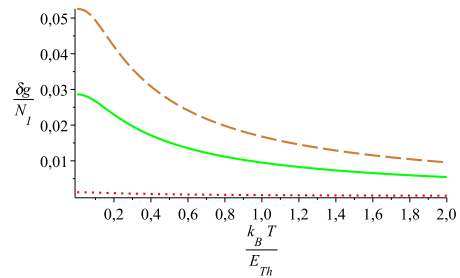}
 \caption{\label{isl_temp}Temperature dependence of the conductance correction for the setup with an superconducting island for $N_2/N_1=0.2$ and $x=0.1$ (dotted line), $x=1$ (solid line) and $x=2$ (dashed line).}
\end{figure}

If we want to allow for non-zero temperature each transmission and reflection coefficient in \eref{nsn_island_current} has to be multiplied by the negative derivative of the Fermi function and integrated over energy. We evaluated these integrals numerically using Gaussian quadrature with a total accuracy $10^{-10}$ and a truncation of the integral at $\en=100\theta$ with $\theta=k_\mathrm{B}T/E_\mathrm{T}$ being the temperature measured in units of the temperature corresponding to the Thouless energy. Doing so we find that the superconducting island obeys a monotonic temperature dependence: The conductance correction has its maximum at $T=0$. For higher temperature it is damped due to the mixing with higher energies for which the side tree contributions become smaller because of the loss of coherence of the electrons and holes. As the temperature tends to infinity the conductance correction vanishes slowly. In \fref{isl_temp} we plotted the conductance correction of the setup with one superconducting island versus the temperature.

\section{Conductance with Superconducting leads}
\label{lead_sec}
Next we consider superconductors with externally controlled chemical potential. In particular we will consider superconductors lying on the same chemical potential as one of the two normal conducting leads, say lead 2. Such a setup is schematically shown for the case of one superconducting lead in \fref{lead_setup}.
The current in lead $i$ may be calculated by the Landauer-type expression \cite{generalized_landauer,ref:landauer_type_multi}
\begin{align}
 I_i=\fr{2e}{\hbar}\sul{j=1}{2}\intg{0}{\infty}{\epsilon}&\left[2N_i\delta_{ij}-T_{ij}^{ee}+T_{ij}^{he}-T_{ij}^{hh}+T_{ij}^{eh}\right] \nonumber\\
  & \cdot\rbr{-\pdiff{f}{\en}}(\mu_j-\mu_S)
 \label{nsn_landauer}
\end{align}
where $f=\cbr{\exp\rbr{-\en/\theta}+1}$ is the Fermi function with the temperature again measured in units of the Thouless energy $E_\mathrm{T}$, $\mu_j$ and $\mu_S$ are the chemical potentials in the normal conducting lead $j$ and in the superconductor, respectively.

Of course we could use the transmission coefficients themselves calculated in \sref{sec:cond_transmission_probs}. However we would like to present a slightly different way to calculate the conductance here which in the case of the superconducting leads simplifies the calculation. For simplicity we will present this way only for the case that the numbers of channels of the superconducting leads are equal since the modifications one would have to do in order to include different numbers of channels are the same as in \sref{sec:cond_transmission_probs}. According to \eref{nsn_landauer} we have to calculate the difference between the Andreev and normal transmission, namely  $\tilde{T}_{ij}^e=T_{ij}^{ee}-T_{ij}^{he}$ and $\tilde{T}_{ij}^{h}=T_{ij}^{eh}-T_{ij}^{hh}$. To do this (considering only the case of an incident electron) we essentially perform the same steps as before and split the diagrams at their first $l$-encounter. In the same way as above we find that the sum over the remaining diagrams again contributes to $\tilde{T}_{ij}^e$ or $\tilde{T}_{ij}^h$ depending on whether the quasiparticle leaving the encounter is an electron or a hole. An additional sign arises as follows: Consider for example a diagram contributing to $T_{ij}^{ee}$ thus contributing to $\tilde{T}_{ij}^e$ with positive sign. Then if the number of $\zeta$-side trees arising from the first encounter is even the remaining diagrams contribute to $T_{ij}^{ee}$, too, and therefore it again contributes with a positive sign to $\tilde{T}_{ij}^e$. However if the number of $\zeta$-side trees is odd the remaining diagrams contribute to $T_{ij}^{eh}$ . Hence it contributes to $\tilde{T}_{ij}^h$.

Taking into account the diagram connecting lead $i$ and lead $j$ directly, which has no Andreev reflection this diagram contributes to $T_{ij}^{ee}$ and therefore to $+\tilde{T}_{ij}^e$. According to the diagrammatic rules this contribution is simply given by $N_iN_j/[N_\mathrm{N}(1+x)]$. The transmission difference $\tilde{T}_{ij}^{e}$ therefore read
\begin{equation}
 \tilde{T}_{ij}^e=\fr{N_iN_j}{N_\mathrm{N}(1+x)}+\sul{l=2}{\infty}A_l^e\tilde{T}_{ij}^e+\sul{l=2}{\infty}B_l^e\tilde{T}_{ij}^h
 \label{leadreceqs}
\end{equation}
with $A_l^e$ and $B_l^e$ given by \eref{eq:encounter_coeff_A} and \eref{eq:encounter_coeff_B}, respectively, and the same equation holds for $\tilde{T}_{ij}^h$ with an exchange of $e$ and $h$ yielding a replacement of $\phi$ by $-\phi$ in the coefficients $A$ and $B$ including an exchange $P^e\leftrightarrow P^h$. If the numbers of channels of the superconductors are the same this also implies a symmetry in the exchange of electrons and holes and $\tilde{T}_{ij}^e=-\tilde{T}_{ij}^h$ which reduces the number of variables by a factor of 2 and therefore makes the arising equations easier to solve.

Since we consider the chemical potential of the superconductors being the same as that of the second lead the contribution to the conductance is given by the transmission coefficient for reflecting an electron entering the cavity from lead 1 back into lead 1 again. The conductance $g$, defined by $I_1=eg\rbr{\mu_1-\mu_S}/\pi\hbar$, therefore reads
\begin{equation}
 g=-2N_1\intg{0}{\infty}{\en}\rbr{1+\left|P\right|^2-\fr{N_1}{N\rbr{1-A+B}}}\pdiff{f}{\en}.
 \label{lead_cond}
\end{equation}
Note that this setup induces an asymmetry due to the channel numbers: Since the chemical potential of the superconductor is the same as that of lead $2$ one cannot exchange the two channels and therefore the solution will not (in general) be symmetric under the exchange of $N_1$ and $N_2$.

\subsection{Low temperatures}
\subsubsection{One superconductor}
Let us consider first the simplest case of no phase difference and the absence of magnetic fields. Moreover we consider sufficiently low temperatures to approximate $-\partial f/\partial\en\approx\delta(\en)$. The contributions of the side trees are therefore given by \eref{simplestside treesol}, and we obtain
\begin{align}
 \delta g= & \fr{N_1\rbr{2N_2x^3+4N_1x^2+8N_2x^2+\kappa}}{2(\rbr{2x+1}\rbr{N_1+N_2}} \nonumber \\
  & -\fr{N_1\rbr{x+1}\kappa}{2\sqrt{1+6x+x^2}\rbr{N_1+N_2}}
 \label{eq:lead_cond_nophase}
\end{align}
with $\kappa=N_2x^2+2xN_1+6xN_2+2N_1+N_2$.

\begin{figure}
  \subfigure[\label{singleleadcorrx_a}]{\includegraphics[width=0.45\columnwidth]{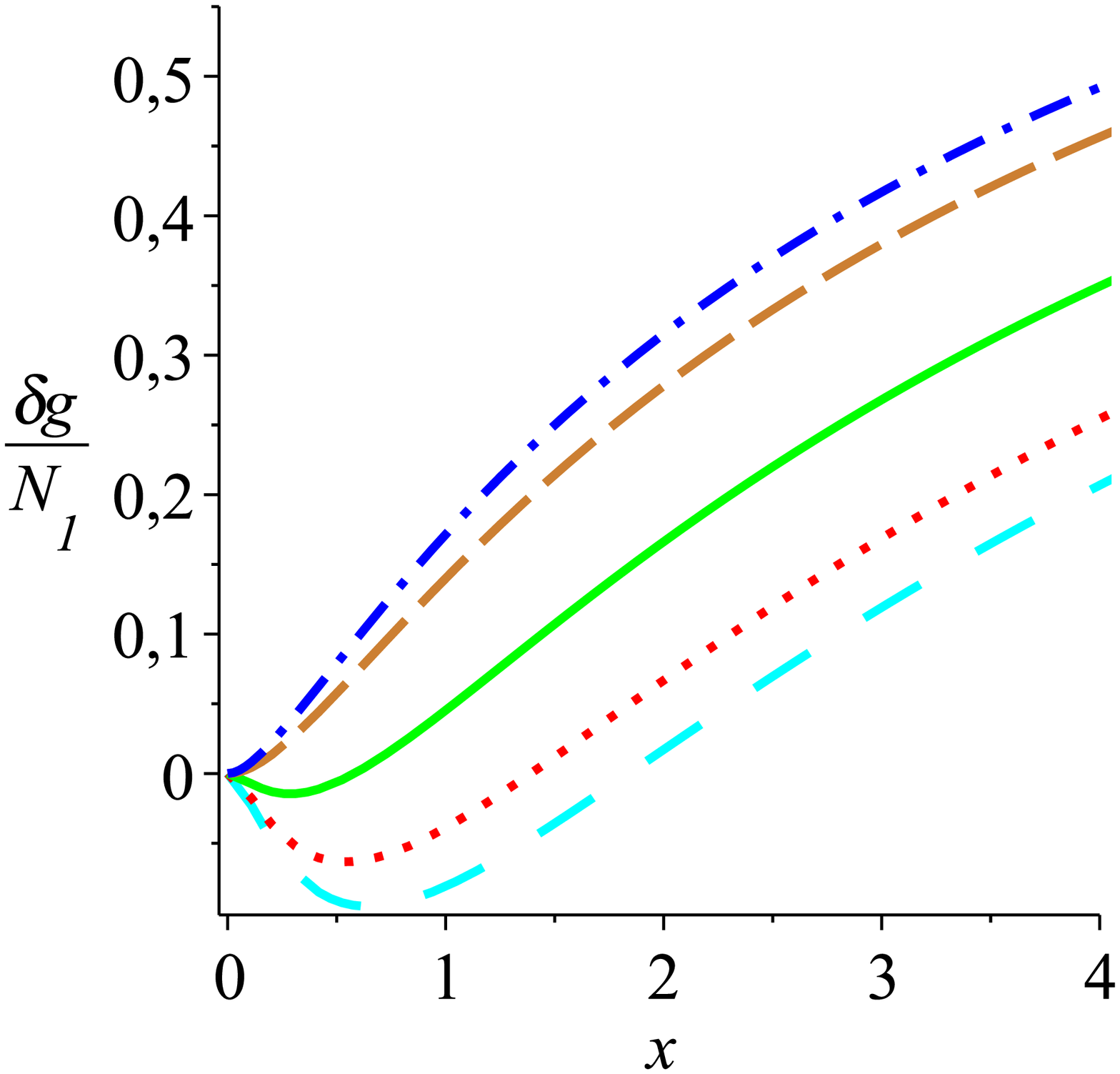}}
  \rule{0.033\columnwidth}{0pt}
  \subfigure[\label{singleleadcorrx_b}]{\includegraphics[width=0.45\columnwidth]{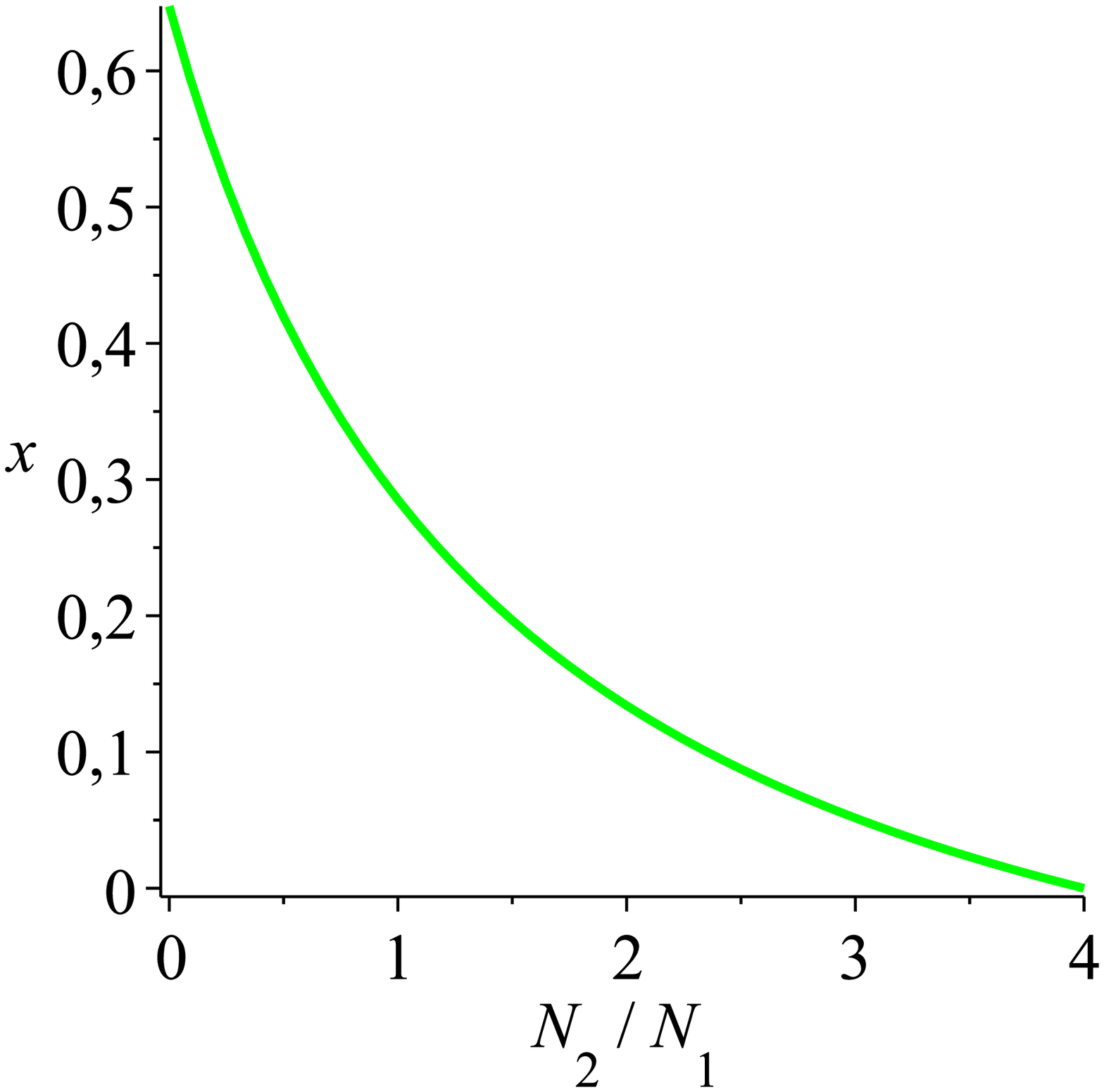}}
 \caption{\label{singleleadcorrx} (a) The quantum correction to the conductance of an Andreev billiard with one superconducting and two normal conducting leads (see \fref{lead_setup}; with the superconducting chemical potential being the same as that of normal lead 2) as a function of the number of channels of the superconductor, $x=N_\mathrm{S}/N_\mathrm{N}$, for $N_2/N_1\rightarrow0$ (space dashed line), $N_2/N_1=0.2$ (dotted line), $N_2/N_1=1$ (solid line), $N_2/N_1=7.2$ (dashed line) and $N_2/N_1\rightarrow\infty$ (dashed dotted line). (b) The size of the superconductor for which the conductance becomes minimal as a function of the ratio $N_2/N_1$.
}
\end{figure}

We can easily compare our result for the conductance correction $\delta g=g-g_{cl}$ to the limiting case found for small superconductors in Ref.~\onlinecite{nsntrans} by expanding our result in a Taylor series in $x=N_\mathrm{S}/N_\mathrm{N}$. The first non-vanishing term in the Taylor expansion is exactly the contribution found by Jacquod and Whitney \cite{nsntrans}:
\begin{equation}
 \delta g^{(2)}=\fr{N_1\rbr{N_2-4N_1}N_\mathrm{S}^2}{N_\mathrm{N}^3}.
 \label{smallsclsollead}
\end{equation}
Therefore for small superconducting leads the correction becomes negative if lead $1$ carries a sufficiently large amount of modes compared to lead $2$. The ratio $N_2/N_1$, for which the conductance has no longer a minimum and hence the conductance correction becomes no longer negative for small $x$, is approximately $N_2/N_1\approx4$ as found by Whitney and Jacquod \cite{nsntrans}. Interestingly, \fref{singleleadcorrx_a} shows moreover a second change in sign of $\delta g$, namely when increasing $x$, for fixed $N_1>N_2/4$. This characteristic crossover is beyond the treatment of Ref.~\onlinecite{nsntrans}and arises from higher order diagrams as calculated here. The size of the superconductor for which the conductance becomes minimal is shown as a function of $N_2/N_1$ in \fref{singleleadcorrx_b}.
Moreover we again found a doubling of the conductance for $N_\mathrm{S}/N_\mathrm{N}\rightarrow\infty$, \textit{i.e.} $g=2N_1$, independently of the ratio $N_1/N_2$ in alignment with previous results for quantum dots with only one normal conducting lead\cite{cond_doubling}.

\subsubsection{Two superconductors with a phase difference}
\begin{figure}
 \subfigure[\label{lead_phase_a}]{\includegraphics[width=0.45\columnwidth]{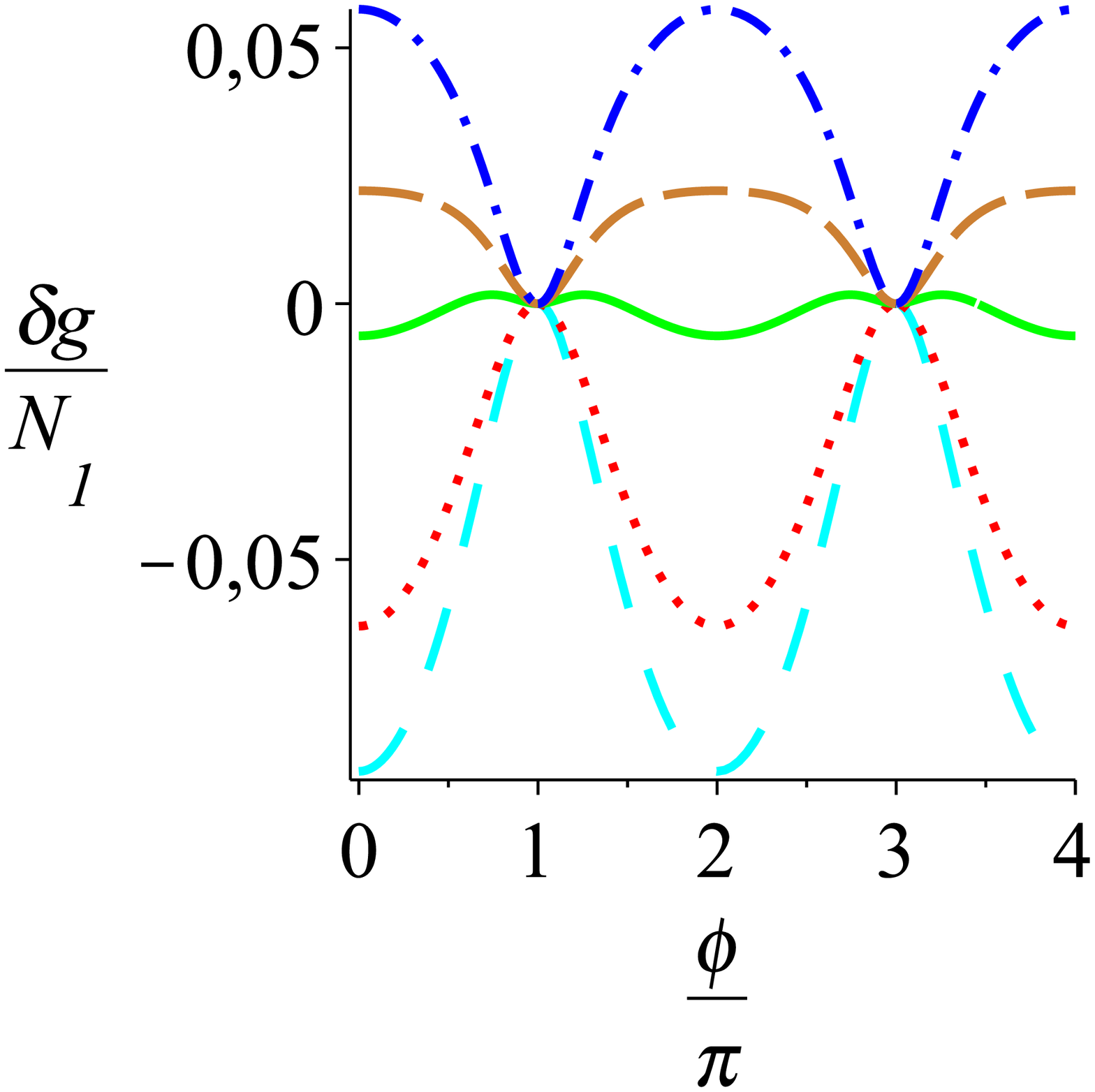}}
 \rule{0.066\columnwidth}{0pt}
 \subfigure[\label{lead_phase_b}]{\includegraphics[width=0.45\columnwidth]{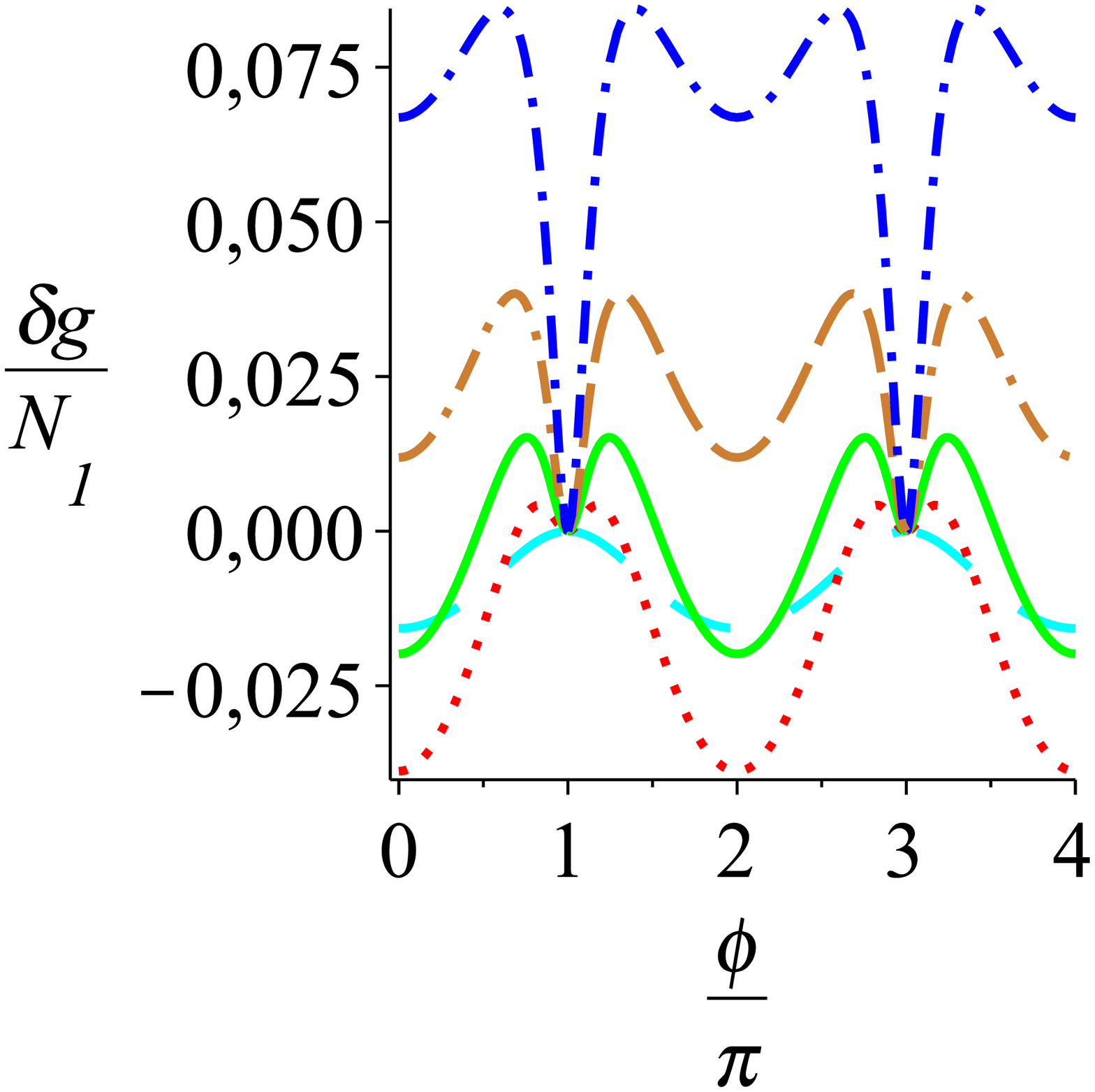}}
 \caption{\label{lead_phase}Phase dependence of the conductance correction. (a) $x=0.5$ and $N_2/N_1\rightarrow0$ (space dashed line), $N_2/N_1=0.2$ (dotted line), $N_2/N_1=1$ (solid line), $N_2/N_1=2$ (dashed line) and $N_2/N_1=7$ (dashed dotted line). (b) $N_2/N_1=0.2$ and $x=0.1$ (space dashed line), $x=1$ (dotted line), $x=1.2$ (solid line), $x=1.5$ (dashed line) and $x=2$ (dashed dotted line).}
\end{figure}

The effect of a phase difference between two superconducting leads depends sensitively on the ratios $x$ and $N_2/N_1$. The result is fairly similar to the phase dependence of the conductance of an normal conducting region with one normal conducting lead and two superconducting leads with a phase difference $\phi$ found by different approaches \cite{circuit_nss,quasicl_nss}. While for most combinations the effect of the superconductor decreases with increasing phase difference due to destructive interference, in some cases the conductance becomes a non-monotonic function between $\phi=0$ and $\phi=\pi$ as shown in \fref{lead_phase}. The phase difference may even cause a change of the sign of the conductance correction. In \fref{lead_phase} this can be seen for the case $N_2=N_1$ and $x=0.5$ as well as for the cases $N_2=0.2N_1$ and $x=1$ or $x=1.2$.

However if the number of channels of the superconducting leads are equal all possible combinations of $x$ and $N_2/N_1$ have in common that the conductance correction becomes zero for a phase difference $\phi=\pi$ and that the conductance is symmetric and periodic in $\phi$ with period $2\pi$. This can also be seen from \earef{eq:encounter_coeff_A}{,b} and \eref{geq}: The symmetry follows from the symmetry of $B_l$ and $P$ as well as the fact that $A_l$ does not depend on $\phi$ explicitly. If $\phi$ is replaced by $\phi+2\pi$ the side tree contribution changes its sign. $A_l$ and the first part of $B_l$ as well as the last two terms of $B_l$ for even $l$'s are symmetric in $P$ and therefore give the same contribution as before. However for an odd $l$ the last two parts of $B_l$ are antisymmetric in $P$ but when increasing the phase difference by $2\pi$ the phase factors contribute an additional minus sign such that the total contribution stays the same.
\begin{figure}
 \subfigure[\label{lead_channel_diff_a}]{\includegraphics[width=0.45\columnwidth]{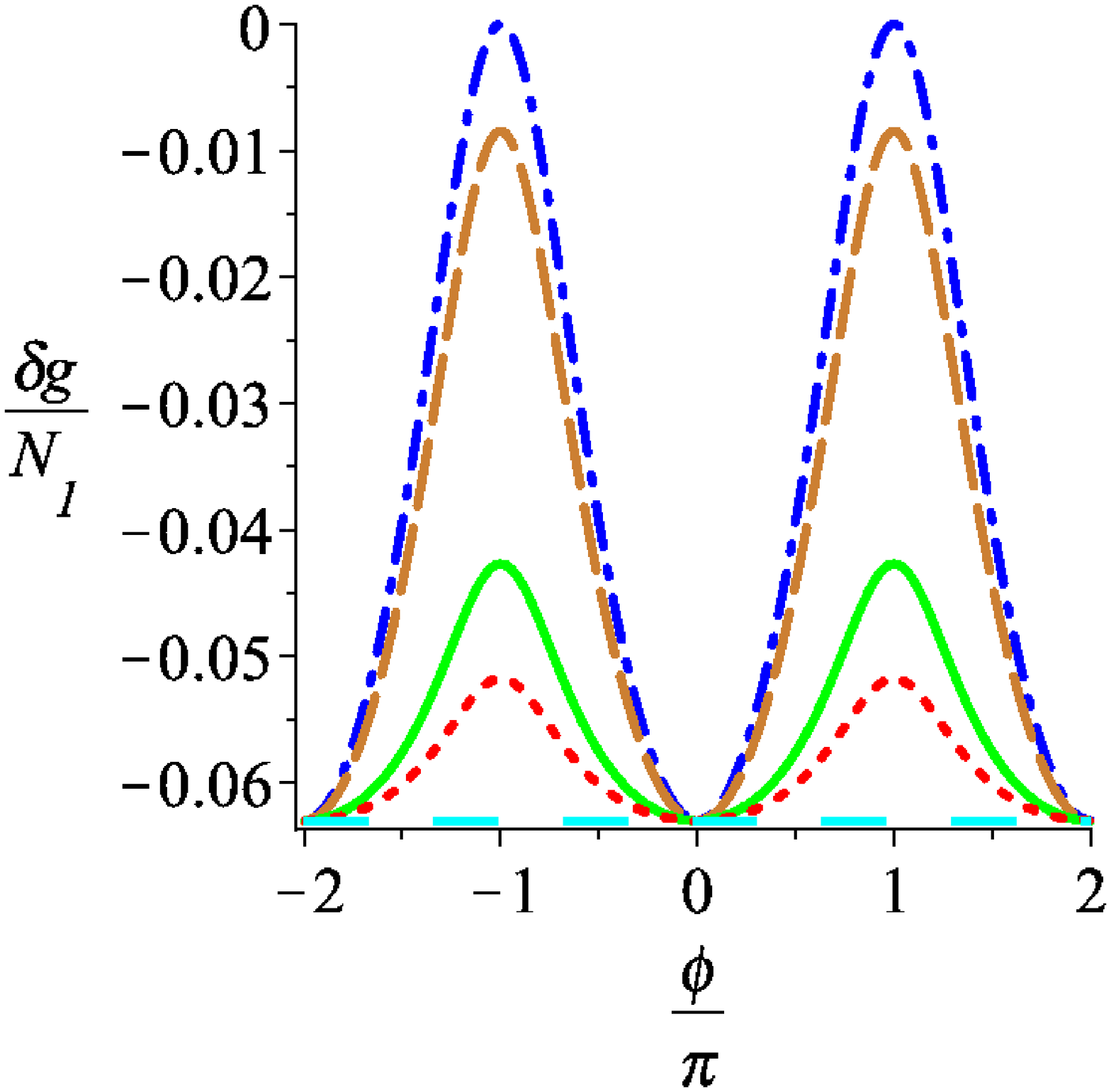}}
 \subfigure[\label{lead_channel_diff_b}]{\includegraphics[width=0.45\columnwidth]{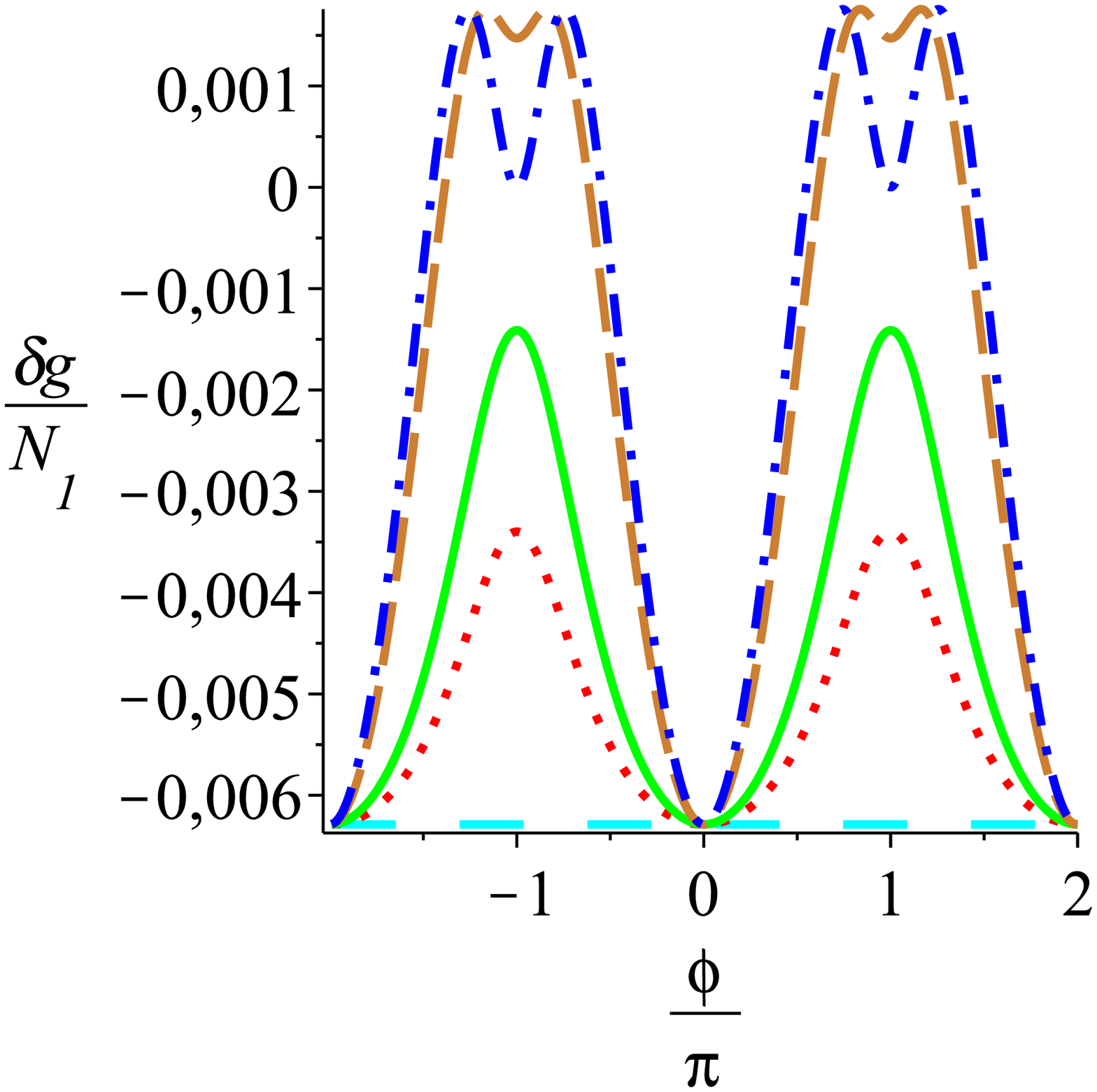}}
\\
 \subfigure[\label{lead_channel_diff_c}]{\includegraphics[width=0.45\columnwidth]{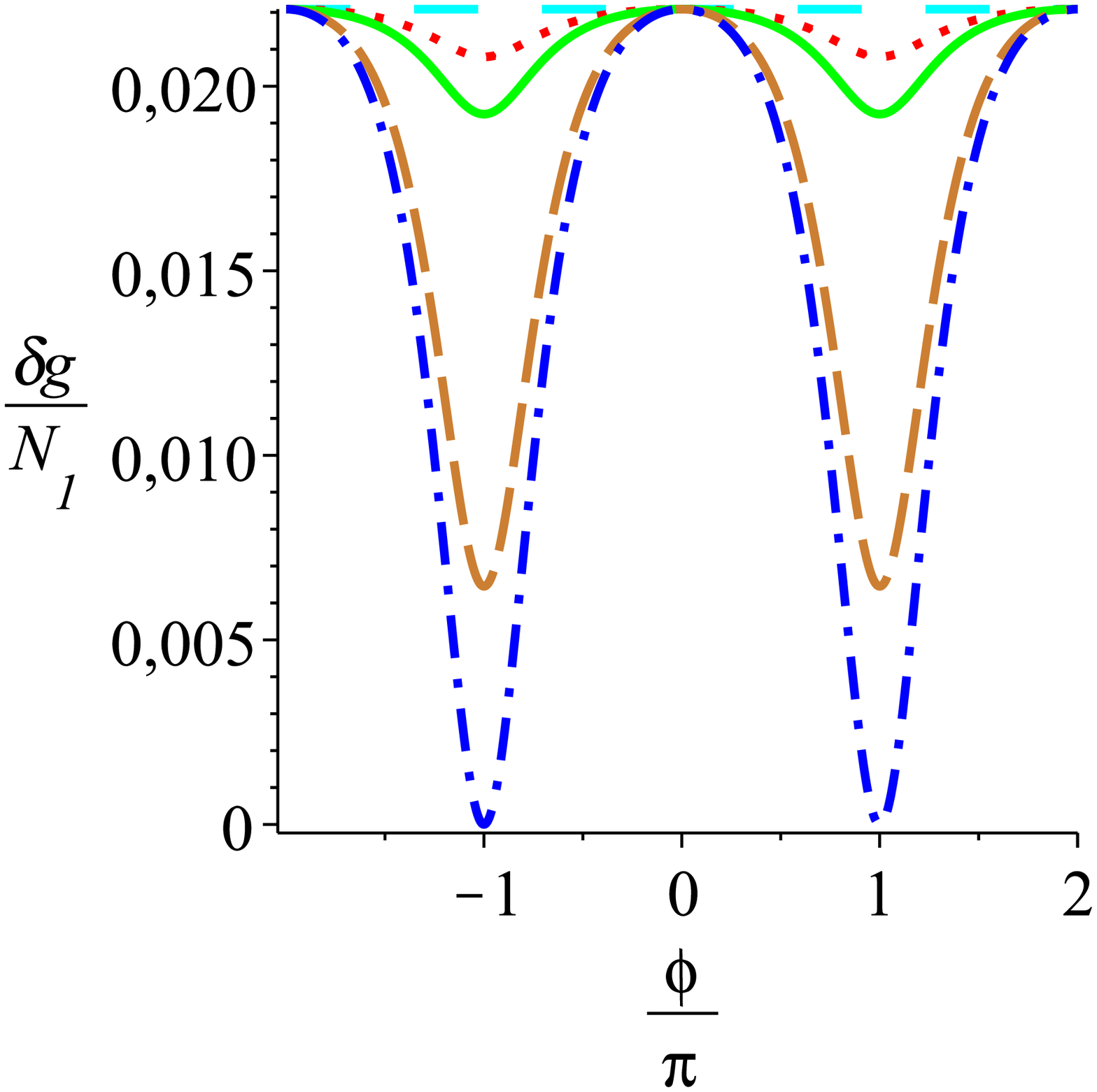}}
 \subfigure[\label{lead_channel_diff_d}]{\includegraphics[width=0.45\columnwidth]{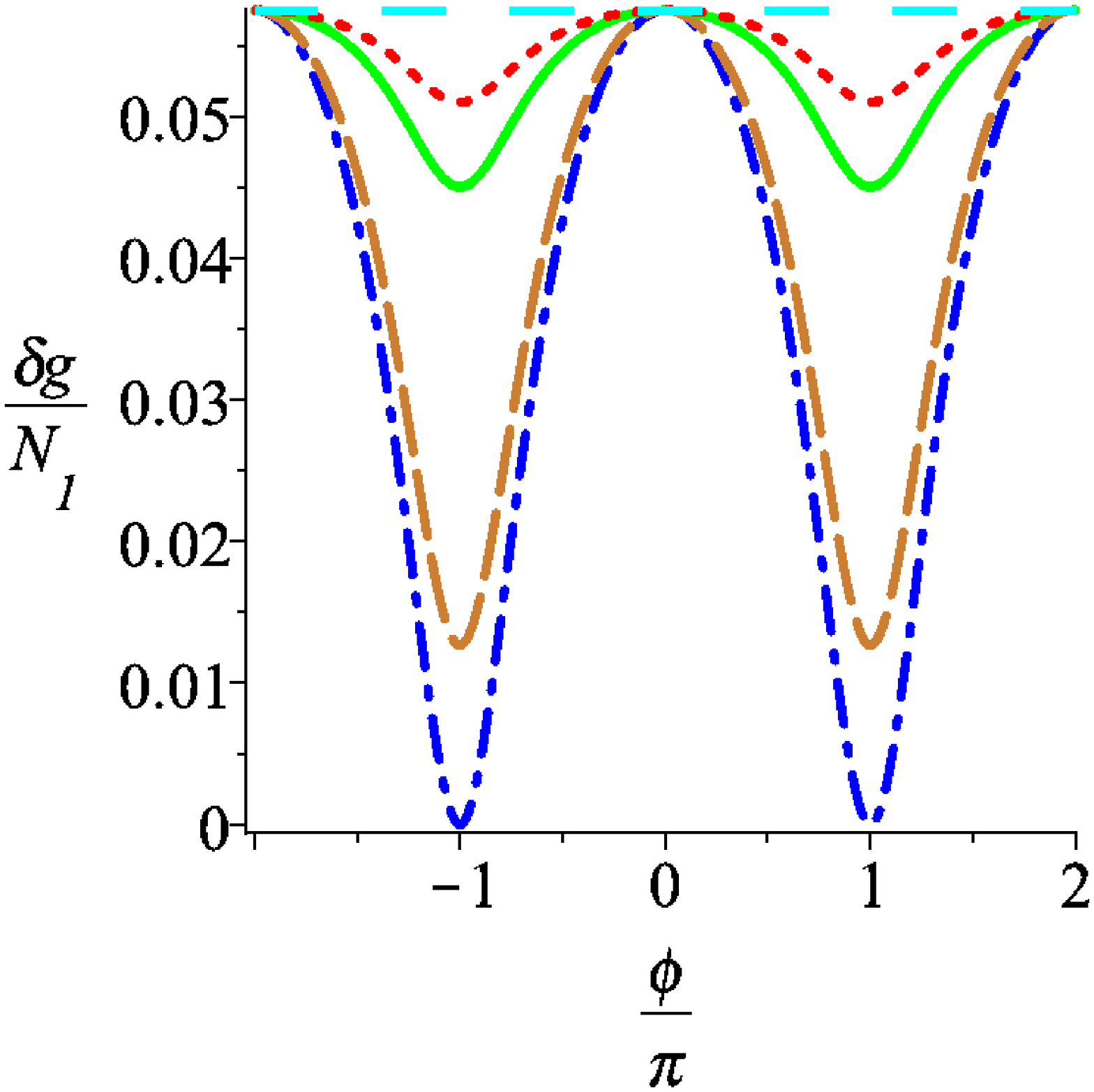}}
 \caption{\label{lead_channel_diff} Dependence on the difference between the numbers of channels of the superconductors with $x=0.5$ and $y=1$ (space dashed line), $y=0.95$ (dotted line), $y=0.9$ (solid line), $y=0.5$ (dashed line) and $y=0$ (dashed dotted line). (a) $N_2/N_1=0.2$, (b) $N_2/N_1=1$, (d) $N_2/N_1=2$ (d) $N_2/N_1=7$.}
\end{figure}

This symmetry towards $\phi$ still holds if the two superconducting leads provide different numbers of channels. We show the crossover from a setup with two superconducting leads having the same number of channels to the setup with just one superconducting lead for different cases in \fref{lead_channel_diff}. As in \eref{eq:encounter_coeff_B} $y=(N_{\mathrm{S}_1}-N_{\mathrm{S}_2})/N_\mathrm{S}$ is the difference between the numbers of channels in the superconducting leads such that $y=0$ corresponds to the symmetric case $N_{\mathrm{S}_1}=N_{\mathrm{S}_2}$ and $y=\pm1$ to the case of just one superconducting lead. We found that in the cases where the conductance correction had a dip at $\phi=\pm\pi$ first of all this dip vanishes and after that the conductance correction tends to the $\phi=0$ result for all $\phi$ if $y$ is increased. If there is no dip the result converges monotonically to the result for the case with just one superconducting lead.

\subsubsection{Weak magnetic field}
\begin{figure}
 \subfigure[\label{lead_mag_a}]{\includegraphics[width=0.45\columnwidth]{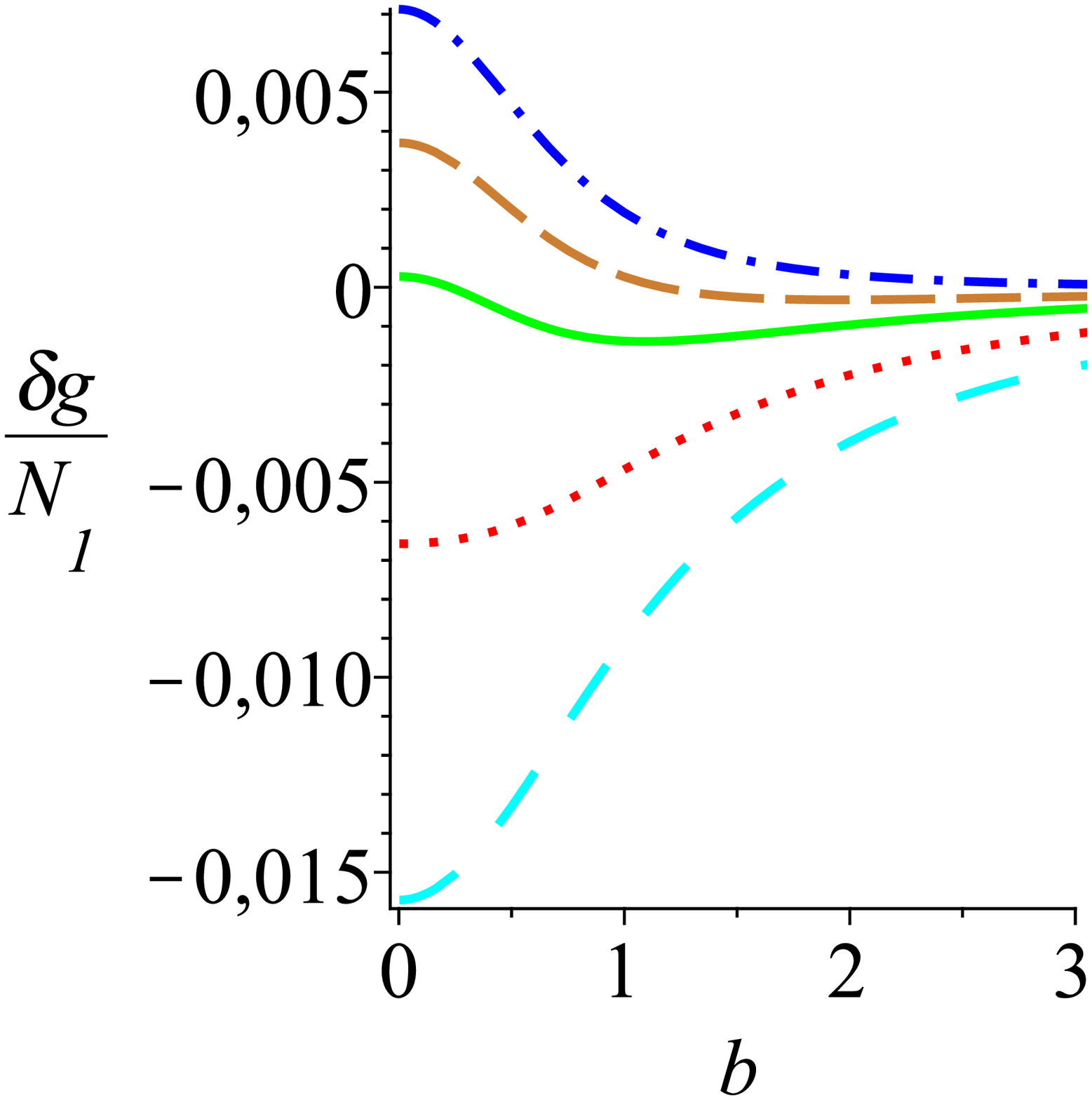}}
 \subfigure[\label{lead_mag_b}]{\includegraphics[width=0.45\columnwidth]{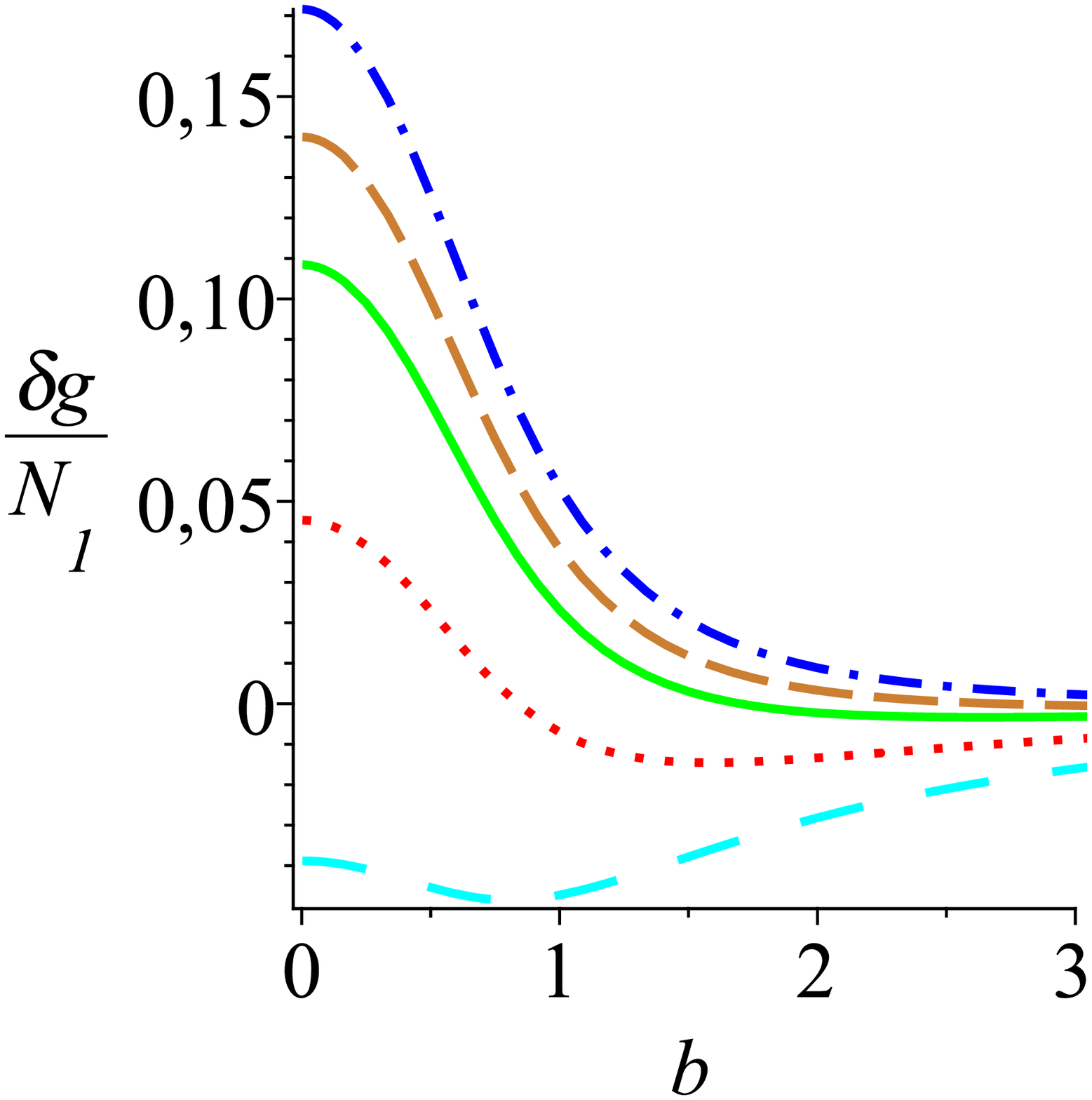}}
 \caption{\label{lead_mag}Magnetic field dependence of the conductance correction for (a) $x=0.1$ and (b) $x=1$ with $N_2/N_1=0.2$ (space dashed line), $N_2/N_1=1$ (dotted line, $N_2/N_1=3$ (solid line), $N_2/N_1=7$ (dashed line) and $N_2/N_1\rightarrow\infty$ (dashed dotted line)}
\end{figure}

If a magnetic field is applied time reversal symmetry is broken. Therefore building side trees becomes less likely and their contribution vanishes as $b\rightarrow\infty$ as can be seen in \fref{lead_mag}.
In particular for small superconducting channel numbers Whitney and Jacquod \cite{nsntrans} predicted that the conductance correction decreases at zero temperature with $(1+b^2)^{-2}$. However, we find a nonmonotonic behaviour for the dependence of the conductance correction on the magnetic field similar to that found for the case of one normal conducting lead \cite{mag_nonmonotonic} and for the magnetic field dependence of the excess current \cite{cond_doubling}. It may even happen that the conductance correction is negative for $b\neq0$ although it is positive at $b=0$.
Since the diagrammatic rules depend on $b^2$ rather than $b$ the conductance is a symmetric function of $b$ and therefore satisfies the Onsager relation \cite{ref:onsager_relation,ref:onsager_relation2} for a two terminal setup $g(b)=g(-b)$.

\subsection{Temperature dependence}
\begin{figure}
\subfigure[\label{lead_temp_a}]{\includegraphics[width=0.9\columnwidth]{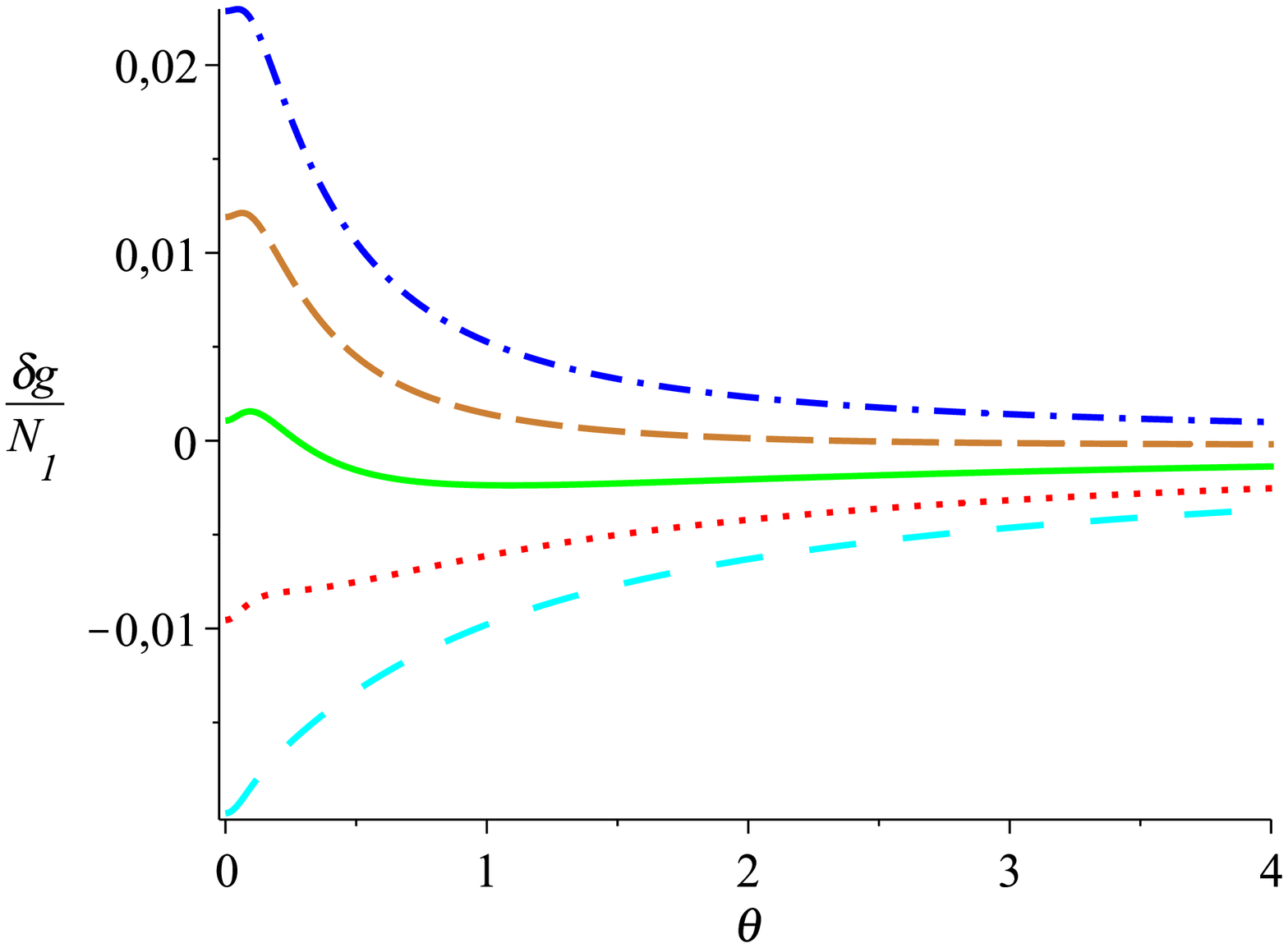}}
\\
\subfigure[\label{lead_temp_b}]{\includegraphics[width=0.45\columnwidth]{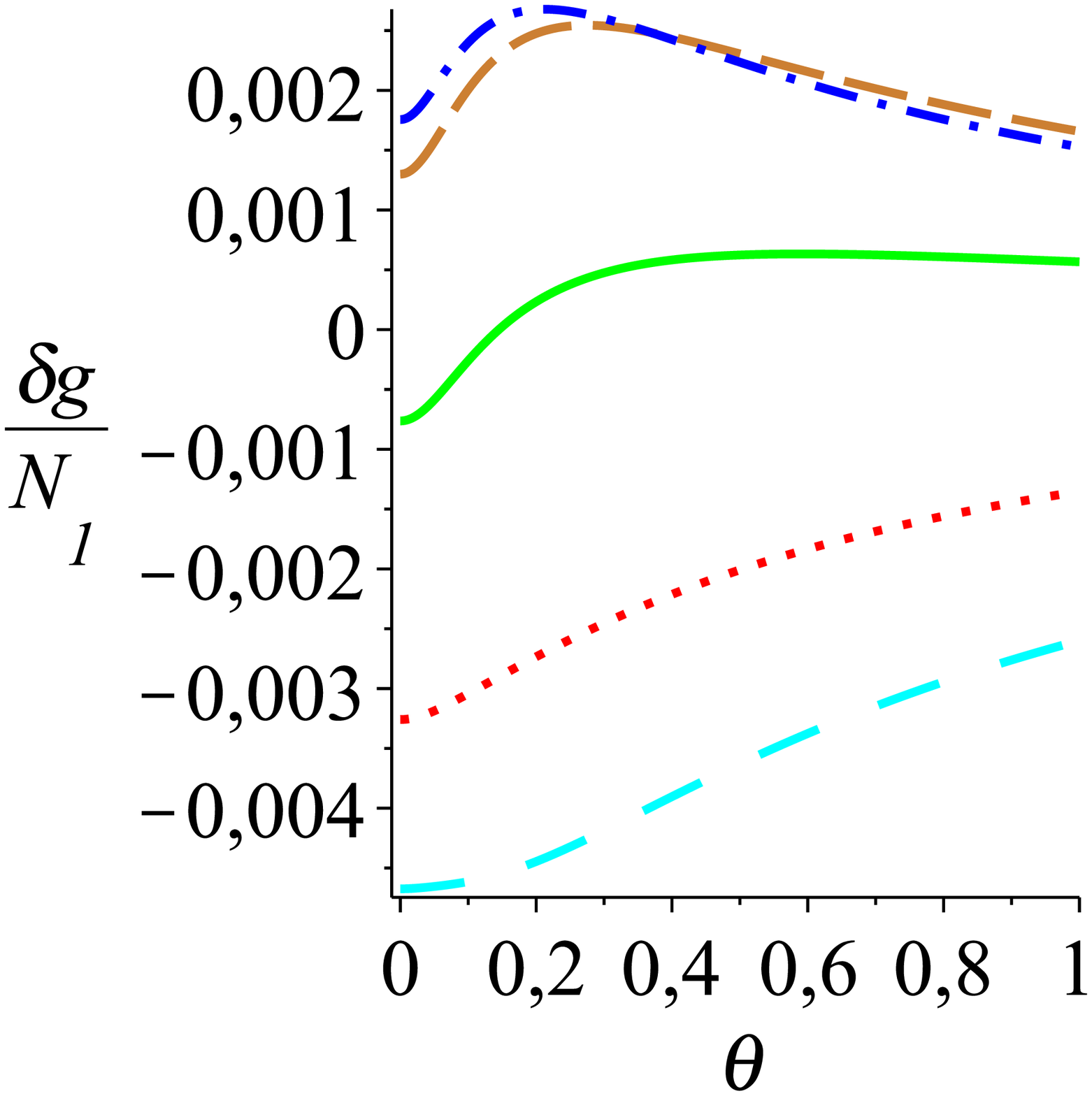}}
\subfigure[\label{lead_temp_c}]{\includegraphics[width=0.45\columnwidth]{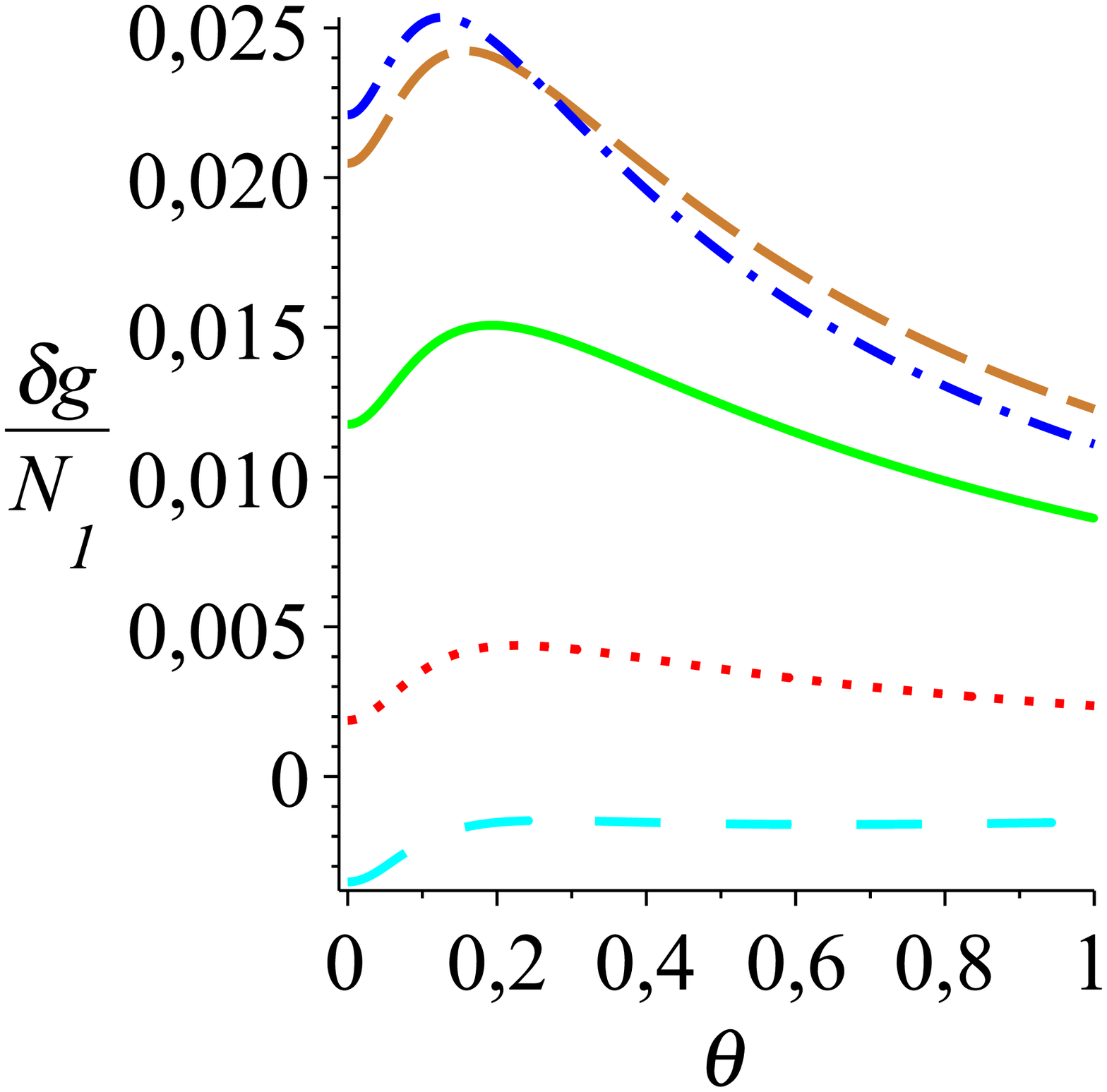}}
\caption{\label{lead_temp} The conductance correction as a function of the temperature $\theta=k_\mathrm{B}T/E_\mathrm{T}$. (a)$N_2/N_1=0.2$, $\phi=0$ and $x=1.2$ (space dashed line), $x=1.3$ (dotted line), $x=1.4$ (solid line), $x=1.5$ (dashed line) and $x=1.6$ (dashed dotted line). (b) $\phi=\pi/4$ (space dashed line), $\phi=7\pi/20$ (dotted line), $\phi=\pi/2$ (solid line), $\phi=13\pi/20$ (dashed line) and $\phi=3\pi/4$ (dashed dotted line), $N_2/N_1=1$ and $x=0.5$. (c) $N_2/N_1=0.2$ and $x=1.3$.}
\end{figure}

To include the effect of finite temperature we have to include the energy dependence of the side tree contribution and the central encounters. For zero phase the sixth order equation for the side tree contribution factorizes and one has to solve a quartic equation. At leading order in $N_\mathrm{S}/N_\mathrm{N}$ the temperature dependence is given by the generalised zeta function. However as can be seen from \fref{lead_temp} when including higher order terms strong derivations from this behaviour may occur again.

The integral in \eref{nsn_landauer} of course can not be performed analytically. Therefore we calculated the integral numerically using gaussian quadrature with a total accuracy of $10^{-10}$. The integral has been truncated at $\en=100\theta$ where $\theta=k_\mathrm{B}T/E_\mathrm{T}$ is again the temperature measured in units of the Thouless energy. In \fref{lead_temp} we plotted the conductance correction versus the temperature.

As one might expect from the Sommerfeld expansion the conductance correction has a local extremum at $\theta=0$ but not necessarily a global one as can be seen from the solid and the dash dotted curves in \fref{lead_temp_a}. For a certain range of combinations of the ratios $x$ and $N_2/N_1$ the conductance correction increases with increasing temperature although it is positive for $T=0$. A similar effect known as the reentrance of the metallic conductance has previously found in NS-structures \cite{reentrance_exp1,reentrance_exp2}. Moreover it may happen that the conductance correction changes its sign when the temperature is increased such as in the case $N_2/N_1=0.2$ and $x=1.4$ shown by the solid line in \fref{lead_temp_a}.

When including a phase difference again this affects the temperature dependence. It may again cause a nonmonotonic temperature dependence even if it is monotonic at $\phi=0$. Moreover in contrast to the case $\phi=0$ where the sign of the conductance correction with increasing temperature may only change from positive to negative it can be the other way around for $\phi\neq0$. The temperature dependence of the conductance correction for different phases is shown in \fref{lead_temp}(b),(c).

In all cases for large temperatures the conductance correction tends to zero. However this limit is approached only very slowly.

\section{Conductance of a double dot setup}
\label{sec:parallelogram_conductance}

For the double dot model shown in \fref{fig:parallelogram}, where the quasiparticles stay on average a time $\delta\tau\cdot\tau_\mathrm{D}$ in the neck connecting the two dots, the necessary modifications are more substantial than in the previous case. First of all we have to find a way to calculate the transmission probabilities at all orders in $x$ and in $n=N_\mathrm{n}/N_\mathrm{N}$ where $N_\mathrm{n}$ is the number of channels of the neck connecting the two different dots. For simplicity we additionally assume that the two dots have the same dwell time. Since we consider the neck to be represented by an ideal lead every electron in the left dot entering the neck at channel $i$ leaves the neck at the same channel into the right dot and vice versa such that a traversal of the neck yields a factor $N_\mathrm{n}$.

The idea of the calculation is as follows: We first of all split the whole setup into two distinct parts each having two normal leads (one of them being the neck) and one superconducting lead as it has been done in Ref.~\onlinecite{ref:semiclassical_anderson_localization}. To find a recursion relation for the transmission we cut the backbone not only at the first encounter but also at the first traversal of the neck, \textit{i.e.}~the diagram is cut when the first path pair enters an encounter or traverses the neck without having entered an encounter. Since traversing the neck contributes a factor $N_\mathrm{n}$ we will assign this factor to a path pair hitting the neck. Due to the splitting of diagrams at the neck we thus get path pairs starting at a neck which do not contribute a factor $N_\mathrm{n}$. Thus a diagram starting at the neck in dot $j$ contributes to $T_{ij}^{\alpha\beta}/N_j$. In order to simplify the handling of the contributed numbers of channels here we define a `normalised' transmission coefficient $\tilde{t}_{ij}^{\alpha\beta}=T_{ij}^{\alpha\beta}/N_j$.
 
Consider a diagram starting in lead $1$ with an electron, thus contributing to $\tilde{t}_{j1}^{\alpha e}$ with $j\in\{1,2\}$ and $\alpha\in\{e,h\}$. If the first path pair enters an encounter without having traversed the neck the diagrams are treated in the same way as above: The remaining part after cutting the diagram right after the first encounter contributes to $\tilde{t}_{j1}^{\alpha e}$ if the number of $\zeta$-side trees emerging from the encounter is even and to $\tilde{t}_{j1}^{\alpha h}$ otherwise. If the number of $\zeta$-side trees is even the contribution of the first path pair, the first encounter and its side trees is given by
\begin{align}
A_{11,l}^e=-\sul{p=0}{l-1}&\cbr{1+\rmi\rbr{2p-l+1}\en} \nonumber \\
&\cdot\rbr{P_1^eP_1^h}^{p}\cbr{\rbr{P_1^eP_1^h}^*}^{l-p-1},
\label{eq:double_dot_first_even_encounter_contribution11}
\end{align} 
where the subscript `$11$' indicates that the encounter lies fully in dot 1. $P_i^\alpha$, $i\in\{1,2\}$, $\alpha\in\{e,h\}$, is the contribution of side trees starting with an $\alpha$-type quasi particle in dot $i$ and will be derived below.

\begin{figure}
\subfigure[\label{fig:encounter_neck_touch_a}]{\includegraphics[width=0.45\columnwidth]{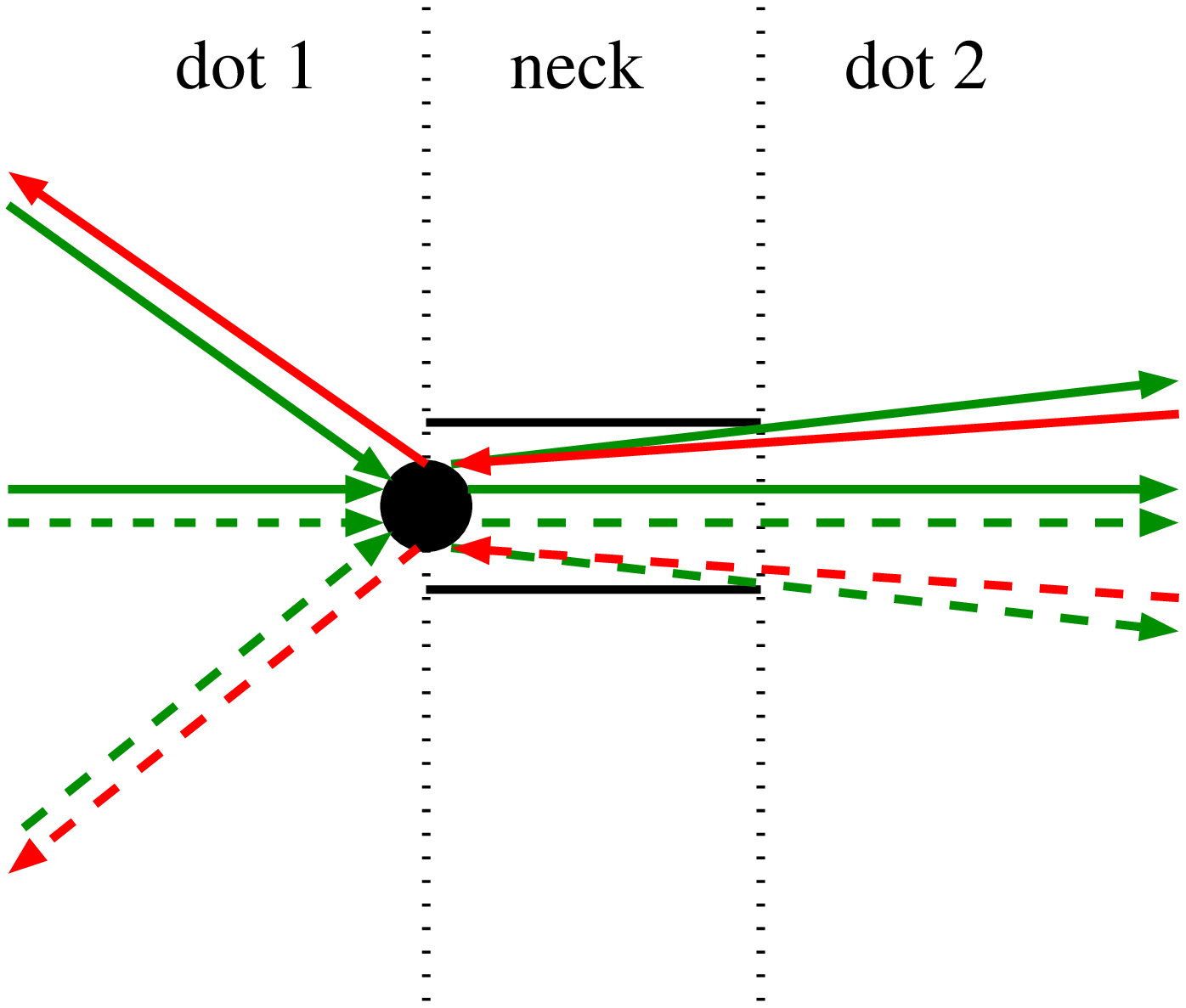}}
\rule{0.033\columnwidth}{0pt}
\subfigure[\label{fig:encounter_neck_touch_b}]{\includegraphics[width=0.45\columnwidth]{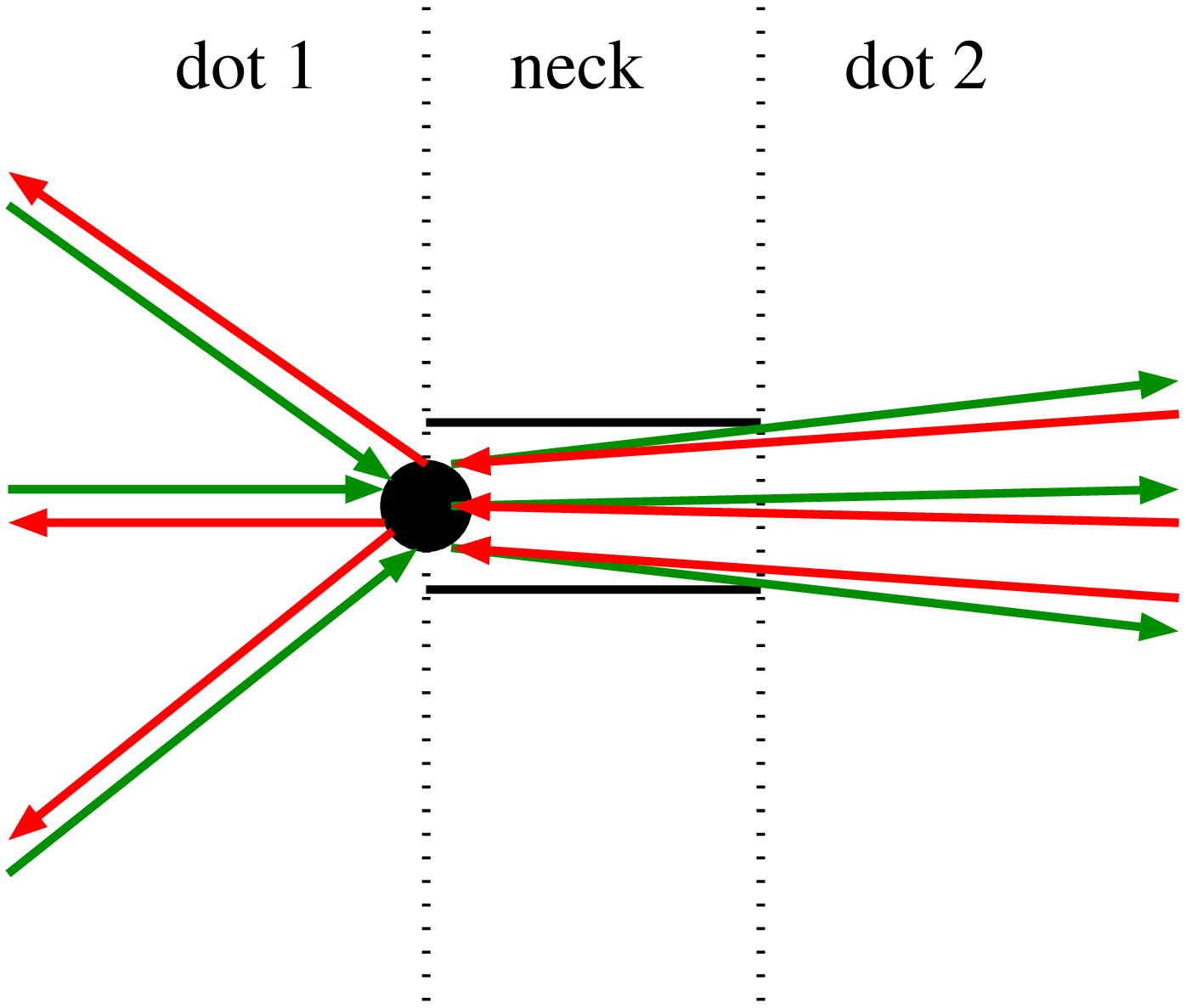}}
\caption{\label{fig:encounter_neck_touch} (a) An 3-encounter of the backbone touching the neck. (b) An 3-encounter of a side tree touching the neck.}
\end{figure}
If the e-\es~path pair leaving the first encounter and the odd numbered side trees attached to the first encounter traverse the neck before entering another encounter the first encounter can touch the neck. As shown in \fref{fig:encounter_neck_touch_a} the odd numbered side trees and the e-\es~path pair leaving the encounter then start in dot 2, since we treat the neck as a ballistic connection between the two dots, while the total number of side trees stays the same. Hence the contribution becomes
\begin{align}
 A_{21,l}^e=\sul{p=0}{l-1}&\fr{N_\mathrm{n}}{N_1+N_{\mathrm{S}_1}+n_\mathrm{n}}\cbr{1+\rmi\rbr{2p-l+1}\en} \nonumber \\
 &\cdot\rbr{P_2^e\rme^{\rmi\en\delta\tau}P_1^h}^p\cbr{\rbr{P_2^e\rme^{\rmi\en\delta\tau}P_1^h}^*}^{l-p-1}.
 \label{eq:double_dot_first_even_encounter_contribution12}
\end{align}
Since the e-\es~path pair now starts in dot 2 the remaining part of the diagram in this case contributes to $\tilde{t}_{i2}^{\alpha e}$. The additional phase factor $\rme^{\rmi\en\delta\tau}$ is the phase accumulated by an e-\es~path pair which traverses the neck.

For the encounters with an odd number of side trees built by $\zeta$ we have to be even more careful. This is because the encounter now can be moved into either the neck or into the superconductor coupled to the dot the encounter lies in.
The fact that when an encounter touches the neck the odd numbered side trees have to traverse the neck means that the path pairs of the backbone entering and leaving the first encounter have both to be in the same dot after traversing the encounter. The $l-2$ in total side trees still starting in the left dot are the originally even numbered side trees and therefore start with a hole.

If again the first encounter is in the left dot the contribution of the encounters with an odd number of side trees built by $\zeta$ is
\begin{widetext}
\begin{align}
B_{11,l}^e=-\sul{p=0}{l-2}\Bigg\{&\cbr{1+\rmi\rbr{2p-l+2}\en}\rbr{P_1^e}^{p+1}\rbr{P_1^h}^p\cbr{\rbr{P_1^e}^*}^{l-p-1}\cbr{\rbr{P^h_1}^*}^{l-p-2} \nonumber \\
 & -\fr{N_{\mathrm{S}_1}}{N_1+N_{\mathrm{S}_1}+N_\mathrm{n}}\rme^{-\rmi\phi\rbr{2p-l+2}/2}\rbr{-\rmi P_1^h}^p\cbr{\rmi\rbr{P_1^h}^*}^{l-p-2} \nonumber \\
 & -\fr{N_\mathrm{n}}{N_1+N_{\mathrm{S}_1}+N_\mathrm{n}}\rme^{\rmi\rbr{2p-l+2}\en\delta\tau}\rbr{P_1^h}^p\cbr{\rbr{P_1^h}^*}^{l-p-2}\rbr{P_2^e}^{p+1}\cbr{\rbr{P_2^e}^*}^{l-p-1}\Bigg\}.
\label{eq:double_dot_first_odd_encounter_contribution11}
\end{align}
Similar formulae as above hold for the remaining coefficients $A_{ij,l}^\alpha$ and $B_{jj,l}^\alpha$, $i,j\in\{1,2\}$, $\alpha\in\{e,h\}$.

If the e-\es~path pair however hits the neck without having traversed an encounter we cut the diagram at the neck such that the first part contributes $N_\mathrm{n}/\rbr{N_1+N_{\mathrm{S}_1}+N_\mathrm{n}}$ and the second part contributes to $\tilde{t}_{j2}^{\alpha e}$. Therefore the `normalised' transmission coefficient is given by

\begin{figure*}
\includegraphics[width=\textwidth]{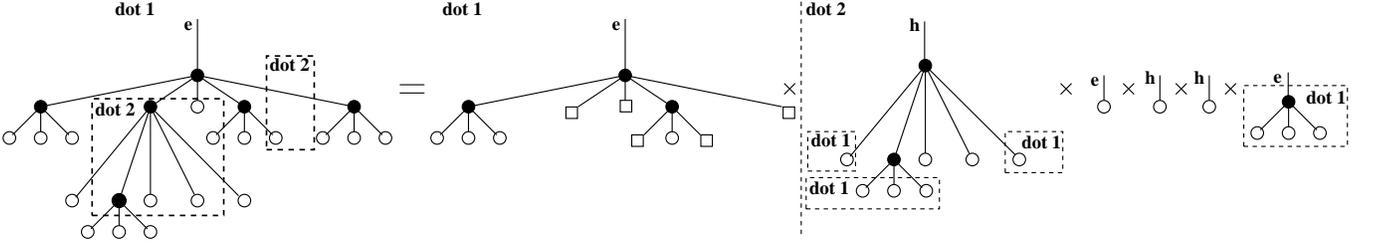}
\caption{\label{fig:side_tree_double_dot} A side tree starting in dot 1 may also have parts lying in dot 2. If a quasiparticle hits the neck we will treat this as an electron being retroreflected as its counter-particle indicated by the empty boxes and include an additional factor of the side tree starting in the other dot for each traversal of the neck. `e' and `h' denote that the side tree starts with an electron or hole, respectively.}
\end{figure*}
\begin{align}
\tilde{t}_{ij}^{\alpha\beta}=\fr{N_i\delta_{ij}}{N_j+N_{\mathrm{S}_j}+N_\mathrm{n}}+\sul{l=0}{\infty}A_{jj,l}^\beta\tilde{t}_{ij}^{\alpha\beta}+\rbr{\fr{N_\mathrm{n}}{N_j+N_{\mathrm{S}_j}+N_\mathrm{n}}+\sul{l=0}{\infty}A_{\bar{j}j,l}^{\beta}}\tilde{t}_{i\bar{j}}^{\alpha\beta}+\sul{l=0}{\infty}B_{jj,l}^\beta\tilde{t}_{ij}^{\alpha\bar{\beta}}
 \label{eq:double_dot_transmission_system_11ee}
\end{align}
\end{widetext}
where $\bar{j}=3-j$ as well as $\bar{\beta}=h$ if $\beta=e$ and vice versa.

This gives in general a 16 dimensional system of linear equations which decomposes into four independent systems of linear equations.
The complication of treating the double dot setup is that the side tree contributions are also different. We may calculate their contribution though following the steps in \sref{cond_sidetrees} with slight changes. First of all it is no longer enough just to consider the generating functions $f$ and $\hat{f}$ for side trees starting with an electron or a hole, respectively, but we have to consider the generating functions $f_1$, $\hat{f}_1$, $f_2$ and $\hat{f}_2$ for side trees starting with an electron (without the hat) or a hole (with the hat) in the left (subscript `1') or in the right dot (subscript `2'). Here we will only consider $f_1$ for the side trees starting with an electron in the left dot explicitly since the derivation of the remaining generating functions is similar and needs only simple replacements.

First of all we have to slightly modify the way we look at the side trees. In general they may consist of parts lying in dot 1 and parts lying in dot 2 as indicated in \fref{fig:side_tree_double_dot}. By cutting a tree starting in dot 1 at the neck we find that an electron traversing the neck builds a tree starting in dot 2 and finally comes back as a hole since the number of Andreev reflections provided by a tree is odd. Thus every time a link of a tree starting in dot 1 hits the neck we cut the tree at the corresponding channel and insert a retro-reflection there contributing an additional factor given by the side tree starting in dot 2, hence a factor $f_2$ if an electron hits the neck and $\hat{f}_2$ if a hole hits the neck. Moreover we have to include the additional phase $\en\delta\tau$ due to the time a quasiparticle spends in the neck. Therefore we may write a diagrammatic rule for a path pair in dot 1 hitting the neck:
\begin{itemize}
\item An electron hitting the neck contributes a factor $N_\mathrm{n}\rme^{\rmi\en\delta\tau}f_2$.
\item A hole hitting the neck contributes a factor $N_\mathrm{n}\rme^{\rmi\en\delta\tau}\hat{f}_2$.
\end{itemize}

We may then start with \eref{eq:generating_func_general_el} and adapt it to our new problem since the steps leading to this can be used here in exactly the same way. The second term in the first line and the last term in the second line which previously corresponded to a tree whose first encounter hits S$_2$ is now given by a tree with its first encounter hitting the neck and therefore building a side tree in dot 2 yielding a contribution $f_2$. Therefore we replace the contribution $z_{o,l}^{(2)}$ of an $l$-encounter touching S$_2$ by the contribution $z_{o,l}^{(\mathrm{n},1)}$ of an $l$-encounter touching the neck. Thus
\begin{align}
f_1=&-\rmi\fr{N_{\mathrm{S}_1}}{N^{(1)}}\rme^{-\rmi\phi/2}+\fr{N_\mathrm{n}}{N^{(1)}}\rme^{\rmi\en\delta\tau}f_2 \nonumber \\
 &+\sul{l=2}{\infty}\cbr{x_{1,l}f_1^l\hat{f}_1^{l-1}+\rbr{z_{o,l}^{(1)}+z_{o,l}^{(\mathrm{n},1)}}\hat{f}_1^{l-1}}
\label{eq:generating_func_general_double_dot}
\end{align}
where $N^{(i)}=N_i+N_{\mathrm{S}_i}+N_\mathrm{n}$, $i\in\{1,2\}$.

Due to the assumption that the neck may be represented by an ideal lead an $l$-encounter may touch the neck even if the odd numbered side trees do not have zero characteristic. The only restriction to them is that they traverse the neck before having an encounter or hitting a superconductor. Thus when sliding an $l$-encounter into the neck we get the situation depicted in \fref{fig:encounter_neck_touch_b}: The odd numbered side trees, which start with an electron, now start in dot 2 instead of dot 1. Moreover there are $l$ path pairs traversing the neck each giving a phase $\en\delta\tau$. Therefore we get
\begin{equation}
z_{o,l}^{(\mathrm{n},1)}=\fr{N_\mathrm{n}}{N^{(1)}}\rme^{\rmi l\en\delta\tau}f_2^l\tilde{r}^{l-1}.
\label{eq:parallelogram_generating_neckhit}
\end{equation}
$x_{1,l}$ and $z_{o,l}^{(1)}$ are obtained in the same way as in \sref{cond_sidetrees} and are given by
\addtocounter{equation}{1}
\begin{align}
x_{1,l}=&-\fr{\rbr{1+\rmi l\en}\tilde{r}^{l-1}}{\rbr{1+\rmi\en}^l},
\tag{\theequation a}
\label{eq:parallelogram_generating_encounter_in_dot} \\
z_{o,l}^{(1)}=&\fr{N_{\mathrm{S}_1}}{N^{(1)}}\rbr{-\rmi}^l\rme^{-\rmi l\phi/2}\tilde{r}^{l-1}.
\tag{\theequation b}
\label{eq:parallelogram_generating_superconductorhit}
\end{align}

The remaining steps then are again the same as in \sref{cond_sidetrees}. For a side tree starting with an electron in dot 2 we have to exchange the labels `1' and `2' as well as the phase $\phi$ by $-\phi$. If we consider side trees starting with a hole in dot $i$ instead of electrons we have also to reverse the phase with respect to $f_i$ and exchange $f_j\leftrightarrow\hat{f}_j$, $j\in\{1,2\}$. All in all after performing the sums using geometric series, making the change of variables \eref{eq:variable_change_side_tree} and setting $r=1$ to get the side tree contributions $P_i^\alpha$ out of $g_i$ and $\hat{g}_i$, we obtain
\begin{align}
\fr{\rbr{-P_1^eP_1^h-\rmi\en\rbr{P_1^eP_1^h}^2+2\rmi\en P_1^eP_1^h-1}P_1^eP_1^h}{\rbr{1-P_1^eP_1^h}^2}& \nonumber \\
+\fr{1}{N^{(1)}}\cbr{\fr{N_{\mathrm{S}_1}P_1^h}{\rbr{\rmi P_1^h+\rme^{\rmi\phi/2}}}+\fr{N_\mathrm{n}P_1^hP_2^e}{\rbr{P_1^hP_2^e-\rme^{-\rmi\en\delta\tau}}}}&=0
\label{double_dot_side_tree_contr_P1e}
\end{align}
and similar equations for $P_1^h$, $P_2^e$ and $P_2^h$.

For simplicity in the following we consider zero temperature, \textit{i.e.}~$\en=0$, and equal leads $N_1=N_2=N_\mathrm{N}/2$ and $N_{\mathrm{S}_1}=N_{\mathrm{S}_2}=N_{\mathrm{S}}/2$. Then one finds that $P_2^\alpha=P_1^{\bar{\alpha}}$ where $\bar{\alpha}$ labels a hole if $\alpha$ labels an electron and vice versa. Moreover in this case exchanging electrons and holes is the same as exchanging dot $1$ and dot $2$ and is related to the fact that when exchanging electrons and holes this is essentially an exchange of $\phi\leftrightarrow-\phi$. This also reduces the number of linear equations for the normalised transmission coefficients by a factor of 2 since it yields $A_{ij}^e=A_{\bar{i}\bar{j}}^h$ and $B_{ij}^e=B_{\bar{i}\bar{j}}^h$ where $\bar{i}=2$ if $i=1$ and vice versa. Therefore we have $\tilde{t}_{11}^{ee}=\tilde{t}_{22}^{hh}$ etc. The set of 4 nonlinear equations indicated by \eref{double_dot_side_tree_contr_P1e} decomposes into two copies of sets of two nonlinear equations where the second set is the same as the first one.

\begin{figure}
 \subfigure[\label{fig:parallelogram_conductance_results_a}]{\includegraphics[width=0.45\columnwidth]{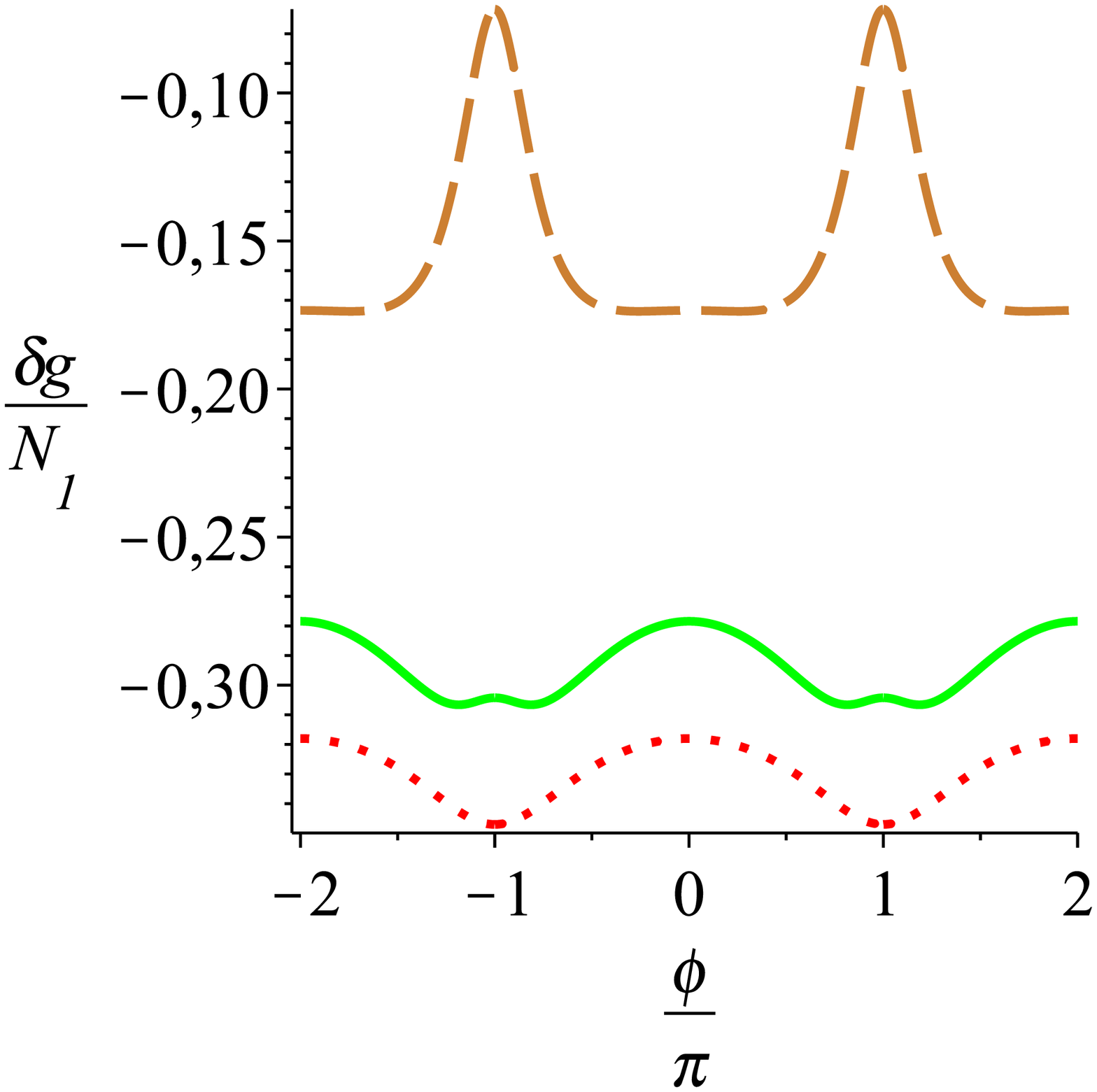}}
 \rule{0.033\columnwidth}{0pt}
 \subfigure[\label{fig:parallelogram_conductance_results_b}]{\includegraphics[width=0.45\columnwidth]{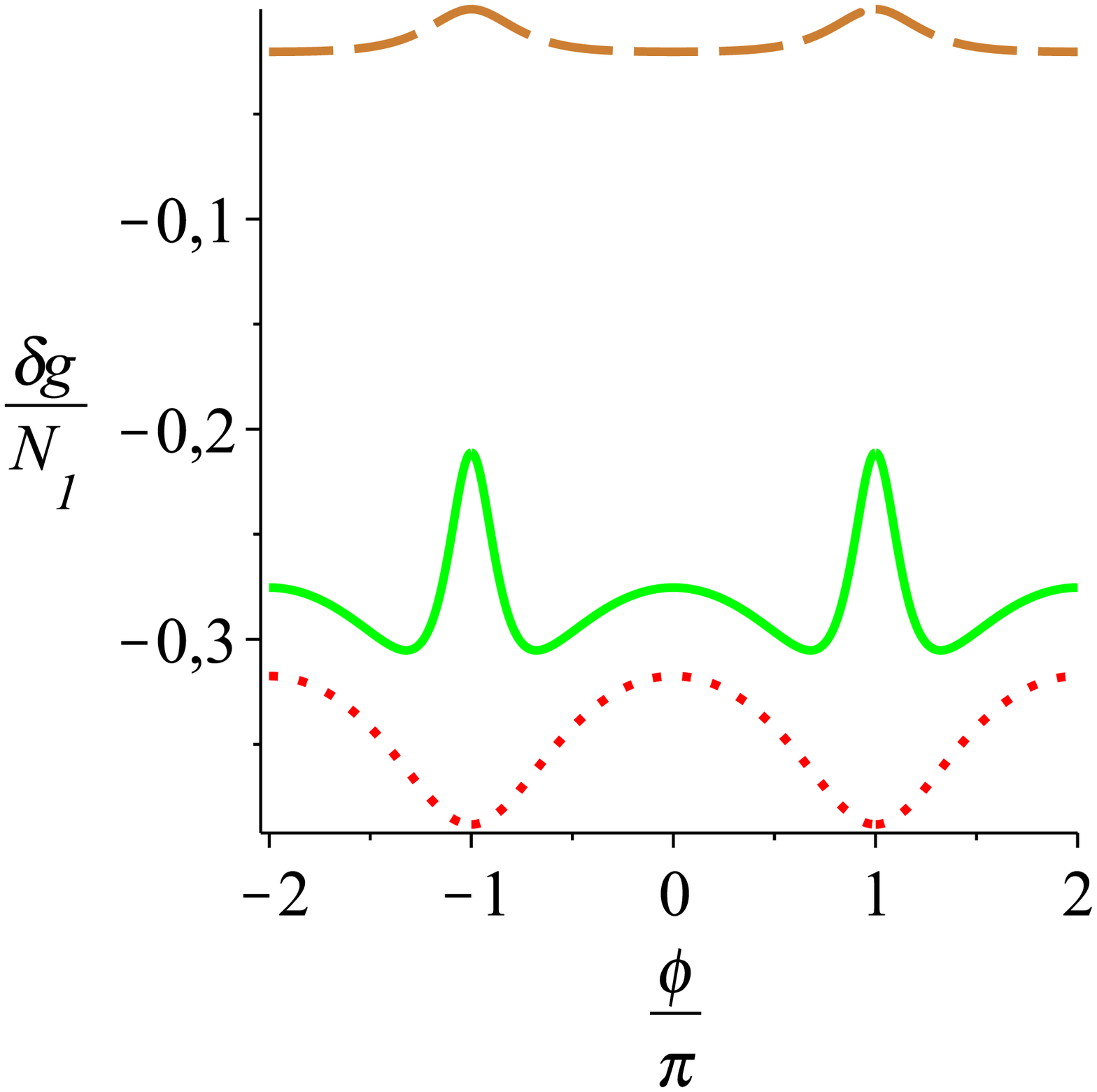}}
 \\
 \subfigure[\label{fig:parallelogram_conductance_results_c}]{\includegraphics[width=0.45\columnwidth]{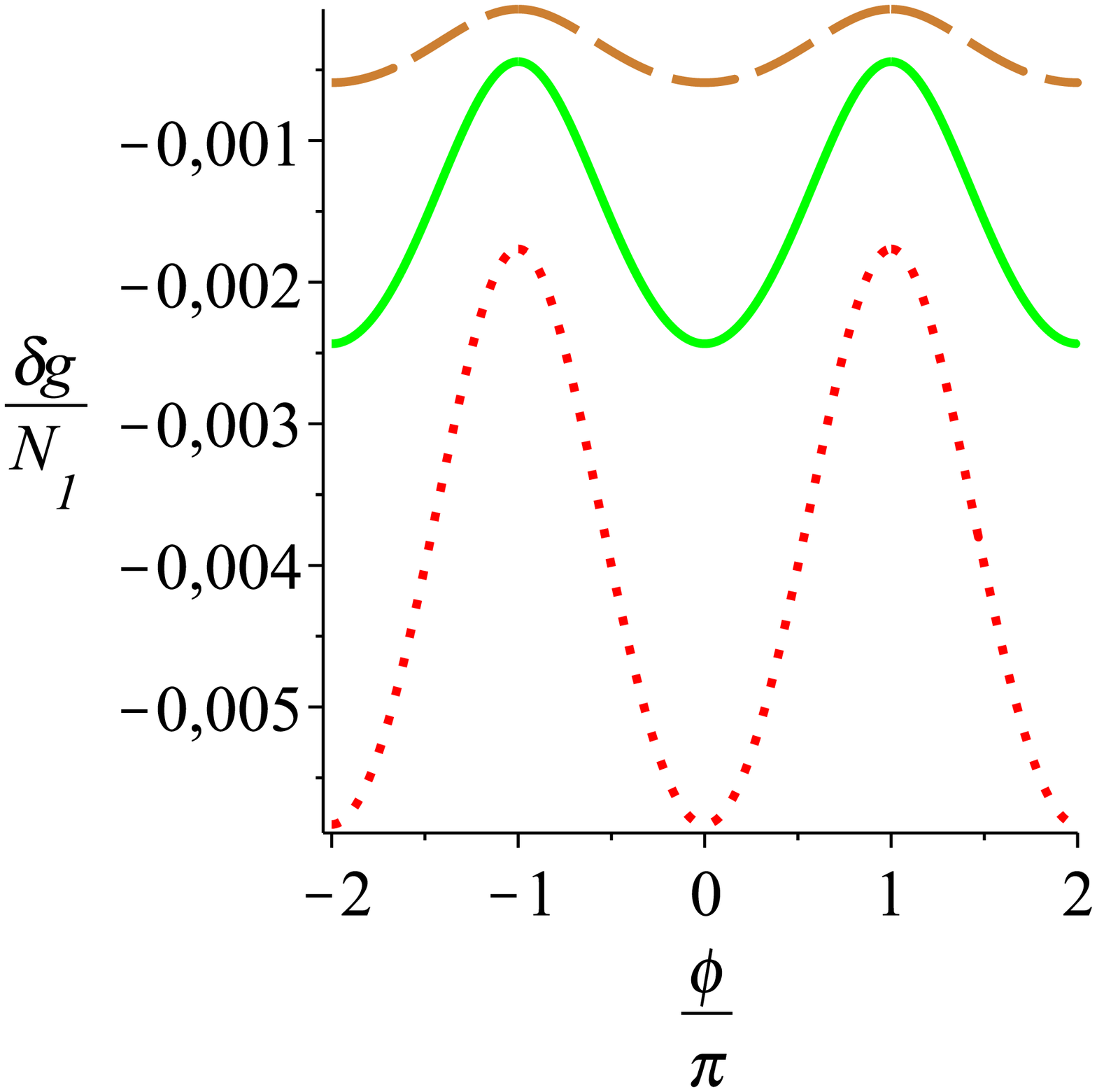}}
 \rule{0.033\columnwidth}{0pt}
 \subfigure[\label{fig:parallelogram_conductance_results_d}]{\includegraphics[width=0.45\columnwidth]{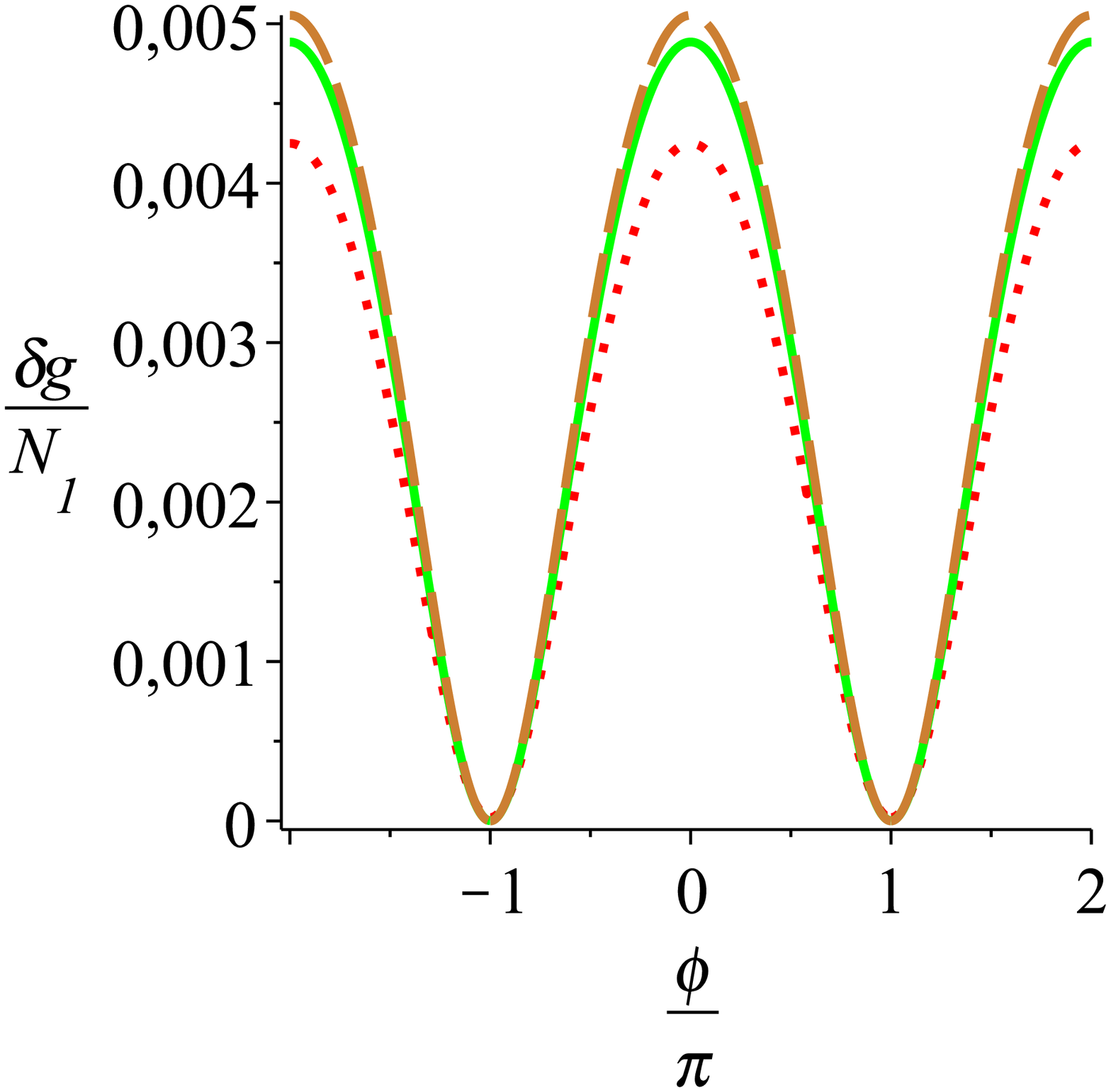}}
 \caption{\label{fig:parallelogram_conductance_results} The conductance correction of the dobule dot setup wit (a) $n=0.5$ and $x=5$ (dashed line), $x=20$ (solid line) and $x=100$ (dotted line). (b) $n=20$ and $x=1$ (dashed line), $x=20$ (solid line) and $x=100$ (dotted line). (c) $x=0.2$ and $n=0.3$ (dotted line), $n=0.5$ (solid line) and $n=0.7$ (dashed line). (d) $n=5$ (dotted line), $n=20$ (solid line) and $n=100$ (dashed line).}
\end{figure}
The classical conductance can be found by using a similar recursion relation for the `normalised' transmission coefficient as above which is for the diagonal diagrams given by
\begin{align}
 \tilde{t}_{ij}^{\alpha\beta}=\fr{N_i\delta_{ij}}{N^{(i)}}+\fr{N_{\mathrm{S}_j}}{N^{(j)}}\tilde{t}_{ij}^{\alpha\bar{\beta}}+\fr{N_\mathrm{n}}{N^{(i)}}\tilde{t}_{i\bar{j}}^{\alpha\beta}.
 \label{eq:parallelogram_diagonal_recursion}
\end{align}
With the classical conductance obtained using this recursion, $g_{cl}=N_\mathrm{N}(2n+x)/(2+4x+8n)$, where $n=N_\mathrm{n}/N_\mathrm{N}$, $x=N_\mathrm{S}/N_\mathrm{N}$ and $N_\mathrm{S}=N_{\mathrm{S}_1}+N_{\mathrm{S}_2}$, at zero temperature we get the conductance correction shown in \fref{fig:parallelogram_conductance_results}. The conductance correction is symmetric in the phase difference due to the symmetry in exchanging the two dots and the symmetry with respect to the exchange of electrons and holes. Moreover we found that the conductance correction is periodic in the phase difference with a period of $2\pi$ as it can be seen from \eref{eq:double_dot_first_even_encounter_contribution11}-\eref{eq:double_dot_transmission_system_11ee} and \eref{double_dot_side_tree_contr_P1e}: If we replace $\phi\rightarrow\phi+2\pi$ the side tree contributions change their sign. However in $A_{ij,l}^\alpha$ and the first and the last term of $B_{ij,l}^\alpha$ in \eref{eq:double_dot_first_odd_encounter_contribution11} this sign cancels since the total number of side tree contributions entering them is always even. An additional minus sign enters the second term of $B_{ij,l}^\alpha$ if $l$ is odd, but when $l$ is odd the phase factor of this term also contributes an additional minus sign. If $l$ is even the total number of side trees entering this term is again even and the phase factor does not contribute an additional minus sign on increasing the phase difference by $2\pi$. Thus the transmission coefficients are symmetric under replacing $\phi$ by $\phi+2\pi$.

Moreover we find that for a small number of channels in the neck the conductance correction is negative. It increases if $n$ is increased and finally becomes positive. For such values of $n$ where the conductance correction is positive we also find that the conductance correction vanishes for $\phi=\pm\pi$. For smaller values of $n$ the changes in the conductance correction when changing $x$ are most pronounced around $\pm\pi$ as it can be seen from figures \ref{fig:parallelogram_conductance_results}a,b.

\section{Thermopower of the symmetric and asymmetric house}
\label{sec:thermopower}
If the normal leads additionally have different temperatures there is also a coupling between the electrical current and the temperature difference. For a two terminal setup with the superconductors being isolated and the two normal leads having a temperature difference $\Delta T$ and a voltage difference $\Delta V$ the electrical and thermal current are given by \cite{ref:general_current_voltage_temperature_dependence}
\begin{equation}
 \rbr{\begin{array}{c}
  I \\ Q
 \end{array}}=
 \rbr{\begin{array}{cc}
       G & B \\
       \Gamma & \Xi
      \end{array}}
\rbr{\begin{array}{c}
      \Delta V \\ \Delta T
     \end{array}},
\label{eq:reduced_currents_island}
\end{equation}
where $G=2\rme^2g/h$ is again the conductance essentially given by \eref{nsn_island_current} and
\addtocounter{equation}{1}
\begin{align}
B=&\fr{2\rme}{\rmh T}\rbr{\bar{T}_{12}^{ee}-\bar{T}_{12}^{he}-2\fr{\rbr{\bar{T}_{11}^{he}+\bar{T}_{21}^{he}}\rbr{\hat{T}_{11}^{he}+\hat{T}_{12}^{he}}}{\hat{T}_{11}^{he}+\hat{T}_{22}^{he}+\hat{T}_{21}^{he}+\hat{T}_{12}^{he}}}\tag{\theequation a}
\label{eq:thermopower_relevant_coefficients_B}\\
\Gamma=&-\fr{2\rme}{\rmh}\rbr{\bar{T}_{21}^{ee}+\bar{T}_{21}^{he}+2\fr{\rbr{\hat{T}_{11}^{he}+\hat{T}_{21}^{he}}\rbr{\bar{T}_{11}^{he}+\bar{T}_{12}^{he}}}{\hat{T}_{11}^{he}+\hat{T}_{22}^{he}+\hat{T}_{21}^{he}+\hat{T}_{12}^{he}}}\tag{\theequation b}
\label{eq:thermopower_relevant_coefficients_gamma}\\
\Xi=&-\fr{2}{\rmh T}\rbr{\hat{T}_{21}^{ee}+\hat{T}_{21}^{he}+2\fr{\rbr{\bar{T}_{11}^{he}+\bar{T}_{21}^{he}}\rbr{\bar{T}_{11}^{he}+\bar{T}_{12}^{he}}}{\hat{T}_{11}^{he}+\hat{T}_{22}^{he}+\hat{T}_{21}^{he}+\hat{T}_{12}^{he}}}\tag{\theequation c}
\label{eq:thermopower_relevant_coefficients_xi}
\end{align}
where $\hat{T}_{ij}^{\alpha\beta}=-\int_{-\infty}^{\infty}\rmd\en\rbr{\partial f/\partial\en}T_{ij}^{\alpha\beta}$ and $\bar{T}_{ij}^{\alpha\beta}=-\int_{-\infty}^{\infty}\rmd\en\,\en\rbr{\partial f/\partial\en}T_{ij}^{\alpha\beta}$. Note that in Ref.~\onlinecite{ref:general_current_voltage_temperature_dependence} these coefficients have been written down only for low temperatures using the Sommerfeld expansion.

An estimate of the thermo-electric coupling is for example provided by the thermopower
\begin{equation}
 S=-\left.\fr{1}{\rme}\pdiff{\mu}{T}\right|_{I=0}=-\left.\fr{\Delta V}{\Delta T}\right|_{I=0}=-\fr{B}{G}.
\label{eq:thermopower_final}
\end{equation}

The calculation of the thermopower is therefore closely related to the electrical transport since due to \eref{eq:thermopower_final} we have to evaluate the same transmission coefficients. We will first consider the two cases shown in \fref{fig:thermo_setups_sym} and \fref{fig:thermo_setups_asym} which are topological equivalent to the symmetric (\fref{fig:thermo_setups_sym}) and asymmetric house (\fref{fig:thermo_setups_asym}) \cite{ref:thermo_exp_house_parallelogram1,ref:thermo_exp_house_parallelogram2}.
We will restrict ourselves to Andreev billiards with two normal leads and two isolated superconducting leads with a phase difference $\phi$. The results presented here are again in leading order in the channel numbers and for the numbers of channels of the two normal leads being equal, since these numbers only enter by a prefactor $N_1N_2/(N_1+N_2)$. It is further assumed that no magnetic field is applied. Note that we consider the superconductors to be isolated thus adjusting their chemical potential such that the net current through the superconducting leads is zero.


\subsection{Symmetric house}
\label{subsubsec:symmetric_house}
We start with the setup Jacquod and Whitney called the symmetric house \cite{ref:nsnthermopower} shown in \fref{fig:thermo_setups_sym}. They treated the transmission coefficients perturbatively in the ratio $x=N_\mathrm{S}/N_\mathrm{N}$ up to second order. Within this approximation they argued that for the symmetric house with equal numbers of superconducting channels the thermopower vanishes in second order in $x$ since $B$, defined in \eref{eq:thermopower_relevant_coefficients_B}, is antisymmetric in exchanging electrons and holes which yields an exchange $\en\rightarrow-\en$ and in leading order in $N$ is equivalent to reversing the superconducting phase $\phi\rightarrow-\phi$. Since additionally the electrical conductance $G$ is symmetric as already seen in \sref{island_sec} the thermopower $S=-B/G$ is antisymmetric in the phase difference. On the other hand the result has to be symmetric under exchanging the superconducting leads which is again equivalent to reversing the sign of $\phi$. Thus the thermopower has to be zero. With our approach we find that this argument holds to all orders and may also be seen from the discussion in \sref{sec:cond_transmission_probs}: There we stated that as long as the superconducting leads both provide the same number of channels the transmission probability is symmetric in exchanging electrons and holes which when inserted in \eref{eq:thermopower_relevant_coefficients_B} gives $B=0$.

\begin{figure}
 \subfigure[\label{fig:sym_house_results_a}]{\includegraphics[width=0.3\columnwidth]{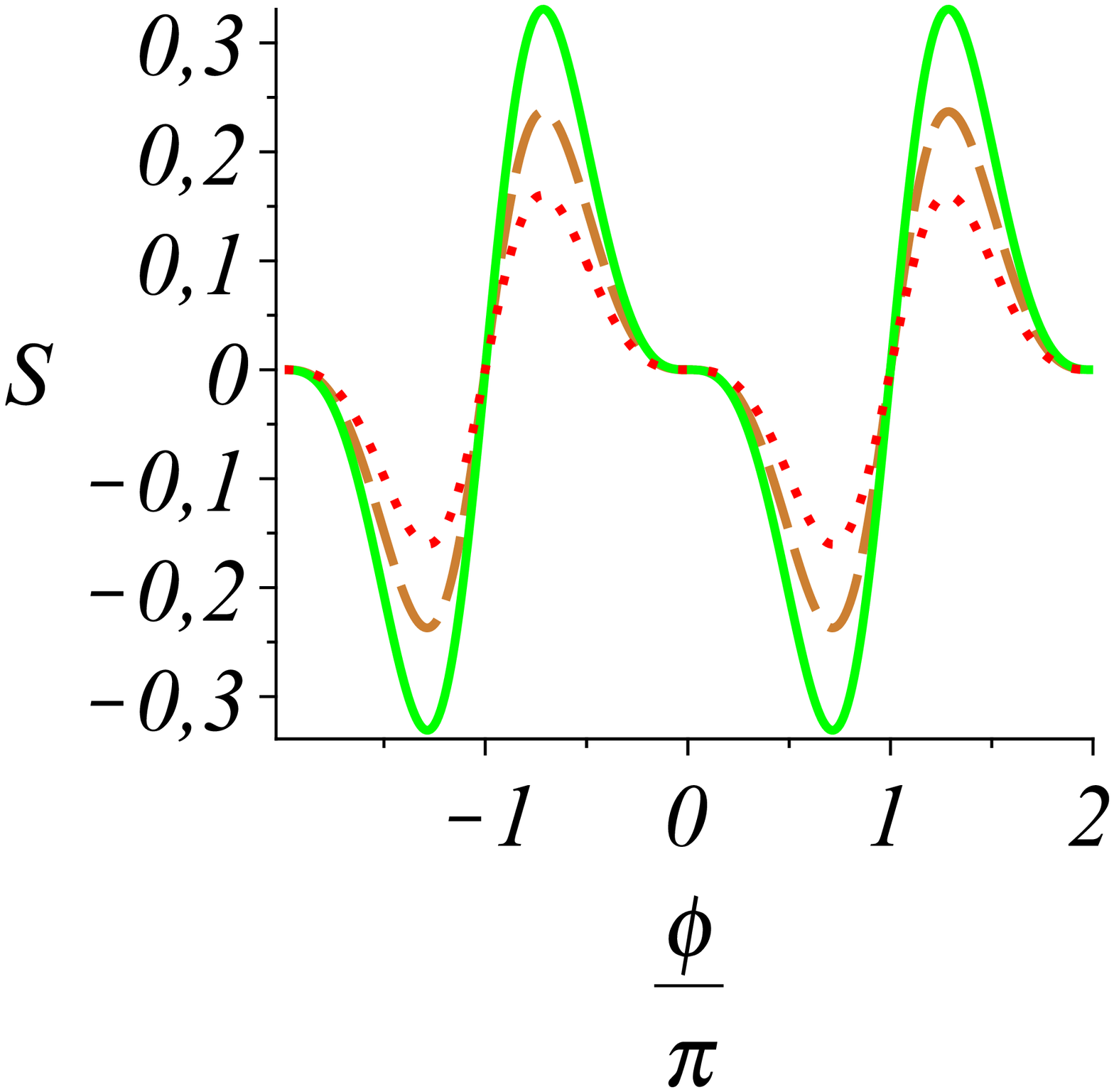}}
 \rule{0.025\columnwidth}{0pt}
 \subfigure[\label{fig:sym_house_results_b}]{\includegraphics[width=0.3\columnwidth]{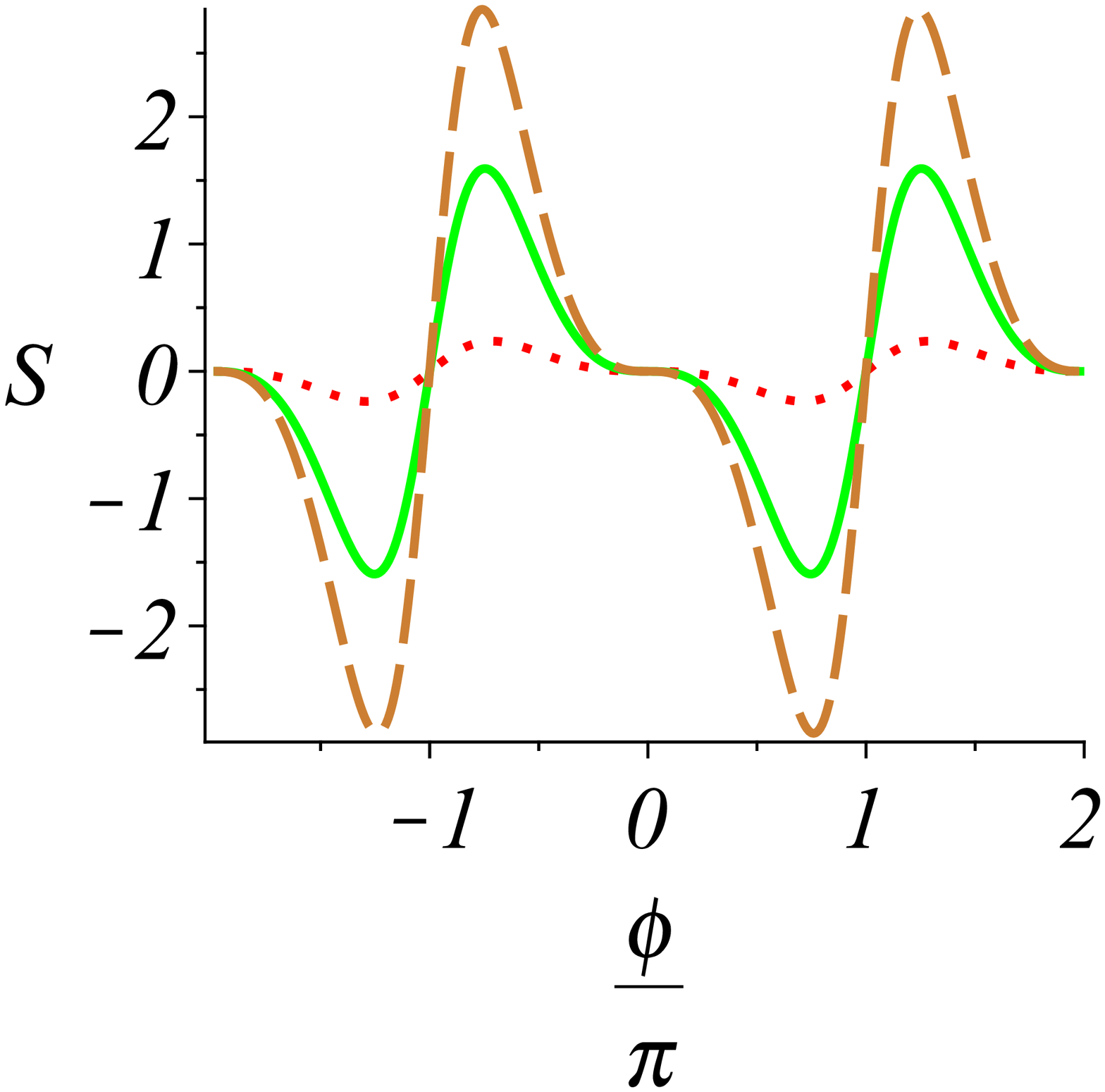}}
 \rule{0.025\columnwidth}{0pt}
 \subfigure[\label{fig:sym_house_results_c}]{\includegraphics[width=0.3\columnwidth]{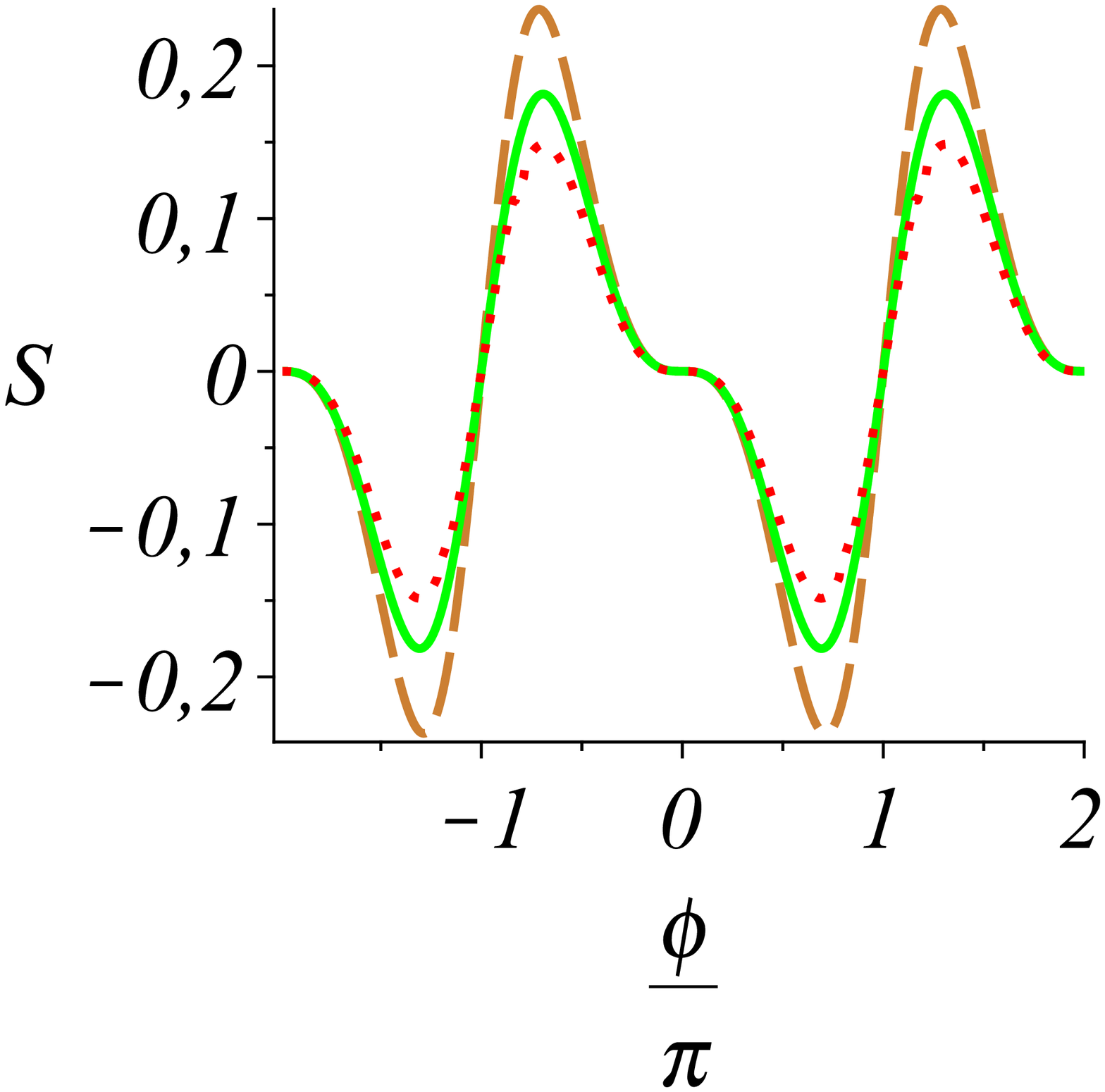}}
 \caption{\label{fig:sym_house_results} The thermopower of the symmetric house in units of $10^{-3}k_\mathrm{B}/\rme$ with different numbers of superconducting channels:
 (a) Dependence of the thermopower on the difference between the numbers of superconducting channels with $x=0.2$, $\theta=0.2$ and $y=0.3$ (dashed line), $y=0.5$ (solid line) and $y=0.9$ (dotted line). (b) Dependence on $x$ with $y=0.3$, $\theta=0.2$ and $x=0.2$ (dotted line), $x=0.5$ (solid line) and $x=0.7$ (dashed line). (c) Dependence on the temperature with $x=0.2$, $y=0.3$ and $\theta=0.2$ (dashed line), $\theta=1.5$ (solid line) and $\theta=2$ (dotted line).}
\end{figure}

However if $N_{\mathrm{S}_1}\neq N_{\mathrm{S}_2}$ the symmetry under exchanging the two superconductors is broken and the thermopower does not vanish. By using the transmission coefficients found in \sref{sec:cond_transmission_probs}, inserting into \eref{nsn_island_current} as already done in \sref{island_sec} and \eref{eq:thermopower_relevant_coefficients_B} and performing the integrals numerically with a total accuracy of $10^{-10}$ we find the results shown in \fref{fig:sym_house_results}.
The thermopower is found to be antisymmetric and $2\pi$-periodic in the phase difference $\phi$. When the difference $y=(N_{\mathrm{S}_1}-N_{\mathrm{S}_2})/N_\mathrm{S}$ is increased starting at zero a non-zero thermopower appears which increases up to about $y=0.5$. If the difference is increased further the resulting thermopower decreases again and vanishes at $y=1$ which corresponds to the case of just one superconductor. Increasing the total number of superconducting channels $x=N_\mathrm{S}/N_\mathrm{N}$ results in an increase of the thermopower while an increase in the temperature $\theta=k_\mathrm{B}T/E_\mathrm{T}$ causes a decrease of $S$. This is due to the fact that the thermopower of Andreev billiards is a pure quantum mechanical property since the diagonal approximation gives $S=0$.

\subsection{Asymmetric house}
\label{subsubsec:asymmetric_house}
Moreover for $N_{\mathrm{S}_1}=N_{\mathrm{S}_2}$ we can also generate a non-zero thermopower by inserting a neck at one of the two superconducting leads, say at S$_1$ as in \fref{fig:thermo_setups_asym}, in which the trajectories spend an additional time $\delta\tau\cdot\tau_\mathrm{D}$. An e-h pair hitting the superconductor S$_1$ picks up an additional phase $\en\delta\tau$. Thus the total phase provided by the neck plus the Andreev reflection at S$_1$ is $(-\phi+2\en\delta\tau)/2$ if an electron is converted into a hole and $(\phi+2\en\delta\tau)/2$ if a hole is converted into an electron. Therefore the electron hole symmetry is broken leading to an asymmetry in the phase difference as well as in the energy. So $T_{ij}^{\alpha\beta}\neq T_{ij}^{\bar{\alpha}\bar{\beta}}$ with $\bar{\alpha}$ ($\bar{\beta}$) labelling a hole if $\alpha$ ($\beta$) labels an electron and vice versa. Note that we treat the neck as an ideal lead such that every quasiparticle entering the neck hits the superconductor before leaving the neck again \cite{ref:semiclassical_anderson_localization}. When redoing the steps of sections \ref{cond_sidetrees} and \ref{sec:cond_transmission_probs} for the calculation of the transmission coefficients again one finally finds the following changes: The variable $\beta=\cos(\phi/2)$ has to replaced by $\beta^e=\cos[(\phi-\en\delta\tau)/2]$ for a side tree starting with an electron and by $\beta^h=\cos[(\phi+\en\delta\tau)/2]$ for a side tree starting with a hole. Furthermore the side trees starting with an electron are multiplied by $\rme^{-\rmi\en\delta\tau/2}$ and those starting with a hole by $\rme^{\rmi\en\delta\tau/2}$. However since in $A_l^\alpha$ and $B_l^\alpha$ each factor $P^\alpha$ is paired either with a factor $P^{\bar{\alpha}}$ or with a factor $\rbr{P^\alpha}^*$ these additional factors cancel. Additionally if the incident quasiparticle is an electron the phase $\phi$ in the second term of \eref{eq:coeff_definition_Be} also has to replaced by $\phi-2\en\delta\tau$. Again for an incident hole the phase $\phi$ has to be replaced by $-\phi$.

\begin{figure}
 \subfigure[\label{fig:asym_house_result_a}]{\includegraphics[width=0.3\columnwidth]{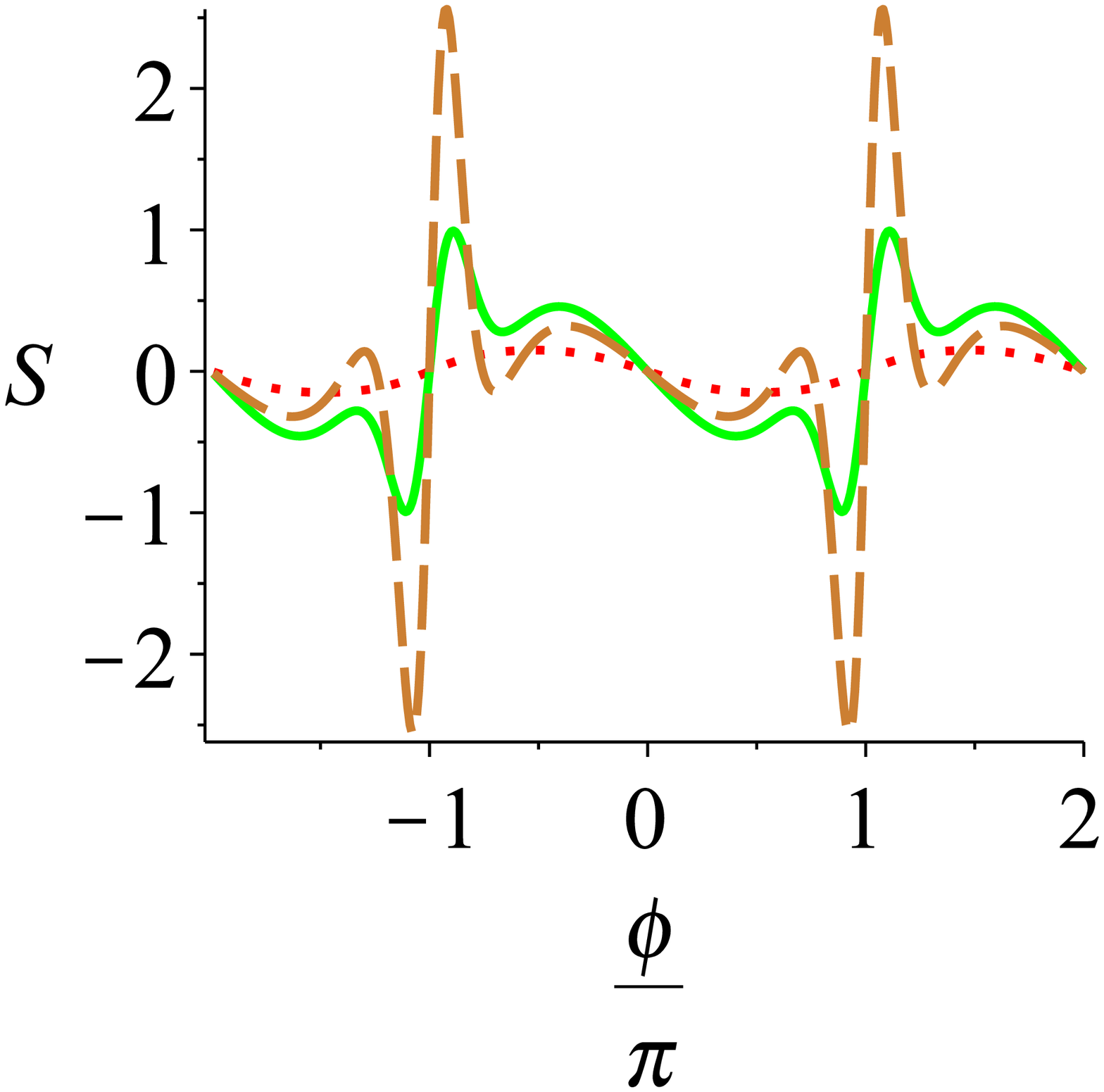}}
 \rule{0.025\columnwidth}{0pt}
 \subfigure[\label{fig:asym_house_result_b}]{\includegraphics[width=0.3\columnwidth]{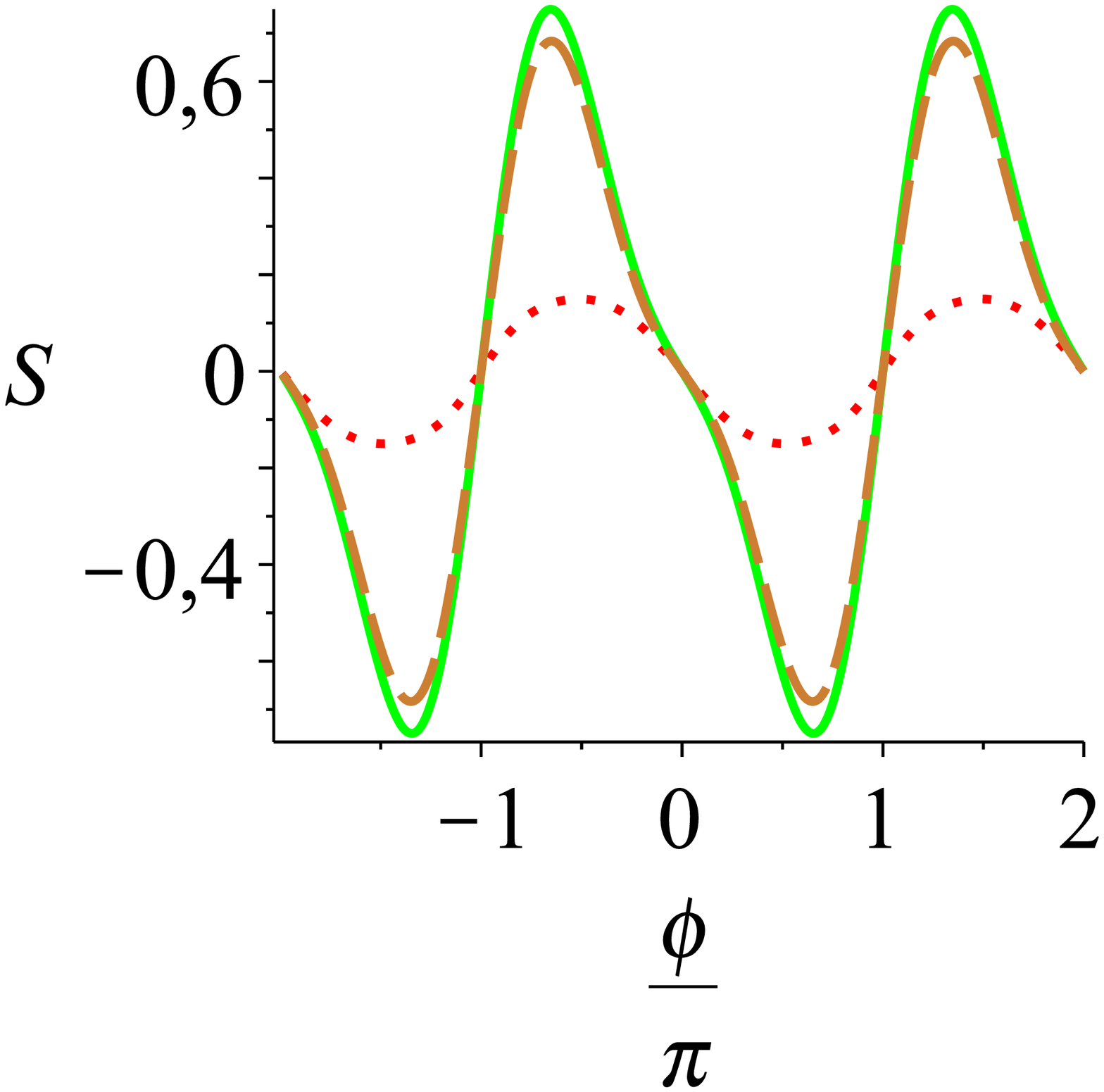}}
 \rule{0.025\columnwidth}{0pt}
 \subfigure[\label{fig:asym_house_result_c}]{\includegraphics[width=0.3\columnwidth]{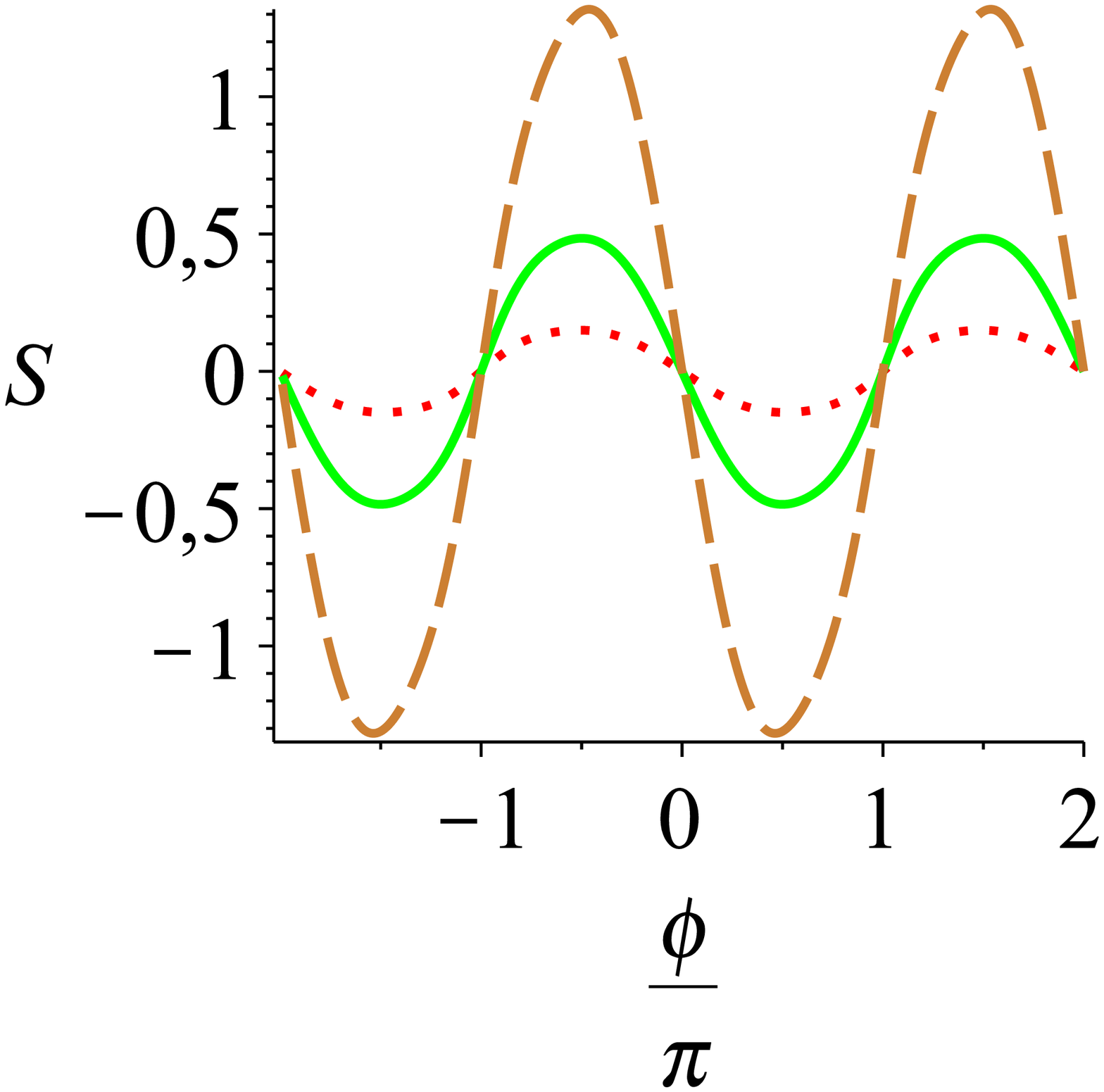}}
 \caption{\label{fig:asym_house_result} The thermopower in units of $10^{-3}k_\mathrm{B}/\rme$ as a function of the phase difference $\phi$. (a) $\delta\tau=0.03$, $\theta=K_\mathrm{B}T/E_\mathrm{T}=0.2$ and $x=0.2$ (dotted line), $x=2$ (solid line) and $x=10$ (dashed line). (b) $\delta\tau=0.03$, $x=0.2$ and $\theta=0.2$ (dotted line), $\theta=4$ (solid line) and $\theta=10$ (dashed line). (c) $\theta=0.2$, $x=0.2$ and $\delta\tau=0.03$ (dotted line), $\delta\tau=0.1$ (solid line) and $\delta\tau=0.3$ (dashed line).}
\end{figure}

All in all we have to solve a $4\times 4$ system of linear equations rather than a $2\times 2$ system of linear equations as it was for the conductance of the symmetric version. Using $N_1=N_2=N_\mathrm{N}/2$ the four equations are
\addtocounter{equation}{1}
\begin{align}
T_{ij}^{ee}=&\fr{N_\mathrm{N}^2}{4N}+\sul{l}{\infty}A_lT_{ij}^{ee}+\sul{l}{\infty}B_l^e T_{ij}^{eh},
\label{eq:thermopower_transmission_ee}
\tag{\theequation a}\\
T_{ij}^{he}=&\sul{l}{\infty}A_lT_{ij}^{he}+\sul{l}{\infty}B_l^e T_{ij}^{hh},
\label{eq:thermopower_transmission_he}
\tag{\theequation b}\\
T_{ij}^{hh}=&\fr{N_\mathrm{N}^2}{4N}+\sul{l}{\infty}A_lT_{ij}^{hh}+\sul{l}{\infty}B_l^h T_{ij}^{he},
\label{eq:thermopower_transmission_hh}
\tag{\theequation c}\\
T_{ij}^{eh}=&\sul{l}{\infty}A_lT_{ij}^{eh}+\sul{l}{\infty}B_l^h T_{ij}^{ee},
\label{eq:thermopower_transmission_eh}
\tag{\theequation d}
\end{align}
with $A_l$ being the same as in \eref{eq:coeff_definition_Ae} but with $P^e$ and $P^h$ redefined to depend on $\beta^e$ and $\beta^h$, respectively, and $b=0$. Analogously one finds
\begin{widetext}
\addtocounter{equation}{1}
\begin{align}
B_l^e=-\sul{p=0}{l-2}\Bigg[&\rbr{1+\rmi\rbr{2p-l+2}\en}\rbr{P^e}^{p+1}\rbr{P^h}^p\cbr{\rbr{P^e}^*}^{l-p-1}\cbr{\rbr{P^h}^*}^{l-p-2} \nonumber \\
&-\fr{x\rbr{-\rmi\rbr{P^h}}^p\rbr{\rmi\rbr{P^h}^*}^{l-p-2}}{2(1+x)}\rbr{\rme^{-\rmi\rbr{2p-l+2}\rbr{\phi-2\en\delta\tau}/2}+\rme^{\rmi\rbr{2p-l+2}\phi/2}}\Bigg],
\label{eq:Bfactors_thermopower_e}
\tag{\theequation a}\\
B_l^h=-\sul{p=0}{l-2}\Bigg[&\rbr{1+\rmi\rbr{2p-l+2}\en}\rbr{P^h}^{p+1}\rbr{P^e}^p\cbr{\rbr{P^h}^*}^{l-p-1}\cbr{\rbr{P^e}^*}^{l-p-2} \nonumber \\
&-\fr{x\rbr{-\rmi\rbr{P^e}}^p\rbr{\rmi\rbr{P^e}^*}^{l-p-2}}{2(1+x)}\rbr{\rme^{\rmi\rbr{2p-l+2}\rbr{\phi+2\en\delta\tau}/2}+\rme^{-\rmi\rbr{2p-l+2}\phi/2}}\Bigg].
\label{eq:Bfactors_thermopower_h}
\tag{\theequation b}
\end{align}
\end{widetext}
By inserting the transmission coefficients into \eref{nsn_island_current} and \eref{eq:thermopower_relevant_coefficients_B} we find the thermopower $S=-B/G$. To this end we again integrated the transmission coefficients numerically using Gaussian quadrature with a total accuracy of $10^{-10}$. The results for the thermopower for different values of $\delta\tau$, $x$ and different temperatures is shown in \fref{fig:asym_house_result}.

We find that the antisymmetry in phase found by Whitney and Jacquod in second order in $x$ holds up to all orders in $x$. However this antisymmetry is in contradiction to previous experimental measurements \cite{ref:thermo_exp_house_parallelogram2,ref:thermo_exp_house_parallelogram1} on diffusive normal metal regions coupled to two superconductors. Moreover one can see from \fref{fig:asym_house_result} that the thermopower is $2\pi$-periodic in $\phi$. The arguments for the periodicity of the conductance in \sref{island_sec} also apply to $B$ and thus the thermopower is periodic in the phase difference $\phi$ with period $2\pi$. A period of $2\pi$ had also been found previously in Ref.~\onlinecite{ref:theo_thermopower_osc}. The antisymmetry and the periodicity may also be obtained by \earef{eq:Bfactors_thermopower_e}{,b} combined with \eref{nsn_island_current} and \eref{eq:thermopower_relevant_coefficients_B}: Due to the summation over $p$, $B_l^e$ and $B_l^h$ are symmetric under an simultaneous exchange $\phi\leftrightarrow-\phi$ and $\en\leftrightarrow-\en$ and satisfy $B_l^h=B_l^e\vert_{\en\rightarrow-\en}$. Thus replacing $\phi$ by $-\phi$ is the same as replacing $\en$ by $-\en$. Now if we replace $\en$ by $-\en$ we get an additional minus sign in \eref{eq:thermopower_relevant_coefficients_B} and therefore a minus sign in the thermopower. In contrast to the symmetric case however the symmetry due to the exchange of the leads is broken and thus the thermopower is non-zero but antisymmetric in $\phi$. Furthermore for specific combinations of $x$ and $\delta\tau$ the thermopower as a function of the phase difference may show additional oscillations with period smaller than $2\pi$ (see \fref{fig:asym_house_result_a}). However these additional oscillations are smoothed out if the temperature is increased.

Over a pretty wide range of increasing the temperature the thermopower also increases before it shrinks again. This increase of the thermopower with increasing temperature is related to the decrease of the thermal resistance found in Ref.~\onlinecite{ref:thermo_exp_house_parallelogram2}. \Fref{fig:asym_house_result_c} also shows that the thermopower increases if the time a quasiparticle spends in the neck is increased. This is due to the fact that with increasing $\delta\tau$ the electron-hole symmetry is increasingly broken.

It should be noted that the thermopower again arises solely from the non-diagonal diagrams since for the diagonal diagrams the energy differences of the links are always zero and thus the neck does not play any role.

\section{Conclusions}
We have shown in this article that the transport properties of chaotic cavities are strongly affected by the introduction of superconducting leads.
In particular, using a semiclassical framework, we have extended the work of Ref.~\onlinecite{nsntrans} to all orders in the ratio $N_\mathrm{S}/N_\mathrm{N}$ of the sizes of the superconducting and normal leads. With this we could reproduce the large-$N$ limit of the random matrix theory result \cite{rmttrans} for two isolated superconductors which shows in the limit $N_\mathrm{S}/N_\mathrm{N}\rightarrow\infty$ the conductance doubling of the N-S interface consisting of one normal conducting and one superconducting lead \cite{btktheory}. Our result shows a $2\pi$-periodicity and symmetry with respect to the phase difference of the superconductors which have already been found by other approaches \cite{islandconductance,ref:island_symmetry}. If a magnetic field is applied or the temperature is increased the conductance correction will decrease towards zero.

An Andreev billiard with two normal leads and two superconducting leads having the same chemical potential as one of the two normal leads shows more interesting features than the setup with superconducting islands. Depending on the numbers of channels, these may be similar to those already found previously in the conductance of N-S junctions. Again the conductance is doubled in the limit $N_\mathrm{S}/N_\mathrm{N}\rightarrow\infty$ and is $2\pi$-periodic and symmetric with respect to the phase difference $\phi$ which has been observed for N-S junctions with two superconducting leads in several approaches \cite{circuit_nss,ref:lead_periodicity,ref:lead_periodicity2,nsntrans}. However the conductance shows a non-monotonic behavior similar to the conductance through quantum dots with one normal and two superconducting leads \cite{circuit_nss}. The phase difference may even cause a change of the sign of the conductance correction.  Furthermore the magnetic field dependence also inherits a non-monotonic behavior from the N-S junction \cite{cond_doubling,mag_nonmonotonic} and again the sign of the conductance correction may be changed by increasing the magnetic field. This has already been found by Whitney and Jacquod \cite{nsntrans} in their consideration of the contribution up to second order in $N_\mathrm{S}/N_\mathrm{N}$. A non zero temperature may even cause an increase of the conductance correction similar to the reentrance in the case of the N-S junction \cite{reentrance_exp1,reentrance_exp2}.

Separating the Andreev billiard into two dots each consisting of one normal and one superconducting lead and connecting them by a neck necessitates a splitting of the diagrams when traversing the neck. A small number of channels in the neck causes a reduction of the conductance while for a large neck the superconductors enhances the conductance as long as the phase difference between the two superconducting leads $\phi\neq\pi$. The conductance of this setup is again symmetric and periodic in $\phi$ with a period of $2\pi$.

The investigations of the thermopower in Ref.~\onlinecite{ref:nsnthermopower} have also been extended to all orders in the number of superconducting channels for the symmetric house and the asymmetric house. For these two setups we could show that the antisymmetry of the thermopower towards the phase difference $\phi$ holds in all orders in $N_\mathrm{S}/N_\mathrm{N}$ and that the thermopower of the symmetric house with equal numbers of superconducting channels is identically $0$. In the case of the asymmetric house this antisymmetry however is in contradiction with some previous experimental results \cite{ref:thermo_exp_house_parallelogram2,ref:thermo_exp_house_parallelogram1}, which found a symmetric thermopower for diffusive normal regions. But the thermopower oscillates in both cases with a period $2\pi$ in agreement with experimental results \cite{ref:thermo_exp_house_parallelogram1,ref:thermo_exp_house_parallelogram2}. Additionally the thermopower increases with increasing temperature over a pretty wide range. With the treatment represented in \sref{sec:parallelogram_conductance} where we calculated the conductance of the double dot we would in principle be able to calculate also the thermopower of this setup. However including the necessary energy difference and the difference between the numbers of channels of the two superconductors increases the complexity of the equations to such an extent that we have not got a reasonable solution of the four coupled nonlinear equations for the side tree contributions.

Like for the the density of states \cite{my_dos}, the considerations in this article show that in the case of normal metal-superconductor hybrid systems, a semiclassical treatment based on the diagonal approximation is not sufficient to describe such systems. Instead one has to consider all diagrams consisting of path pairs and encounters contributing in leading order in the channel number. Importantly it shows that the effect of superconductors on attached normal regions is of the order of the total number of channels and the leading order corrections in $N_\mathrm{S}/N_\mathrm{N}$ compete with higher order terms for larger numbers of superconducting channels.

However up to now there are neither confirming nor disproving experimental data. Thus it would be interesting to see results of measurements upon the conductance of Andreev billiards. While the realization of ballistic Andreev systems based on InAs has been experimentally shown \cite{ref:mag_field_dep_andreev_refl}, corresponding transport measurements have yet to be performed.

The results presented here are only valid in leading order in the inverse channel numbers, and further investigations could include the calculation of subleading order terms, especially the weak localisation correction. Doing this first of all requires a way of systematically finding all the diagrams contributing in the next order in $1/N$. The key problem then is to find a suitable recursion since the structures found here break down. Namely the trees have no longer to be trees and the diagrams may not necessarily have a backbone. However the types of recursion relations presented here would prove useful in such a case. We may also wonder about the effects of superconductors on the noise and other finer transport statistics.

Here we considered Andreev billiards coupled to at most two superconductors. However extending our calculation to a higher number of superconductors seems (maybe up to solving the equation found for the side tree contributions) straight forward. We also restricted ourselves to zero Ehrenfest time. However it has been shown that the Ehrenfest time plays a crucial role in Andreev billiards \cite{ln98}. Allowing non-zero Ehrenfest time yields a simple replacement for the side trees \cite{waltneretal10,my_dos} but the effect on the backbone has to be investigated further.

\begin{acknowledgements}
 We thank Philippe Jacquod for useful conversations. The work was supported by the Deutsche Forschungsgemeinschaft through the Research Group FOR 760 ``Scattering systems with complex dynamics'' (TE, KR) and by the Alexander von Humboldt-Foundation (JK).
\end{acknowledgements}

\bibliography{transport}
 \bibliographystyle{apsrev4-1}
\end{document}